\documentclass[9pt,twocolumn,twoside]{pnas-new} 

\templatetype{pnasresearcharticle}

\title{Vulnerability of democratic electoral systems}

\author[a,b,1]{Tomasz Raducha}
\author[c]{Jaros\l{}aw Klamut} 
\author[d,e]{Roger Cremades}
\author[f]{Paul Bouman}
\author[g,2]{Mateusz Wili\'nski}

\affil[a]{Grupo Interdisciplinar de Sistemas Complejos (GISC), Departamento de Matem\'aticas, Universidad Carlos III de Madrid, 28911 Legan\'es, Spain}
\affil[b]{Instituto de F\'isica Interdisciplinar y Sistemas Complejos IFISC (CSIC-UIB), E-07122 Palma, Spain}
\affil[c]{Faculty of Physics, University of Warsaw, 02-093 Warsaw, Poland}
\affil[d]{Department of Social Sciences, Wageningen University \& Research, 6706 KN Wageningen, the Netherlands}
\affil[e]{Fondazione Eni Enrico Mattei, 30135 Venice, Italy}
\affil[f]{Erasmus School of Economics, Erasmus University, 3062 PA Rotterdam, The Netherlands}
\affil[g]{Theoretical Division, Los Alamos National Laboratory, 87544 Los Alamos, USA}

\leadauthor{Raducha}

\significancestatement{
Different countries use different electoral systems (ES) to perform elections. 
Although they are often compared, there is no consensus on the best ES and 
their susceptibility to external influences is mostly neglected. We develop 
new measures of electoral systems’ vulnerability to different means of influence.
We show that plurality voting systems (e.g. first-past-the-post)
are less stable than proportional representation systems (e.g. party list). 
They are also more susceptible to political agitators and media propaganda.
We apply a novel computational methodology to evaluate real-world ES for possible improvements.
Our work provides a new tool for dealing with modern threats to democracy that
could destabilize voting processes. Moreover, our results
add quantitative arguments to the long-standing discussion on the evaluation of ES.
}

\authorcontributions{
TR and MW conceptualized the study, all authors discussed and designed the research, TR, MW, JK, and PB developed the software, TR, JK, and RC collected the data on electoral systems, TR, PB, and MW collected the data on mobility, TR ran the simulations, prepared the figures and wrote the initial manuscript, all authors revised and corrected the final version of the manuscript.
}
\authordeclaration{The authors declare no competing interests.}
\correspondingauthor{\textsuperscript{1}To whom correspondence should be addressed. E-mail: raducha.tomasz@gmail.com\\
\textsuperscript{2}To whom correspondence should be addressed. E-mail: wilinski.mateusz@gmail.com}


\begin{abstract} 
The two most common types of electoral systems (ES) used in electing national legislatures
are proportional representation and plurality voting
\cite{bormann2013democratic,farrell2011electoral}. When they are evaluated,
most often the arguments come from social choice theory and political sciences. 
The former overall uses an axiomatic approach including a list of mathematical criteria
a system should fulfill \cite{urken1995classics,sen1995judge}. The latter 
predominantly focuses on the trade-off between proportionality of apportionment 
and governability \cite{monroe1994disproportionality,carey2011electoral}. 
However, there is no consensus on the best ES \cite{bowler2005expert,farrell1999british},
nor on the set of indexes and measures that would be the most important 
in such assessment \cite{pennisi1998disproportionality}.
Moreover, the ongoing debate about the fairness of national elections neglects 
the study of their vulnerabilities \cite{blau2004fairness,plescia2020people}. 
Here we address this research gap
with a framework that can measure
electoral systems’ vulnerability to different means of influence.
Using in silico analysis we show that plurality voting systems are less 
stable than proportional representation. They are also more susceptible 
to political agitators and media propaganda. A review of real-world ES 
reveals possible improvements in their design leading to lower susceptibility.
Additionally, our simulation framework allows computation of popular indexes,
as the Gallagher index or the effective number of parties, in different scenarios.
Our work provides a new tool for dealing with modern threats to democracy 
that could destabilize voting processes \cite{hyde2020democracy}. Furthermore, 
our results add an important argument in a long-standing discussion on evaluation of ES.
\end{abstract}

\dates{This manuscript was compiled on \today}
\doi{\url{www.pnas.org/cgi/doi/10.1073/pnas.XXXXXXXXXX}}

\begin{document}

\maketitle
\thispagestyle{firststyle}
\ifthenelse{\boolean{shortarticle}}{\ifthenelse{\boolean{singlecolumn}}{\abscontentformatted}{\abscontent}}{}



\dropcap{D}emocracies across the globe are facing multiple pressing issues,
ranging from the erosion of public trust and persistent corruption scandals
to the pervasive spread of disinformation.
The vulnerability of democratic processes has become evident in the aftermath of
scandals related to Cambrige Analytica -- 2016 US elections, the Brexit referendum,
and elections in Kenya
\cite{persily20172016,laterza2018cambridge}.
The deceptive use of social 
media increased social and political polarization across world regions, as visible
in the United States, the European Union and several Asian countries
\cite{tan2020electoral,goodman2017new,grinberg2019fake,havertz2019right,yang2017challenges,pildes2011center,morgan2018fake,tucker2017liberation,dubois2019political}.
Furthermore, the threat of authoritarianism materializes through unequivocal
instances of fraud, exemplified by the Crimea referendum and the Belarus elections. 
These challenges are eroding democracy, the most frequent source of 
governmental power, and raise urgent questions about its vulnerabilities
\cite{bennett2018disinformation}. One of them is its susceptibility 
to external influence, which has been systematically overlooked in electoral systems research.

The phenomenon of democratic backsliding, or reverse democratic transition, 
has recently attracted research attention \cite{waldner2018unwelcome,druckman2023correcting}.
Organizations like The Economist Intelligence Unit and Freedom House
report worrying declines in democratic indexes \cite{demo_intex2019,house2017freedom}.
However, there is no agreement on the requirements of a good democracy, other than institutional.
As a result, some researchers argue that the global state of democracy is stable
and has improved considerably since the 1990s \cite{levitsky2015myth}.
Either way, there are many potential threats to democratic stability that can impact
civil liberties, functioning of government, political participation, political culture, or electoral processes.
When analyzing the latter aspect, fairness and stability of national 
elections are often pivotal \cite{wiesner2018stability,eliassi2020science}.
Indeed, fair elections lay at the core of representative democracy, however, 
the definition of fairness is unclear in this context \cite{blau2004fairness}.

Democratic states have countless ways of performing elections 
based on different electoral systems (ES).
Their design is a critical component of democratic governance, 
because it directly influences citizen representation and decision-making processes within society.
Electoral systems define the rules by which citizens elect their 
representatives and, consequently, determine the composition and 
functioning of legislative bodies. The need for systems that best
reflect the will of the people and foster effective governance is broadly recognized.
Yet, the stability of national elections and their security in 
the context of electoral systems is not well studied. It is in 
the public interest to study and understand how different ES
relate to different vulnerabilities and contemporary challenges.
We address this issue using a new
framework that enables us to simulate and analyze the influence of media
and zealots within different electoral systems.


\subsection*{Electoral systems and their evaluation}

In this study we focus on the two most common types of ES used to elect candidates to national
legislatures, namely proportional representation (PR) and plurality voting (PV) 
\cite{bormann2013democratic,farrell2011electoral} (see Supplementary Information 
for more detailed description of different ES). Proportional representation is 
the most frequently used voting system world-wide. It is normally associated with 
a party list, which can be either open or closed. PR systems usually
assume multiple multi-member constituencies (e.g. in elections to the Poland's Sejm), but
one nation-wide electoral district can also be encountered 
(e.g. in elections to the Israel's Knesset). The method of 
seat allocation within districts is a crucial part of PR, which may give
an advantage either to bigger or smaller parties. Often a global entry threshold is also specified.
Ranked versions of PR exist, for example single transferable vote (STV), but are rarely applied.

The second most common ES in national parliaments is plurality voting.
In this system the candidate or the party with the highest score wins, with no requirement to obtain a certain fraction
of votes. It usually involves dividing the country into single-member constituencies 
(e.g. elections to the India's House of the People or Poland's Senate) 
and is then called single member plurality (SMP) or first-past-the-post voting (FPTP). 
Multi-member constituencies are also possible in PV, for instance in 
party block voting (PBV) or single non-transferable vote (SNTV).
Although it is less common than FPTP, PBV is still used in some countries
(the most well known example
being the United States Electoral College).

The study of electoral systems dates back at least to the XVIII century and
the work of Nicolas de Condorcet,
who discovered the paradox of cyclic collective preferences \cite{young1988condorcet}.
Together with much later ideas of Kenneth Arrow \cite{arrow1950difficulty},
it laid the foundation of the modern social choice theory \cite{urken1995classics}.
It is concerned with aggregation of individual preferences into a single outcome and
provides a general mathematical understanding of such process. Many
reasonable criteria were established within social choice theory to be fulfilled by electoral systems,
but no system can satisfy all of them. This fact is usually described by
different \textit{impossibility theorems}. For example, the Arrow's impossibility theorem showed that no
aggregation mechanism can exist that satisfies all of some intuitively
desirable criteria. Later, the Gibbard-Satterthwaite and Gibbard theorems followed, showing
that single-outcome voting systems either have a dictatorial voter, are limited to two choices,
or are susceptible to strategic voting, meaning that casting a sincere ballot is not necessarily
aligned with the preferences of the voter.

This spawned a lot of further research, where alternative criteria were considered
and other impossibility theorems were developed.
For example, it has been argued that Arrow's impossibility theorem has little relevance when
there is a large group of voters, as a simple majority rule typically gives a good approximation
of desirable outcomes \cite{tullock1967general}.
Another example is the development of \emph{majority judgment}, an electoral system that aims to circumvent the impossibility theorems
by letting voters judge the suitability of the choices and a median rule is
used to limit strategic considerations for the voters \cite{balinski2007theory}.
Nevertheless, a big part of the work in social choice theory considers exclusively ranked voting, which
is still rare in national legislatures \cite{bormann2013democratic}. It is, 
therefore, not surprising that it does not provide an answer to which electoral
system is better. Furthermore, when arguing against or in favor of a given 
voting system, it is often possible to find mathematical criteria that
will back up the preferences \cite{sen1995judge}.
Ultimately, the theoretical study of voting systems and the empirical analysis
are two important and complementary approaches.

Without a doubt, the evaluation of electoral systems is a prominent 
topic in political sciences \cite{blais1996electoral}. A substantial part
of the research concentrates on disproportionality (also known as malapportionment),
i.e. the discrepancy between the fraction of votes gained and the fraction of seats assigned 
\cite{monroe1994disproportionality,pennisi1998disproportionality,taagepera2003mapping,koppel2009measuring}.
Several indexes have been proposed to evaluate malapportionment in electoral
systems, for example the Gallagher index 
\cite{gallagher1991proportionality,gallagher1992comparing,benoit2000electoral} 
or the Loosemore-Hanby index \cite{loosemore1971theoretical,fry1991note},
but there is no agreement on which index, or set of indices, is the most relevant 
\cite{pennisi1998disproportionality,shugart2001electoral}.
PR systems generally minimize the disproportionality between
the votes and seats gained, but it is not the only measure of
electoral system's quality (disproportionality also depends on
the seat assignment method, see Supplementary Information).

Many authors highlight
the trade-off between proportionality and \textit{governability}
-- the elected body should 
have enough power to be able to act \cite{sen1995judge}.
Perfectly proportional representation
of a diverse party list would result in no party having the ability
to rule and would force broad coalitions, which are often ineffective.
This political fragmentation can be measured by the index called effective number of parties
\cite{laakso1979effective}, which is higher for PR systems.
That is the main argument for FPTP or more generally plurality voting in the literature
\cite{blau2004fairness,pinto1999send}. Other advantages of PV commonly mentioned by scholars are:
having a more direct effect on the expulsion of prime
ministers and cabinets (it's easier to follow political activity of a single
chosen candidate in the local district and, perhaps, don't vote for them again)
and the translation of votes to power, which
is arguably more important than the translation of votes to seats
(i.e. the proportionality as usually defined).
On the other hand, PR usually results in a better minority representation.
Additionally, under PV so-called electoral inversion can happen --
the party with the highest number of votes nationwide might not win the most seats. 
Hence, it comes as no surprise that there is no consensus among experts about
what electoral system is the best \cite{bowler2005expert},
and people have difficulties with choosing one preferred system \cite{farrell1999british}.
It is, therefore, essential to provide new tools for evaluation of electoral systems.


\subsection*{The perspective of complexity science}

The electoral system together with the underlying voting processes
and opinion dynamics can be seen as a complex system
\cite{gell2002complexity,ladyman2013complex,san2023frontiers,bianconi2023complex}.
Complexity science had a significant impact on social and economic sciences.
It provided a new paradigm in the analysis of macroscopic phenomena emerging
from microscopic rules and interaction of many elements \cite{castellano2009statistical}. 
Concepts from complexity science and complex networks have been applied
in the description of opinion formation and voting processes,
as well as in analyses of the influence of media or terrorist attacks on them
\cite{fortunato2007scaling,sobkowicz2016quantitative,moya2017agent,fieldhouse2016cascade,siegenfeld2020negative,lee2022effect}.

Notably, applications can also be found in the evaluation of electoral systems.
For instance, the Sznajd model \cite{sznajd2021review} and cellular automata
\cite{bandini2001cellular} have been used to simulate the dynamics of voter opinions
and subsequently conduct elections, followed by evaluating the outcomes
using metrics such as the Gallagher index \cite{gwizdalla2012dynamics,gwizdalla2008gallagher}.
Others have explored the influence of district magnitude and seat allocation methods
on the proportionality of election results, assuming a log-normal distribution of
vote share \cite{pierzgalski2018balancing}.
These studies provide valuable insights into the mechanisms that
can either increase or decrease disproportionality in electoral outcomes.
Furthermore, a complexity science perspective has been outlined to analyze
the overall stability of democracy \cite{wiesner2018stability,eliassi2020science}.
In the authors' words:
,,\textit{scholars of democracy need
to move away from arguments based on static equilibrium to more dynamic
frameworks which are better suited to understanding how stable the
equilibria are to perturbations}''.

A remarkable contribution of complexity science to understand the dynamics of voting processes
was the development of the voter model 
\cite{granovsky1995noisy,carro2016noisy,redner2019reality,raducha2018coevolving,raducha2020emergence,jedrzejewski2020nonlinear}.
A group of voters in the model is represented by a network -- a set of nodes connected 
by links, which allow for an interaction. In its basic formulation there 
are two possible states representing two opposite opinions.
These opinions translate into votes that will be cast, hence they can also represent two parties
or candidates participating in the elections. 
It is straightforward to generalize the model into many possible parties -- indeed,
the multi-state version of the voter model has been studied
\cite{herrerias2019consensus,nowak2021discontinuous}.
It has been shown that, as in the 
basic binary formulation, an order-disorder phase transition is induced by increasing
noise.

There are two basic actions in the noisy voter model -- state copying and independent state update.
State copying between neighbors represents social influence, 
which is the core of social interactions and opinion dynamics. 
Independent state changes due to noise are interpreted as free choices
of the individuals \textit{making up their own mind}. Therefore, 
the time evolution can be seen as a competition between two attitudes
-- conformism and independence \cite{nyczka2013anticonformity,abramiuk2021discontinuous}.
It has been shown that such a relatively simple model has a great 
explaining power when predicting vote share distribution and vote share spatial correlations
\cite{fernandez2014voter,braha2017voting}.
Note, that the voter model can cover different underlying strategies 
of voters, since the outcome of any strategy for one voter is always 
a single ballot. The state of a node tells us how the voter would fill this ballot at a given time
(see Supplementary Information for more information on the voter model).


\subsection*{Modeling electoral systems}

There has been little focus, as mentioned before, on electoral
systems' stability and vulnerability in presence of external influences.
Here, we study electoral systems in a dynamical framework in order to explore these issues.
Our approach goes far beyond two-point volatility already seen in the literature 
\cite{pedersen1979dynamics,taagepera2003mapping,koppel2009measuring}. We analyze the vulnerability
of ES based on a long run of opinion dynamics with many elections
performed during the evolution. Given the strong empirical support for the noisy voter 
in the context of elections \cite{fernandez2014voter,braha2017voting},
we use it as the underlying opinion dynamics model in the simulations.
This allows us to construct a probability
distribution of vote share and seat share under every electoral system.
Then, we consider a system less stable if it has larger variance of 
the election results, and more vulnerable if it magnifies external influences, like those of extremism and media propaganda.

Using computer simulations in order to produce vote statistics is common in the literature
\cite{gwizdalla2012dynamics,gwizdalla2008gallagher,tideman2014voting,green2016statistical,green2014strategic,kenny2023widespread}.
After all, ,,\textit{in order to calculate theoretical averages one has to make some kind of distributional assumption}''
\cite{schuster2003seat}. In our case we assume the dynamics of
the voter model in order to produce the distribution of votes.
Numerical approach has two main advantages over the analysis of empirical vote distributions.
First of all, it is impossible to obtain a sample of thousands of elections, as we do, from the
observational data of any country. Secondly, using simulations
allows us to plant distortions and biases already at the level
of opinion dynamics and observe their macroscopic effects under different ES.
In particular, we introduce zealots in the population -- 
voters that never change their mind \cite{khalil2018zealots,khalil2021zealots}.
They can be political agitators,
internet bots, or just very ardent individuals. We also add skewed noise 
resulting in biased independent choices. In this way we can 
model asymmetrical media coverage or straightforward
propaganda \cite{moya2017agent,gonzlez-avella2007,gonzalez2012model}.
Finally, we analyze zealot susceptibility and media susceptibility
in stylized and real-world electoral systems, providing an understanding of their (in)stability.

\begin{figure}[hb!]
\centering
\includegraphics[scale=0.99]{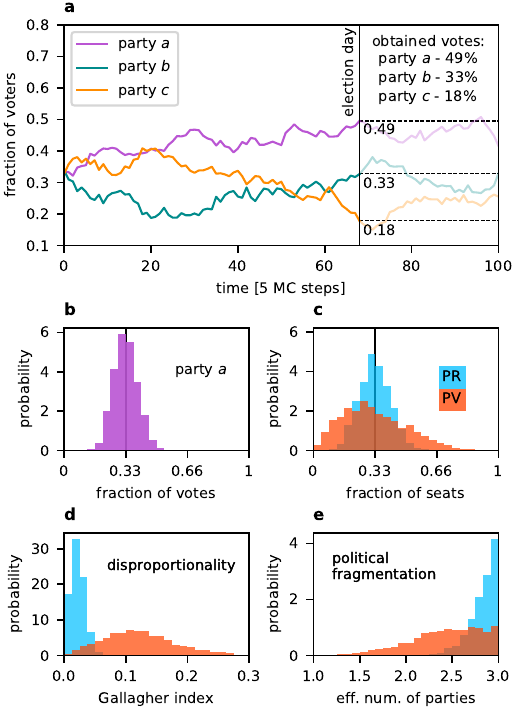}
\caption{\textbf{Simulation of elections and basic differences between electoral systems.}
(a) Changes over time of the fraction of voters supporting each of 3 parties in the simulation,
which directly translates
into the fraction of votes. An exemplary day of election is indicated together
with the fraction of votes each party would obtain on that day.
(b) Vote share distribution of party $a$ obtained over 4000 simulated elections,
the average fraction of votes is equal $0.332\pm0.065$.
(c) Seat share distribution of party $a$ obtained in the same elections
under PR (blue) and PV (red) ES. The average fraction of seats is equal
$0.331\pm0.085$ for PR and $0.33\pm0.17$ for PV.
(d) Histogram of the Gallagher index for the two ES, the average value
is equal $0.022\pm0.011$ for PR and $0.124\pm0.058$ for PV.
(e) Histogram of the effective number of parties for the two ES,
the average value is equal $2.83\pm0.15$ for PR and $2.43\pm0.38$ for PV;
All histograms obtained over 4000 elections.
Simulations were ran for $\varepsilon=0.005$, $k=12$, $r=0.002$,
$N=10^4$ divided into 100 equal communities.
Each electoral district corresponding to a community
has 5 seats assigned, giving 500 seats in total.
See Supplementary Information Fig.~S1-S3 for different ES and the Loosemore–Hanby index.
}
\label{fig:basic_traj_hist}
\end{figure}

We perform the simulations as follows. First, we construct a network of $N$ voters
with an average degree $k$ using the stochastic block model or its spatial version
(see Materials and Methods for details on the simulation setup). The blocks in the network
(topological communities) correspond to electoral districts (sometimes several communities
can be merged into one electoral district). Each individual in the network is then assigned
one of $n$ states $s_i \in \{a, b, c, ... \}$ at random with a uniform probability.
The parameter $n$ denotes the number of parties running in elections,
a node in state $a$ will vote for party $a$ etc. If there are zealots,
they are marked and their state is set to the biased state $s_b$ shared between
all zealots. If there is media propaganda, the probability of independent change
into the biased state $s_b$ is set to a value $p_m \in [0, 1]$ and the probability
for other states to $(1 - p_m) / (n-1)$. Without media propaganda probability
is equal for all states and $p_m = 1/n$, therefore the media bias can be directed
in favor as well as against a given party.

After preparing the network structure the simulation of opinion dynamics is ran
with voters progressively changing their preferred candidates --
see Fig.~\ref{fig:basic_traj_hist}a. The noise rate parameter $\varepsilon$
controls the competition between influence of the peers and 
independent choices. Note, that those choices are independent from the social influence, 
but not from the media influence. Every given number of time steps the process is stopped and 
elections are performed using several electoral systems which will be compared 
(when analyzing zealot susceptibility the network is initialized again before each election). 
The vote share and the seat share are saved and the process is continued until 
a desired sample of election results is collected.

Finally, based on the collected sample, we build vote share distribution 
(Fig.~\ref{fig:basic_traj_hist}b) and seat share distributions for all considered ES
(Fig.~\ref{fig:basic_traj_hist}c). Note, that unless indicated otherwise, by seat share
distribution we mean the probability distribution
of fraction of seats obtained by a given party,
not how seats are distributed among parties in a legislature.
The former can be constructed after collecting statistics of many elections,
the latter is a result of just one election. If there was any external
influence included in the simulation, we can estimate its consequences under
a given ES by analyzing how much the seat share changes compared to 
the neutral run. In this manner we can study zealot and media susceptibility.


\section*{Results}


\subsection*{Basic differences between electoral systems}

We initially focus on two stylized electoral systems denoted in the figures by:
PR -- a proportional representation system with a division into 100 constituencies
and the Jefferson seat assignment method (also known as the D'Hondt method),
and PV -- a plurality voting system with a division
into the same 100 constituencies, which is either FPTP or PBV depending on the number of seats per
district. For comparison we also include a PR system with one nation-wide constituency
and the Jefferson seat assignment method, denoted by PR-G (see Supplementary Information for more ES).
Here, we do not consider an entry threshold and all systems have the same number of 500 seats available
(5 per district for PR and PV), as well as the same value of other parameters of the simulation:
$N=10^4$, $k=12$, $r=0.002$, $\varepsilon=0.005$, and $n=3$ parties.

The systems we begin the analysis with, although possessing realistic elements,
are more theoretical due to the simplified district size and magnitude distributions, as well as
no geographical structure of constituencies and connections between them.
These idealized ES serve multiple purposes.
Firstly, they allow us to elucidate the fundamental differences between the main categories of ES.
Secondly, we can explore the significance of key parameters and their influence on
the stimulation outcomes. Furthermore, our goal is to ensure that the results obtained
align with well-established findings in the field of political sciences, 
thereby confirming the validity of our framework.

The vote share distribution of party $a$ obtained over 4000 elections is presented in
Fig.~\ref{fig:basic_traj_hist}b. As expected from the symmetrical design of the simulation
in respect to all parties, it is centered around the average value of $1/n=1/3$.
For the same reason the average fraction of seats won is the same across all ES,
but the characteristics of the distribution can and do vary, as visible in Fig.~\ref{fig:basic_traj_hist}c.
Note, that corresponding results for party $b$ and $c$ are roughly the same in absence
of external influences (see Supplementary Information Fig.~S2).

Comparing the distribution of votes with those of seats in Fig.~\ref{fig:basic_traj_hist},
we can see that PR system produces a seat distribution with a higher similarity
to the vote distribution, while PV produces much more volatile results. Even in the simplest case without
external influences, the contrast between plurality voting and proportional representation is 
clearly visible. This confirms the known intuition about the two systems,
but also allows to quantify it -- in the analyzed case the standard deviation
of PV election results is twice as high as for PR (0.17 and 0.085 respectively).

The difference is also significant in the distribution of disproportionality
and political fragmentation indexes of PR and PV, as we can see in Fig.~\ref{fig:basic_traj_hist}d
and \ref{fig:basic_traj_hist}e.
PV produces much less proportional results with the average value of the Gallagher index
$G=0.124\pm0.058$, which is order of magnitude bigger than for PR scoring $G=0.022\pm0.011$.
Additionally, the latter does not exceed $0.08$ in any election, while the former can reach values over $0.34$.
In the Supplementary Information we include analysis of the Loosemore-Hanby index,
with equivalent conclusions.
Similarly, there is a discrepancy in the effective number of parties, although not as extensive.
PV promotes lower fragmentation with $E=2.43\pm0.38$ on average, while PR obtains $E=2.83\pm0.15$.
Nevertheless, there is a big difference in the standard deviation of the effective number of parties
and the general shape of the distribution.

The possibility of computing popular indexes measuring performance of ES in
different scenarios creates a unique tool for evaluation of ES.
It allows us to study their statistical properties over samples of thousands 
electoral results, which is impossible to obtain solely from historical elections
of a given country. Moreover, it provides an additional meaning to the observed values.
If a country obtains a certain value of the Gallagher index in a given election,
what does it say about its electoral system? After all, disproportionality indexes
depend on the vote distribution, therefore the score might be significantly different
in the next elections. On the other hand, if we observe persistent patterns
across many elections with different vote distributions, and across
varying parameters, we can be confident that they describe
the underlying ES.

The case of (unrealized) electoral reform in Canada well illustrates the need
for a precise application of ES measures.
In 2016 the House of Commons of Canada created a special committee
to review different possibilities and provide a guidance in the potential reform.
The first recommendation put forward by the committee was that
,,\textit{the government should seek to design a system that achieves a Gallagher score of 5 or less}''
\cite{ourcommonsCommitteeReport} (where 5 means 5\% translating to 0.05 in the normalized
scale that we use). It is not clear, however, what the authors want to achieve.
Should it be always lower than 0.05 for
any conceivable distribution of votes? Or maybe the median or the average value of the
index should be below 0.05? If so, over what sample this average value should be calculated?
Our framework provides a possibility of computing the indexes under different conditions
and defining precise requirements for their values.


\subsection*{Effects of simulation parameters}

The above considerations raise a question about how the details of elections
(e.g. the number of parties running) and parameters of the model
(e.g. the noise rate $\varepsilon$) influence observed dependencies.
We provide an in-depth analysis of the effects of different parameters in the
Supplementary Information text and figures~S2-S11. Here we summarize the main findings.

With an increasing number of parties $n$, the standard deviation of
the seat share decreases for PR and PV and their seat share distributions become more similar.
The two systems produce more similar results in terms of the seat share distribution also for
a bigger number of districts $q$, fewer seats available in each district $qs$, higher noise rate
$\varepsilon$, smaller network size $N$, and lower interconnectedness ratio $r$.
Not surprisingly, disproportionality measured by the Gallagher and the Loosemore-Hanby indexes
is higher, and political fragmentation measured by the effective number of parties
is lower under PV compared to PR, across all tested parameters' values. Different parameters,
however, can have divergent effects on the two ES. For example, the average value of the Gallagher
index under PR will increase after increasing $n$ or $q$, or after decreasing $qs$, $\varepsilon$, or $N$.
The disproportionality of PV, in comparison, increases with growing $r$,
while $n$, $q$, and $\varepsilon$ have the opposite effect.
Notably, the average degree $k$ has negligible influence on the resulting seat share
and values of the indexes for both systems.
PV and PR display identical behavior only after decreasing the number of seats available in each
constituency to one -- in such case both system effectively become FPTP voting.


An interesting effect can be observed for a low noise rate $\varepsilon$,
low number of districts $q$, or high interconnectedness ratio $r$: a shift
from unimodal to bimodal seat share distribution appears for PV. 
Moreover, the two peaks
of the bimodal distribution are placed on the extreme values of 0 and 1.
Put simply, landslide victories of one party become highly probable, while
no such effect is obtained for PR in the explored parameters' range.
This effect is a manifestation of an inherent property of PV systems.
In PV even the smallest plurality equally distributed across all districts is
enough to secure a total victory, whereas the same support
concentrated in a few constituencies may result in losing the
leading position in the legislature. This fact is well known and
utilized in the manipulative practice of gerrymandering
\cite{stewart2019information,kenny2023widespread}. In the context of opinion dynamics,
it means that more isolated districts provide less opportunity for
the dominance of a single party in PV, since its support cannot
easily spread across them.



\subsection*{Zealots and media influence}

\begin{figure}[ht!]
\centering
\includegraphics[scale=0.99]{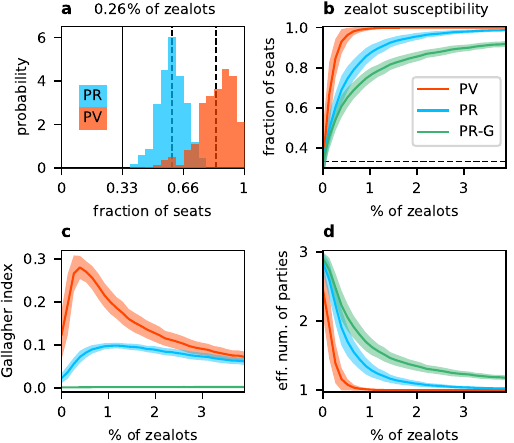}
\caption{
\textbf{Influence of zealots on electoral outcomes.}
(a) Seat share distribution of the zealot party $a$ under PR (blue) and PV (red) ES for 
$0.26$\% of zealots in the population. The solid line indicates the average fraction
of seats obtained in the neutral
case with no zealots (1/3), while dashed lines indicate average values of shifted distributions:
$0.602\pm 0.072$ for PR and $0.845\pm0.097$ for PV.
(b) The average seat share of the zealot party $a$ vs percentage of zealots in the population for
three ES: PV, PR, and PR-G. The colored shade indicates results within
one standard deviation from the average. The dashed line marks the average
fraction of seats obtained with no zealots (1/3).
(c) The average value of the Gallagher index
vs percentage of zealots in the population for the three ES.
(d) The average value of the effective number of parties
vs percentage of zealots in the population for the three ES;
All results obtained over 500 elections with
$N=10^4$, $q=100$, $qs=5$, $k=12$, $r=0.002$, $\varepsilon=0.005$, and $n=3$ parties.
See Supplementary Information Fig.~S12 and~S13 for other values.
}
\label{fig:basic_zealot_sus}
\end{figure}

The central consideration of this study is a scrutiny of ES
under influence of political agitators and media propaganda.
To analyze the effect of the former on elections we
introduce a specific number of zealots in the population and asses the
resulting change in the distribution of seat share. 
In Fig.~\ref{fig:basic_zealot_sus}a we present the effect
of 0.26\% of voters being zealots of party $a$. Comparing it
to the neutral case in Fig.~\ref{fig:basic_traj_hist}c
we can see how much one party can gain with the help of
political agitators. But the most striking result is
the magnitude of advantage that zealots can produce
in different ES -- PV amplifies the
influence when translating the votes into seats resulting
in $0.845\pm0.097$ of all seats gained, compared to
$0.602\pm0.072$ for PR. In this sense
we can say that PV is a more vulnerable system than PR, at least
in respect to zealots.

In order to comprehensively investigate the vulnerability of
electoral systems to zealot influence, we examine the concept of
zealot susceptibility. It measures the extent to which a system
tends to gravitate towards the zealot's state on average, when
the number of zealots varies. We present the zealot susceptibility
of both ES in Fig.~\ref{fig:basic_zealot_sus}b, together with a comparison
to the PR-G system. The disparities between the systems are substantial.
We observe a gradual increase in the mean election
result of the zealot's party towards a complete dominance in each case. However,
under PR even with almost 4\% of voters being zealots, the zealot
party does not achieve a land-slide victory. Conversely, in PV we observe
a much more rapid increase in the zealot's party average seat share, leading to
almost all seats taken with less than 1\% of zealots. Consequently,
PV system necessitates considerably less effort and resources to
exert effective influence. PR-G with one global
district, on the other hand, is the least zealot susceptible system.
Indeed, any local influence that zealots might exert does not matter
under PR-G unless it is strong enough to shift the opinion of
the whole population.

\begin{figure}[ht!]
\centering
\includegraphics[scale=0.99]{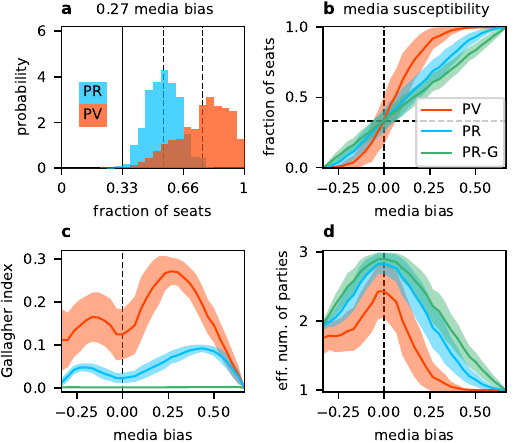}
\caption{
\textbf{Influence of biased media on electoral outcomes.}
(a) Seat share distribution of the media party $a$ under PR (blue) and PV (red) ES for 
a media bias of $0.27$. The solid line indicates the average fraction
of seats obtained in the case of neutral media (1/3),
while dashed lines indicate average values of shifted distributions:
$0.558\pm0.088$ for PR and $0.77\pm0.14$ for PV.
(b) The average seat share of the media party $a$ vs media bias for
three ES: PV, PR, and PR-G. The colored shade indicates results within
one standard deviation from the average. The vertical dashed line
indicates the simulation with neutral media, and the horizontal one marks the average
fraction of seats obtained that case (1/3).
(c) The average value of the Gallagher index
vs media bias for the three ES.
(d) The average value of the effective number of parties
vs media bias for the three ES;
All results obtained over 1000 elections with
$N=10^4$, $q=100$, $qs=5$, $k=12$, $r=0.002$, $\varepsilon=0.005$, and $n=3$ parties.
See Supplementary Information Fig.~S14 and~S15 for other values.
}
\label{fig:basic_media_sus}
\end{figure}

What is crucial, these results are robust under
various parameter choices.
The basic differences between ES zealot susceptibility remain true
after varying the average degree $k$, noise rate $\varepsilon$, interconnectedness
ratio $r$, or number of parties $n$ (see Supplementary Information Fig.~S12-S13).
Notably, higher $r$ hardly changes susceptibility of PR, but visibly increases
it for PV. Bigger $\varepsilon$ lowers the susceptibility for both
systems, but more so for PR.

In Fig.~\ref{fig:basic_zealot_sus}c and~\ref{fig:basic_zealot_sus}d we also
show how the Gallagher index and the effective number of parties are influenced by
varying percentage of zealots. Clearly, PV displays the highest disproportionality
and lowest fragmentation, followed by PR and PR-G. The last system can achieve
virtually perfect proportionality of apportionment even under high influence of zealots.
Remarkably, the disproportionality indexes of PV and PR reach the maximal value
for intermediate numbers of zealots. For around 0.4\% of zealots
in the population, election results become the most disproportional for
PV with the average Gallagher index reaching a value of $G=0.280 \pm 0.027$.
For PR, the maximum is obtained
at approximately 1.2\% of zealots with the index equal $G=0.0982\pm 0.0066$, while
PR-G reaches at most a value of $G=0.00157 \pm 0.00066$.
It shows that the division into districts will inherently increase
disproportionality (although it depends on the seat assignment method,
see Supplementary Information).
It also demonstrates that not always vast resources are necessary to damage
proportionality of elections and sway the result in a preferred direction.

The second influence of electoral processes that we study is media propaganda
modeled by skewed noise. The seat share distribution obtained after
increasing the media bias towards party $a$ to 0.27 is
shown in Fig.~\ref{fig:basic_media_sus}a.
Similarly to zealots, biased media can significantly
shift the average number of seats gained by a party,
what is clearly visible after comparing Fig.~\ref{fig:basic_media_sus}a
to Fig.~\ref{fig:basic_traj_hist}c. Once again, the average seat share
of party $a$ is much more increased under PV than PR for the same amount
of influence exerted in voters. In this sense we can say that PV system
is more susceptible to biased media than PR.

Media propaganda, in contrast
to zealots, can be also directed against party $a$.
In Fig.~\ref{fig:basic_media_sus}b we present media susceptibility
of three ES for the whole range of possible media bias values.
The biggest difference can be observed between PV, which is the most
susceptible, and any proportional system. Noticeably, there is only
a minor difference in media susceptibility between PR and PR-G --
they fall within one standard deviation from each other.
This indicates that the division into many constituencies is
not a decisive factor in response to media propaganda. Indeed,
the influence of media is global, while the influence of zealots was local.

Even after altering the values of the average degree $k$, noise rate $\varepsilon$,
interconnectedness ratio $r$, or the number of parties $n$, the
qualitative distinctions among the considered ES
remain consistent (see Supplementary Information Fig.~S14-S15).
In each case, PV exhibits the highest media 
susceptibility, followed by PR, and PR-G at the end.
Larger values of $\varepsilon$ significantly reduce the
standard deviation of media susceptibility for all systems.
Additionally, an increase in $r$ has a minimal impact on the susceptibility
of PR, yet clearly increases it for PV.

In Fig.~\ref{fig:basic_media_sus}c and~\ref{fig:basic_media_sus}d,
we present the variations in disproportionality and fragmentation
indexes in response to varying media bias. The electoral systems
can be arranged in ascending order of the Gallagher index and
descending order of the
effective number of parties: PR-G, PR, and PV. Nonetheless,
the relative differences between the systems vary.
In particular, the disproportionality indexes reach their maximum
for intermediate values of negative or positive media bias.
For a positive bias of approximately 0.27 the election results for
PV become the most disproportional, with the average Gallagher index
reaching $G=0.271 \pm 0.032$.
For PR, the maximum disproportionality occurs at around 0.43, with
the average index value $G=0.0918\pm 0.0085$.
For the negative bias, PV obtains the maximum at $-0.17$ with $G=0.165 \pm 0.057$,
and PR at $-0.2$ with $G=0.0472 \pm 0.0097$.
PR-G is able to maintain near-perfect proportionality
even in presence of considerable media bias with $G=0.00169\pm 0.00064$ in
the most disadvantageous result. The effective number of parties
decreases for extreme negative bias down to 2 (or slightly below for PV),
and down to 1 for extreme positive bias. This is because negative bias
can eliminate one of three parties and positive bias can eliminate two,
by forging a landslide victory of one party.

These findings demonstrate the widely recognized fact
that obtaining high media coverage and using it in favor or against
a political party can drive the election results. Moreover,
they show how disruptive it can be to the proportionality of the
electoral system.
The impact of media propaganda and zealots is
especially powerful in PV for relatively small deviations from
the neutral case -- the impact of any influence per its unit is significantly
higher than in PR (see Supplementary Information for marginal susceptibilities).
It is also worth noting that with increasing media bias or number of zealots we
observe diminishing returns of both means of influence.

\begin{figure*}[ht!]
\centering
\includegraphics[scale=1]{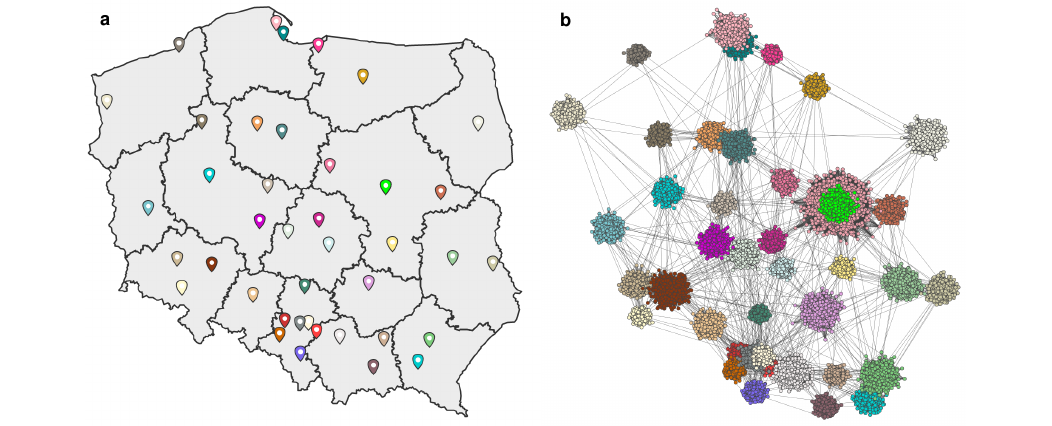}
\caption{
\textbf{Constructing a network for simulations of the Poland's Sejm.}
(a) Map of Poland with division into 16 voivodeships, i.e. the biggest administrative districts of the country.
The capital of each of 41 electoral districts applied
in elections to the Sejm is indicated with a colored marker.
(b) Network of $41 \times 10^3$ nodes generated for simulations of elections to the Poland's Sejm.
Nodes are centered around the capital of the constituency.
The bigger the population of a constituency, the more spread are the nodes.
}\label{fig:pl_info}
\end{figure*}

\begin{figure}[ht!]
\centering
\includegraphics[scale=0.99]{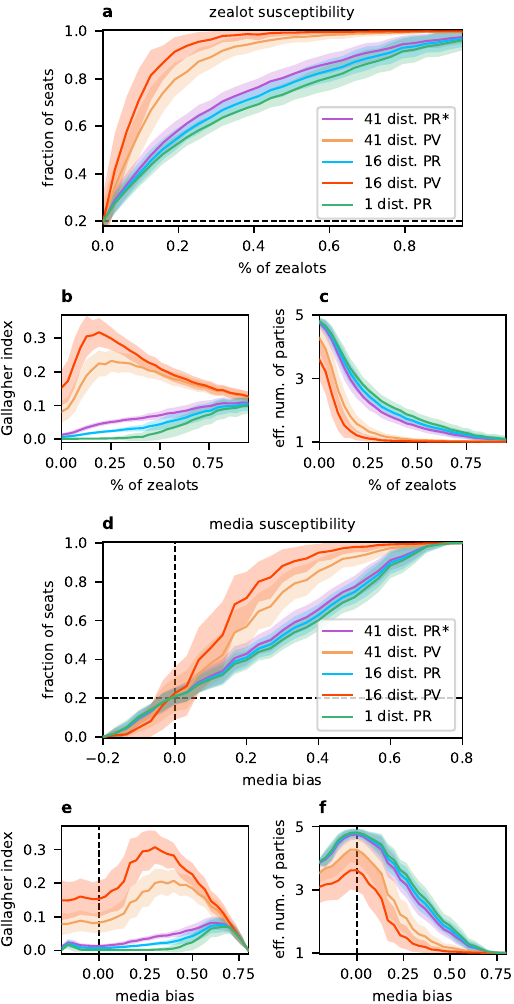}
\caption{
\textbf{Influence of zealots and biased media on the elections to the Poland's Sejm.}
(a) The average seat share of the zealot party $a$ vs percentage of zealots in the population for
the five considered ES. The colored shade indicates results within
one standard deviation from the average. The dashed line marks the average
fraction of seats obtained with no zealots (1/5).
(b) The average value of the Gallagher index
vs percentage of zealots in the population for the five ES.
(c) The average value of the effective number of parties
vs percentage of zealots in the population for the five ES;
All results with zealots were obtained over 200 elections.
(d) The average seat share of the media party $a$ vs media bias for
the five considered ES. The vertical dashed line
indicates the simulation with neutral media, and the horizontal one marks the average
fraction of seats obtained that case (1/5).
(e) The average value of the Gallagher index
vs media bias for the five ES.
(f) The average value of the effective number of parties
vs media bias for the five ES;
All results with media bias were obtained over 1000 elections.
The network has $41 \times 10^3$ nodes, with the average degree
$k=12$, divided into 41 communities corresponding to the actual electoral districts.
There is 460 seats to gain in total, $n=5$ parties, and the noise rate $\varepsilon=0.002$.
See Supplementary Information Fig.~S16-S18 for more details and results for the Poland's Sejm.
}
\label{fig:pl_sus}
\end{figure}

\subsection*{Case study: Poland's Sejm}

After exploring differences between the two main categories of ES and basic dependencies
between parameters, we study real-world ES applied in Poland and India
(see Supplementary Information for Israel and the Senate of Poland).
The first system we scrutinize is the one used in elections to the Poland's Sejm,
which is the lower house of the bicameral parliament of Poland. It is a party-list
proportional representation with 460 seats in total to be assigned across 41 constituencies,
whose capitals are presented on a map in Fig.~\ref{fig:pl_info}a.
The number of seats per one electoral district varies from 7 to 20,
and they are assigned using the Jefferson seat assignment method, after excluding
parties below the threshold of 5\% of votes nation-wide.
In the 2019 elections to the Poland's Sejm
there was approximately 30 million people able to vote,
and in every district there were at least 5 political parties registered
\cite{pkwWyborySejmu}.

For the simulation of elections to the Poland's Sejm we create a network with
$N=41 \times 10^3$ nodes divided into 41 topological communities,
as shown in Fig.~\ref{fig:pl_info}b, using the spatial version of the SBM
(see Materials and Methods). Those communities correspond to the real electoral
districts and their relative sizes are based on the actual populations of voters
in the districts. Each community is
assigned a geographical location, more precisely the coordinates of the district's capital
(see Fig.~\ref{fig:pl_info}a).
The connectivity between nodes is based on the work commuting distance,
which means it decreases with the geographical distance as in Eq.~[\ref{eqn:planar_affinity}].
The supplementary datasets and the configuration files in our online code repository
\cite{github_repo} contain the exact data used in the simulations.

In addition to the actual electoral system used in elections to the Poland's Sejm 
(labeled as ,,41 dist. PR*'' in the figures),
we study four alternative systems.
The first alternative system (,,41 dist. PV'') is a PV system with the same division
into districts and the same number of seats
in each district, but the party with plurality in a given district obtains all seats
(effectively it is a PBV system). Plurality voting was considered during 2011 electoral reform
in Poland. That year the system of the Poland's Senate changed from a PR to FPTP voting.
Similar transformation was considered also for the Sejm, but was deemed unconstitutional.

We then consider two alternative systems with electoral districts corresponding to 16
voivodeships, i.e. the highest-level administrative districts of Poland
(see Fig.~\ref{fig:pl_info}a). We merge all electoral districts
within the same voivodeship into one, accumulating the seats into a single pool per voivodeship.
On top of this new division we consider the original PR system (,,16 dist. PR'') and
a PBV system with the winner-takes-all approach in each district (,,16 dist. PV'').
Finally, we examine a PR system where all 460 seats are assigned within a singe nation-wide 
district (,,1 dist. PR''). We do not include the entry threshold in the PV alternative systems,
and the PR alternative systems have the same 5\% threshold
and the same seat assignment method as the original one.

After these initial steps we can analyze zealot and media susceptibility of the
elections to the Sejm in the same manner as for the theoretical systems. 
In Fig.~\ref{fig:pl_sus} we demonstrate this analysis
for all ES considered -- the original one and four alternatives.
Fig.~\ref{fig:pl_sus}a shows the average fraction of seats won by the zealot party
versus the percentage of zealots in the population.
The first striking result is the difference in
zealot susceptibility between any PR and PV systems.
For instance, with 0.22\% of zealots the most susceptible PR systems assigns $0.611\pm0.041$
of seats to the zealot party, while the least susceptible PV system $0.829\pm0.059$.
For the Poland's Sejm, which has 460 seats in total, this makes a differences of 100 seats.
All three PR systems, although having
very different district divisions, are similar
in their response to zealots (within one standard deviation from each other)
and much less susceptible than both PV systems. This observation
holds for any value of the noise rate $\varepsilon$ (see Supplementary Information Fig.~S17).

In Fig.~\ref{fig:pl_sus}b and Fig.~\ref{fig:pl_sus}c we show the Gallagher index
and the effective number of parties in presence of zealots. Under PV systems
elections can become highly disproportional for very small percentage of zealots in the population
-- the maximum value of disproportionality indexes in PV is reached close to 0.2\% of zealots,
while in PR systems disproportionality becomes maximal when the zealot party is close to
a complete domination at almost 1\% of zealots.
For example, for 0.25\% of zealots under ,,41 dist. PV'' system the Gallagher index
is equal $G=0.233 \pm 0.032$, while at the same time $G=0.109 \pm 0.015$
for the original ES.
The political fragmentation decreases with increasing number of zealots,
as expected, and is generally lower for PV systems.

Fig.~\ref{fig:pl_sus}d illustrates the second external influence that we study,
namely the susceptibility to media propaganda, as a function of amount of bias.
The conclusions are similar to the ones in respect to zealots -- PV systems are generally
more susceptible to biased media than PR systems. All PR systems
display a very similar media susceptibility
within one standard deviation from each other.
For example, for a bias of 0.2 PR systems, ordered from 1 to 41 districts,
assign $0.396\pm0.050$, $0.411\pm0.053$, and $0.429\pm0.057$ seats to the media party,
while $0.57\pm0.11$ is assigned under ,,41 dist. PV''.
In terms of the number of seats in the legislature it is 182, 189, and 197 in the PR systems,
and 262 seats in the PV system.
Notably, in PV media bias provides the highest
returns not at the very beginning, like zealots, but after already exerting some influence
-- the maximum of marginal susceptibility is shifted to the right from the zero bias
(see Supplementary Information Fig.~S18).

Media propaganda causes the highest disproportionality for intermediate
values of influence, as visible in Fig.~\ref{fig:pl_sus}e and Fig.~\ref{fig:pl_sus}f.
It happens much earlier, however, in PV systems than in PR ones. Again, the disparity between
the two types of ES is much bigger in terms of proportionality than political fragmentation,
but PR systems maintain higher values of the effective number of parties most of the time.

These results show the applicability of our approach in a potential electoral reform.
If the stability of elections to the Poland's Sejm was the main concern of Polish authorities,
based on our analysis we can say that an electoral reform changing the system to plurality
voting would be detrimental. Conversely, decreasing the number of districts under the
current electoral system would marginally improve the robustness against external influences.
However, the effect is so small that most likely it would not justify a reform.
The current electoral system, therefore, is a prudent choice in the context of vulnerabilities.

\subsection*{Case study: India's House of the People}

\begin{figure*}[ht!]
\centering
\includegraphics[scale=1]{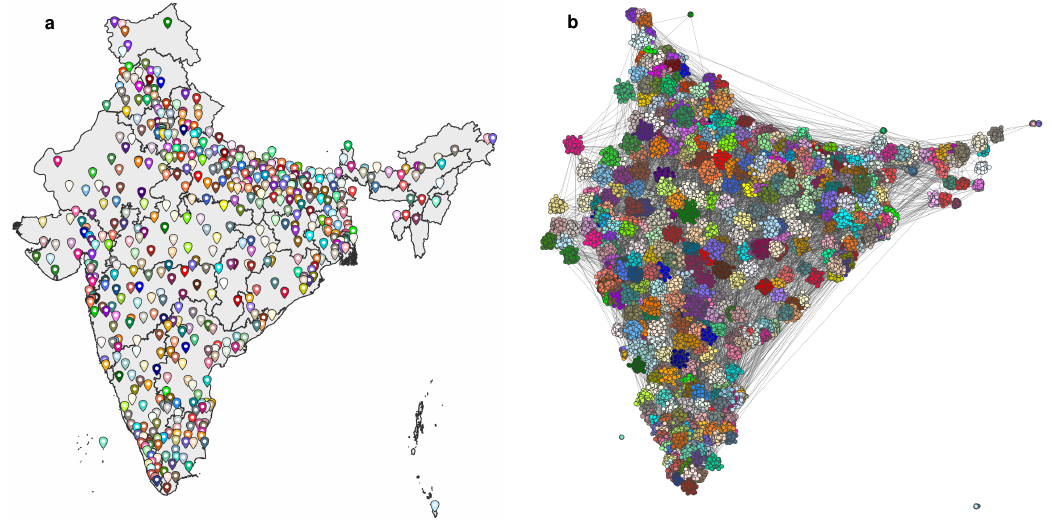}
\caption{
\textbf{Constructing a network for simulations of the India's House of the People.}
(a) Map of India with division into 36 administrative districts of the country
(28 states and 8 union territories).
The capital of each of 542 electoral districts applied in elections to
the House of the People is indicated with a colored marker.
(b) Network of $10^5$ nodes generated for simulations of elections to the India's House of the People.
Nodes are centered around the capital of the constituency.
The bigger the population of a constituency, the more spread are the nodes.
}\label{fig:ind_info}
\end{figure*}

\begin{figure}[ht!]
\centering
\includegraphics[scale=0.99]{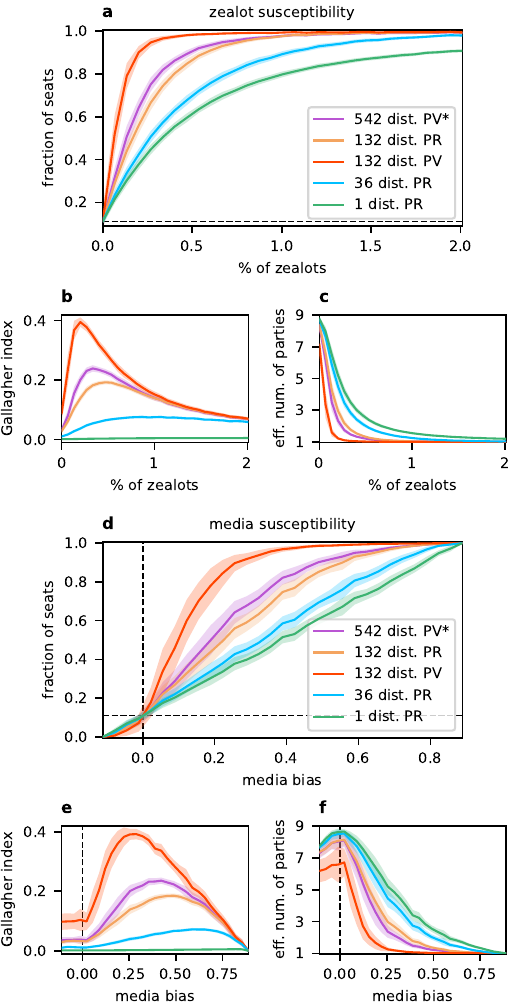}
\caption{
\textbf{Influence of zealots and biased media on the elections to the India's House of the People.}
(a) The average seat share of the zealot party $a$ vs percentage of zealots in the population for
the five considered ES. The colored shade indicates results within
one standard deviation from the average. The dashed line marks the average
fraction of seats obtained with no zealots (1/9).
(b) The average value of the Gallagher index
vs percentage of zealots in the population for the five ES.
(c) The average values of the effective number of parties
vs percentage of zealots in the population for the five ES;
All results with zealots were obtained over 200 elections.
(d) The average seat share of the media party $a$ vs media bias for
the five considered ES. The vertical dashed line
indicates the simulation with neutral media, and the horizontal one marks the average
fraction of seats obtained that case (1/9).
(e) The average value of the Gallagher index
vs media bias for the five ES.
(f) The average value of the effective number of parties
vs media bias for the five ES;
All results with media bias were obtained over 500 elections.
The network has $10^5$ nodes, with the average degree
$k=12$, divided into 542 communities corresponding to the actual electoral districts.
There is 542 seats to gain in total, $n=9$ parties, and the noise rate $\varepsilon=0.002$.
See Supplementary Information Fig.~S25-S27 for more details and results for the India's House of the People.
}
\label{fig:ind_sus}
\end{figure}

The second real-world ES we investigate is the one employed in elections
to the India's House of the People, also known as Lok Sabha, which is the lower house of the bicameral
parliament of India. The system utilized to select its members is a plurality voting system with 543 
single-member districts, presented in Fig.~\ref{fig:ind_info}a.
The candidate with the biggest number of votes in a constituency wins the seat,
i.e. it is FPTP voting.
In the 2019 elections approximately 910 million of Indian citizens were eligible to vote
\cite{indiaGeneralElection}, making it the biggest democracy in the world.
The same year, from among many active political parties, only 9 parties obtained
more than 2\% of votes.

For the simulation of elections to the House of the People we construct a network of 
$N = 10^5$ nodes divided into 542 topological communities, analogous to real electoral 
districts (excluding Vellore, see Supplementary Information).
The relative sizes of these communities are based on the actual populations of voters 
in the districts. Each community is assigned a geographical location, corresponding to the 
coordinates of the district's capital (see Fig.~\ref{fig:ind_info}a).
The connectivity between nodes decreases with the 
geographical distance between them as in Eq.~[\ref{eqn:planar_affinity}]
(see Fig.~\ref{fig:ind_info}b).
The supplementary datasets and the configuration files in our online code repository
\cite{github_repo} contain the exact data used in the simulations.

We study the actual electoral system used in elections to the lower house of
the India's parliament (labeled as ,,542 dist. PV*'' in the figures)
and four alternative systems.
The first alternative system (,,132 dist. PV'') is a PV system with a division into 132 multi-member
constituencies corresponding to India's divisions, i.e. administrative units bigger than districts
and smaller than states. There are 102 divisions in India -- 10 states and 5 union territories
are not divided into divisions. However, for a more homogeneous distribution of district magnitude,
we divide bigger states according to the geographical proximity of districts resulting
in 132 constituencies in total in the alternative system. Under the system ,,132 dist. PV''
all seats in a constituency are assigned to the party with the best score
(effectively it is a PBV system).
Next, we consider a PR system with an identical division of the country (,,132 dist. PR''),
and two more PR systems with electoral districts corresponding to 
the 28 states and 8 union territories of India (,,36 dist. PR'') and with one
nation-wide district (,,1 dist. PR'').
We merge all single-member districts of the original system within the same 
alternative constituency, accumulating the seats into a single pool. 
The alternative systems do not include an entry threshold, and
the PR alternative systems use the Jefferson seat assignment method.

In Fig.~\ref{fig:ind_sus}a we present an analysis of zealot susceptibility across 
all electoral systems considered for the House of the People. It shows the average fraction 
of seats won by the zealot party in relation to the percentage of zealots in the population. 
In general, there is a divergence in zealot susceptibility between PR
and PV systems. However, in contrast to Poland, the three PR 
systems exhibit distinct responses to zealots. The system with one constituency is the least 
susceptible to zealots, followed by the one with 36 constituencies, and the most susceptible 
system is the one with 132 constituencies. The differences in the average seat share are 
significantly larger than one standard deviation.
The PR system with 132 constituencies (,,132 dist. PR'') is surprisingly similar, in terms
of zealot susceptibility, to the real system (,,542 dist. PV*'').
Those observations hold true for any noise rate $\varepsilon$ (see Supplementary Information Fig.~S26).

In Fig.~\ref{fig:ind_sus}b and Fig.~\ref{fig:ind_sus}c we can see the Gallagher index
and the effective number of parties changing in response to zealots.
Again, under PV systems, elections can 
become highly disproportional even with a minuscule percentage of zealots in the population.
Notably, it also the case for the PR system with 132 districts (,,132 dist. PR'').
The maximum level of disproportionality in PV systems is reached for 0.2\% and 0.34\% of zealots,
depending on the number of districts,
while for 0.47\% in the ,,132 dist. PR'' system. For the latter the Gallagher index reaches then
$G=0.1925 \pm 0.0066$, and is a bit higher for the original ES in the most disproportional
outcome with $G=0.239 \pm 0.011$.
As expected, an increase in the number of zealots leads to a decrease 
in political fragmentation, with PV systems generally exhibiting lower levels of fragmentation.
The discrepancy between PV and PR in the effective number of parties, however, is much smaller
than in the Gallagher index.

Fig.~\ref{fig:ind_sus}d shows the susceptibility to media propaganda as a function of 
bias magnitude. PV systems tend to be more susceptible to biased media compared 
to PR systems. But in contrast to the zealot susceptibility,
PR systems exhibit more comparable levels of media susceptibility to each other.
In Fig.~\ref{fig:ind_sus}e and Fig.~\ref{fig:ind_sus}f we present the Gallagher index
and the effective number of parties vs media bias.
Media propaganda causes the highest level of disproportionality for 
intermediate levels of influence, just like the zealots. 
Maximal disproportionality arises slightly earlier in PV systems compared to PR systems. The discrepancy 
between the two types of ES is more pronounced in terms of proportionality rather than political 
fragmentation.

The similarity between the actual ES (,,542 dist. PV*'') and the PR system with 132 districts
(,,132 dist. PR'') is noteworthy -- both systems have a similar scoring 
across all measures. The PR system is only marginally less susceptible to media propaganda
(the difference is below one standard deviation),
and only in some cases more proportional and more fragmented.
For example, when media bias is equal to 0.42 the media party obtains $0.841\pm0.036$
of seats in the original ES and $0.769\pm0.044$ in ,,132 dist. PR'' -- a difference
of 39 seats in the Hose of the People.
For 0.34\% of zealots, when the discrepancy in the results of the two systems is the biggest,
the original ES assigns $0.819\pm0.025$ of seats to the zealot party while the other
$0.743\pm0.027$, giving 41 seats of difference from over 500 in total.
These differences are much smaller than for the ,,132 dist. PR'' system
and any other PR system.
Disproportionality and fragmentation in presence of zealots, and zealot susceptibility
of these two systems are in many cases 
within a single standard deviation from each other as well.
This is partially an effect of
a relatively small district magnitude for 132 constituencies -- on average each district
has only 4 seats assigned in this case.
As shown in Fig.~S7 from Supplementary Information,
for lower district magnitude PR systems become similar to FPTP.

As in many countries, there is a discussion in India about possible improvements
of the electoral system used in elections to the House of the People, possibly
transforming it into a proportional representation \cite{misra2018shift}.
If one of the goals of the Indian policymakers was minimizing potential threats to democratic processes,
it would be more advisable to create 36 multi-member constituencies based
on the states, rather than 132 constituencies based on the divisions.
A change to the latter system would yield marginal improvement of zealot
and media propaganda.
A possible reform deciding for a PR system
would decrease the disproportionality, although at the cost of a slightly higher
political fragmentation. Finally, the lowest vulnerability to external influences
is achievable for one nation-wide district,
however such a system is highly unrealistic to be applied
in the biggest democratic country of the world.


\section*{Discussion}

We study the vulnerability of democratic electoral systems (ES) in a dynamic setting, with 
a particular focus on their stability and susceptibility to external influences. Our approach
involves constructing a network of voters and simulating opinion dynamics among them, 
enabling us to explore the effects of various factors at the population level and gain 
insights into the overall behavior of ES with different designs. The numerical
framework employed in our study allows for averaging outcomes across multiple elections under
any ES, which is not feasible when analyzing empirical data from a single 
country.

Initially, we examine stylized ES, specifically emphasizing the differences 
between the two main categories: proportional representation (PR) and plurality voting (PV).
By conducting multiple election simulations, we observe that PV systems exhibit significantly
higher result variance, leading to a bimodal seat share distribution in extreme cases. 
In this sense we can say that PV systems are less stable than the PR systems.
Furthermore, we investigate how parameters influencing the network structure and opinion dynamics, 
such as the network size and noise rate, and details of ES, such as
the district magnitude and number of districts, impact electoral outcomes.
Notably, we find that a higher interconnectedness ratio, 
representing closer proximity between constituencies, increases the probability of landslide
victories in PV systems, whereas no such effect is observed in PR systems.

PV electoral systems consistently demonstrate higher susceptibility to zealots and biased 
media, which we simulate by introducing voters who do not change their opinions and by 
incorporating skewed noise. This implies that the increase in the number of obtained seats 
resulting from political agitators or media propaganda is greater in PV systems than
in PR systems. External influences also exert a stronger impact on the Gallagher index of PV
systems, as relatively minor distortions at the population level can yield highly 
disproportional outcomes. On the other hand, PR systems tend to exhibit greater political 
fragmentation, although the difference in the effective number of parties between the two
systems is relatively smaller.

In addition, our analysis of real-world ES corroborates the main finding 
regarding the susceptibility of PR and PV systems, with the latter being more zealot and media susceptible
across various geographical settings. However, the magnitude of
this difference is not always substantial, suggesting that not every electoral reform
solely changing PV into PR would be justified. For example, in the case of India's
House of the People, merging all single-member districts into constituencies based on India's
divisions, with seat allocation utilizing the Jefferson method, would yield only marginal 
improvements in terms of susceptibility.

The study of ES vulnerability holds paramount importance in democratic 
societies. Our analysis indicates that proportional representation electoral systems 
generally offer a more robust choice than plurality voting systems, particularly in light of 
contemporary threats to democratic processes. However, we acknowledge that there are numerous 
important indicators of ES performance, and our results should be considered 
alongside other insights from political sciences and related fields. Electoral reforms often
encounter challenges as the design of ES is influenced by multiple 
factors, including historical, cultural, and economic considerations. Nevertheless, 
comprehending vulnerabilities helps to identify potential weaknesses and safeguards the 
integrity of democratic processes. Future research could delve into the effects of seat 
assignment methods within the realm of ES susceptibility, as our findings 
suggest that this aspect could be decisive among PR electoral systems.



\matmethods{
To simulate electoral processes, we employ a model comprising of three main components --
generating a network, running opinion dynamics, and performing elections.
In the subsequent paragraphs we describe our simulation framework constructed
to evaluate both theoretical and real-world electoral systems.
The details provided below are sufficient to reproduce our work, however
for a more meticulous explanation see Supplementary Information
and our online code repository \cite{github_repo}.


\subsection*{Network generation}

The network is constructed utilizing the Stochastic Block Model (SBM) \cite{holland1983stochastic},
facilitating the division of network nodes into multiple topological communities.
Inter-community connections are assigned lower probabilities, while intra-community
connections are assigned higher probabilities. This ratio is regulated by
the interconnectivity parameter $r$. Other key parameters such as network size $N$,
number of communities $q$, and average degree $k$ allow customization of the network.
The network structure remains constant throughout the simulation, with topological
communities, or their aggregates, functioning as electoral districts.

For simulations involving abstract electoral systems, the standard SBM is employed
to generate the network. The number of nodes $N$, the average degree $k$, and
the number of communities $q$ are specified, with each community comprising
an equal size of $N/q$ nodes (although varying sizes are also possible).
The total number of links in the network equals $Nk/2$. Connectivity within
and between districts is determined solely by the interconnectedness ratio $r$,
resulting in every pair of districts having an equal probability of being connected.
Geographical or spatial structures are not considered, aside from the division into communities.

In simulations of real electoral systems,
after specifying the network size $N$ and average degree $k$, the network is divided
into communities corresponding to actual electoral districts. Consequently,
the proportions between community sizes mirror the proportions between voter
populations in the respective districts. Furthermore, geographical coordinates
corresponding to the district capitals are assigned to each community and its nodes,
resulting in every node having a geographic location. In this modified version of the SBM,
the probability of connection between two nodes decreases
with the geodesic distance between them as:
\begin{equation}
    p(x) \propto \frac{1}{(x+c)^2}  ,
\label{eqn:planar_affinity}
\end{equation}
where $x$ is the distance between the nodes and $c$ is a parameter we call the planar constant.
Probability $p$ is then normalized according to the specified average degree.
The planar constant $c$ is determined by
fitting [\ref{eqn:planar_affinity}] to empirical data on work commuting distance
(see Supplementary Information Fig.~S16, S22, and S25 for the fitting).
This way, a realistic connectivity between districts is established.


\subsection*{Opinion dynamics}

We utilize the noisy voter model (NVM)
\cite{granovsky1995noisy,carro2016noisy,redner2019reality,raducha2018coevolving,raducha2020emergence,jedrzejewski2020nonlinear}
to simulate opinion dynamics among voters,
wherein nodes represent individuals with $n$ possible states, denoted as 
$s_i \in \{a, b, c, ... \}$ (this version of the model is also called the multi-state noisy voter model
\cite{herrerias2019consensus,nowak2021discontinuous,khalil2021zealots}).
States of nodes represent multiple possible party choices -- a node in state $a$ will vote for party $a$ etc.

After generating the structure of the network,
every node is assigned one
of $n$ states at random with a uniform probability. Then, the simulation of opinion dynamics
starts, with the basic step consisting of four actions:
(i) draw a random node called the active node, (ii) draw a random neighbor of the active node,
(iii) the active node copies the state of the neighbor (iv) with probability $\varepsilon$
the active node adapts one of $n$ states at random.
The noise rate $\varepsilon$ is a parameter of the simulation.
For $\varepsilon=1$ the state updates are fully random 
and state copying is always overridden by noise (in such case the results can 
be described analytically, see the \textit{Binomial approximation} section
is the Supplementary Information and Fig.~S4).
For $\varepsilon=0$ there is no noise and nodes can change their states solely 
due to interaction.

There are two possible phases in the NVM which can be obtain by changing the noise rate:
a consensus phase and a fully-mixing phase.
We perform simulations in the fully-mixing phase in order
to obtain symmetrical results for all parties. Otherwise it would be 
impossible to asses to what degree the superiority of one party is
caused by the zealots and media propaganda, and what is just a typical outcome
of the consensus phase.
The simulation is ran for a fixed number of time steps
until a stationary state is obtained before collecting
election results. After thermalization, elections are performed a given number of times,
each election being separated
a given number of Monte Carlo (MC) time steps from the next one
(typically 50 MC time steps).
When studying zealot susceptibility, however, we always reset the setup
by initializing the states of nodes and
thermalizing the system again before every election. This ensures
that zealots are in a random (and different) part of the network before each election.
A simulation ends after performing desired number of elections.

Zealots are introduced in the network at the beginning
of the simulation after initializing state of the nodes. The nodes for
zealots are randomly selected, they are marked as zealots and they change the state
to the zealot state $s_b$, regardless of what state they had at the beginning.
As mentioned, when the states are reset zealots are randomly chosen again.
Zealots always have the same state $s_b=a$ and they never change
it during the simulation, i.e.
when a zealot is chosen as the active node all actions are skipped and we move
to the next time step.

Media propaganda is simulated by modifying the random state update (action no. iv)
into a biased state update in favor or against a given party.
In the non-biased case every party has the same probability of $1/n$ to be chosen by
the active node.
A bias is introduced by fixing the probability $p_m \in [0, 1]$ of choosing
a particular party. If $p_m<1/n$ (negative bias) the propaganda is directed against the party,
if $p_m>1/n$ (positive bias) the propaganda works in favor of the party.
For simplicity, we always choose party $a$ as the one with the probability $p_m$.
All other parties are equally likely
to be chosen, and the probability for each of them is equal to $(1-p_m)/(n-1)$.
We measure the strength of media bias as $p_m-1/n \in [-1/n, (n-1)/n]$.
This range depends on the number of parties $n$, however a negative or positive value
always indicates a propaganda directed against or in favor of party $a$ respectively.


\subsection*{Applying electoral systems}

Our simulations incorporate elections using different electoral systems,
and each of them may have a distinct division into constituencies.
We determine the voters of each electoral district under each ES at the beginning of the simulation.
In most cases electoral districts
overlap with topological communities of the SBM, but
our framework allows merging multiple topological communities into a single constituency.
Importantly, we can test various divisions
into districts during the same simulation,
ensuring that observed differences are not solely attributed to specific
simulation trajectories.
Furthermore, we can define the number of seats 
available in each district, accommodating both plurality voting systems, 
such as FPTP or PBV, and proportional representation systems.
The model supports the application of an
electoral entry threshold and offers several popular seat assignment methods 
like the Jefferson method and the Webster method.

The election process begins with vote counting for each party, i.e. counting
nodes in each state. If an entry threshold exists, voters 
affiliated with parties falling below the threshold are excluded. Seats 
available in each district are then assigned to parties according to the 
specified method, and the obtained seats are aggregated to determine the
final seat distribution. Vote share and seat share data for all parties 
under different electoral systems are recorded and the simulation can continue.


\subsection*{Disproportionality and political fragmentation indexes}

Let $n$ be the number of parties running in the elections, the fraction of votes obtained
by party $i$ be $v_i\in[0,1]$, and the fraction of seats assigned for the party be $t_i\in[0,1]$.
Then, the Gallagher index  \cite{gallagher1991proportionality,gallagher1992comparing,benoit2000electoral}
can be expressed as:
\begin{equation}
    G = \sqrt{\frac{1}{2} \sum_{i=1}^n (v_i - t_i)^2 } ,
    \label{eqn:gallagher}
\end{equation}
and the effective number of parties \cite{laakso1979effective} is given by:
\begin{equation}
    E = \frac{1}{\sum_{i=1}^n t_i^2 } .
    \label{eqn:eff_num_parties}
\end{equation}
Note, that $G\in[0,1]$ with higher values indicating more disproportionality,
and $E\in[1,n]$ with  higher values indicating more
political fragmentation. In political sciences literature, the Gallagher
index is often computed using percentages of seats (instead of fractions)
and so gives a result which is a percentage as well.
To obtain the percentage one has to
simply multiply the value which we provide by 100.


}

\showmatmethods{} 


\acknow{
This work was supported by the Spanish Ministry of Science and Innovation under the grant number PID2022-141802NB-I00.
TR would like to thank Juan Fern\'andez Gracia for the discussions at the early stage of the research.
RC gratefully acknowledges a discussion with Dr. Divya Pandey about the India’s House of the People.}

\showacknow{} 


\end{document}



\maketitle
\tableofcontents
\listoffigures

\newpage

\SItext


\section{List of abbreviations}
All the abbreviations used in the supporting information text and the main manuscript in alphabetic order:
\begin{itemize}
    \item \textbf{ES} - Electoral System / Systems,
    \item \textbf{FPTP} - First Past The Post, PV electoral system with single-member districts, aka SMP,
    \item \textbf{IRQ} - Interquartile Range, the difference between the 75th and 25th percentiles of the data,
    \item \textbf{IRV} - Instant-Runoff Voting, a majoritarian electoral system,
    \item \textbf{MC} - Monte Carlo, e.g. Monte Carlo time steps,
    \item \textbf{MMP} - Mixed-Member Proportional representation, a mixed electoral system,
    \item \textbf{MV} - Majoritarian Voting, one of the categories of electoral systems,
    \item \textbf{NVM} - Noisy Voter Model, the model we use to simulate opinion dynamics,
    \item \textbf{PBV} - Party Block Voting, PV electoral system with multi-member districts, aka general ticket,
    \item \textbf{PR} - Proportional Representation, one of the categories of electoral systems;
    if not indicated otherwise, we assume the Jefferson method of seat assignment,
    \item \textbf{PR-G} - Proportional Representation with one global, nation-wide constituency,
    \item \textbf{PR-W} - Proportional Representation with the Webster method of seat assignment,
    \item \textbf{PV} - Plurality Voting, one of the categories of electoral systems,
    \item \textbf{SBM} - Stochastic Block Model, the model we use for generating networks,
    \item \textbf{SMP} - Single Member Plurality, PV electoral system with single-member districts, aka FPTP, or SMDP - single member district plurality,
    \item \textbf{SNTV} - Single Non-Transferable Vote, a semi-proportional plurality voting electoral system,
    \item \textbf{STV} - Single Transferable Vote, a proportional electoral system.
\end{itemize}


\section{Types of electoral systems}

Significant differences in electoral systems (ES) exist across countries and regions,
reflecting a combination of historical, cultural, and political factors.
By analyzing the characteristics and outcomes associated with different electoral systems,
we can deepen our understanding of the complex interplay between institutional design,
political representation, and democratic outcomes.
Electoral systems can be broadly categorized into three main types: proportional representation systems (PR),
plurality voting systems (PV), and majoritarian voting systems (MV) \cite{bormann2013democratic,farrell2011electoral}.
Additionally, different kinds of mixed electoral systems exist,
which combine selected properties of the three main categories. Some authors prefer a division into two categories,
where plurality systems and majoritarian systems are treated as two branches of the same category.
In general, PR systems strive to create
good representation of votes distribution in the parliament, with
a diverse range of political perspectives including minorities,
and PV systems aim at producing high governability, i.e. accountable governments with
enough power to introduce new legislation or make important reforms.

Proportional representation is the most common system used to elect
members of national legislatures present in the majority of European and South American countries,
as well as in many Asian and African ones. It is usually associated with a party list,
which can be either open, or closed.
In open list PR voters influence the order of party's candidates entering the parliament --
the candidates with the biggest number of votes gain seats, regardless of their position on the list.
Examples of legislatures using open list are the Poland's Sejm, the Brazil's Chamber of Deputies,
or the National Assembly of the Democratic Republic of the Congo (for most of its constituencies).
In closed list PR voters effectively vote just for a party -- who enters the parliament is decided
by the party by choosing the order of candidates.
Closed list is used in governing bodies such as the Israel's Knesset, the Argentina's Chamber of Deputies,
or the Grand National Assembly of Turkey.

Under PR systems countries are usually divided into multiple multi-member constituencies (electoral districts), but
one nation-wide constituency can also exist (as for the Israel's Knesset).
Voters select a candidate from the party lists available in their district.
A crucial part of PR electoral systems
is a method of seat allocation within districts, which provides an algorithm for rounding the results
to fit them into a discrete number of seats (a fraction of votes won in a district
multiplied by the number of seats is typically not an integer).
Generally, the rounding methods are from two categories: highest average methods (or divisor methods)
and largest remainder methods.
The first category includes the Jefferson-D'Hondt method (we refer to it as Jefferson method for simplicity),
the Webster/Sainte-Lagu\"e method  (we refer to it as Webster method), and the Huntington-Hill method,
to name the main ones. In the second category some popular methods are 
the Hare quota (aka the Hamilton method), the Droop quota, and the Imperiali quota. 
Most of them have a certain amount of bias
towards larger parties \cite{marshall2002majorization}.
Minorities have the biggest disadvantage under the
Jefferson method and the Imperiali quota \cite{schuster2003seat}, however
less disadvantageous extensions of them have been proposed \cite{medzihorsky2019rethinking}.
On the other hand, the Webster method and the Hare quota give an advantage
to small parties. It was also shown that they
tend to be more robust in respect to a set of multiple disproportionally indexes
under varying number of parties and district magnitude
\cite{pennisi1998disproportionality,benoit2000electoral}.
Finally, applying an entry threshold for parties is common in PR systems --
below a certain score on the national level
a party will not obtain any seat and its votes will not be taken into account.
Ranked versions of PR also exist,
for example single transferable vote (STV), but are more rarely applied (one of the few exceptions
can be the House of Representatives in Malta).
STV is a preferential voting method where voters rank candidates in order of preference.
Seats are allocated to candidates who reach a specific quota, and surplus votes from elected
candidates or eliminated candidates are transferred to remaining candidates based on voters'
subsequent preferences until all seats are filled.

The second most common electoral system in national legislatures is plurality voting.
Broadly, the candidate or party who receives the highest number of votes in PV,
regardless of the share of the total vote, wins the election.
It usually involves dividing a country into single-member districts,
where only one candidate per constituency can enter the parliament.
In this case it is referred to as single member plurality (SMP),
or first-past-the-post voting (FPTP).
It is used in legislative bodies such as the House of the People in India,
the House of Commons of Canada, or the Yemen's House of Representatives
(many of former British colonies use FPTP system).
Plurality voting can also be applied to multi-member districts,
where all the available seats are given to the best scoring party,
and is then called general ticket or party block voting (PBV).
Although it is less common and may result in a landslide victory of one party,
PBV is still used in some countries. Perhaps the most well know example
is the United States Electoral College -- presidential electors in all but two states
are chosen using general ticket. It is also used in the Parliament of Singapore
and the National Assembly of Ivory Coast.
Another generalization of FPTP voting into multi-member constituencies
is single non-transferable vote (SNTV). It is also less common, but still used,
for instance in the Legislative Assembly of Puerto Rico.
SNTV is a simple plurality-based voting method, where voters cast a single vote for a candidate
in a multi-member constituency, but it cannot be transfered to other candidates as in STV.
The candidates with the highest number of votes are elected. SNTV can lead to disproportional outcomes,
as parties with concentrated support may win seats,
while others with broader but less concentrated support may be left without representation.
Nevertheless, it leads to more proportional outcomes than PBV.

Majoritarian systems emphasize the representation of a majority of voters,
often resulting in a winner-takes-all approach. 
The candidate has to simply obtain a majority (over 50\%) of votes. These systems are
most widely used in presidential elections. Majoritarian voting can be achieved by
using several rounds of voting or ranked voting, for example instant-runoff voting (IRV).
The multi-round system usually consists of two rounds; in case of no candidate
achieving the majority in the first round, the two candidates with the best results
take part in the second round. Sometimes, the worst-scoring candidate may be
eliminated from the next round until only two remain. We do not include majoritarian systems in our analysis
since they are uncommon in elections to national legislatures (with exceptions of
the Australia's House of Representatives and the Papua New Guinea's National Parliament).

Within these overarching categories, numerous specific types of electoral systems have emerged,
each with its own distinctive features and consequences.
Moreover, hybrid or mixed systems have gained prominence, incorporating elements of both plurality voting
and proportional representation systems.
The simplest form of a mixed system is called parallel voting, where normally one part of the seats
is assigned using FPTP and the other using party-list PR. Votes cast in the former system do not influence
the latter, and vice versa,
because they are used to fill a separate and predefined part of the legislature. In the end, however,
elected candidates serve in the same chamber. Such system is used in the Mazhilis of the Parliament of
the Republic of Kazakhstan, or the House of Representatives of the Philippines for example.
A more complex hybrid system is mixed-member proportional representation (MMP), which
combines seats from single-member districts allocated by first-past-the-post rule
with party-list proportional representation
to achieve proportional results on the national level (or the state level).
MMP is used in the German federal parliament, the Bundestag, and the New Zealand's House of Representatives.
In pallarel voting and MMP voters have two votes, one for a single-member constituency candidate
and one for a party list, but the allocation of seats is done differently.

Some authors suggest that hybrid ES will provide
the ultimate compromise between proportionality and governability
\cite{shugart2001mixed}. Mixed ES certainly can obtain a highly proportional
seat allocation, like in the German federal parliament.
However, mixed systems create their own issues \cite{michalak2016mixed,blais1996mixed}.
They seek to strike a balance between
the need for stable governance and the goal of broad representation, 
but paradoxically they can loose both of these traits. They are also
more complicated than the regular electoral systems, leading
to a potential confusion of voters \cite{bowler1991voter}.

The equilibrium between the proportionality of apportionment
and governability can also be adjusted by changing district magnitude in PR systems.
Constituencies with fewer available seats obviously promote more centralized power,
while districts with a big number of seats can achieve highly proportional
representation. Adjusting district magnitudes can potentially yield
a balance for a good representation of voter preferences and accountable
governments \cite{carey2011electoral}. Experiments show, however, that
this is not an easy task: ,,the most likely outcome is still
having either high representativeness or good accountability,
not both'' \cite{st2016electoral}.


\section{Election simulations -- methods}

To simulate elections, we employ a conceptually simple yet comprehensive
model consisting of multiple interconnected components. The main ones are: the complex network
of social interactions, the opinion dynamics which takes place on top of this network and leads
to a given vote distribution, and the application of electoral systems to assign seats based on the votes.
Each of these components
is meticulously designed to reflect real-world phenomena and structures observed in actual electoral processes.
This flexibility enables us to customize specific elements to suit empirical
examples and conduct detailed analyses of the electoral systems in various countries.
In the following sections, we describe our simulation framework,
constructed to evaluate both abstract and real-world electoral systems.


\subsection{Network generation}

The first step in simulating electoral processes is constructing a population of voters,
who can interact, choose their candidate, and who eventually will cast a vote.
Voters are placed on a network generated using the
Stochastic Block Model (SBM) \cite{holland1983stochastic,Lee2019review}, which enables the division of nodes into
multiple topological communities. In principle, each node represents one person,
but sometimes it is interpreted as a homogeneous group of people.
We treat the topological communities
as electoral districts, assuming that individuals are more likely to interact
within their electoral district than outside of it, i.e. the probability of interaction
decreases with distance between two people. This assumption is not only reasonable, 
but also finds strong
support in the scientific literature, as observed in the case of counties during
the US presidential elections \cite{fernandez2014voter}.
The network in the SBM is therefore constructed by assigning lower probabilities
for inter-community connections and higher probabilities for intra-community
connections.

Utilizing the SBM allows us to parameterize the network based on
key quantities, including the network size $N$, the number of topological communities $q$
(they can be of different sizes),
the average degree $k$ (i.e. the number of connections a node has),
and the interconnectedness ratio $r$, which is the probability of forming a connection between communities
divided by the probability of forming a connection within a community.
Once initialized, the network structure remains
constant throughout the simulation. Our model can reflect the structural characteristics
of real-world electoral processes by adjusting the number of districts, their relative
sizes, and their connectivity.

For simulations of abstract electoral systems we generate the network using
a simplified version of the SBM, sometimes referred to as a planted partition model.
The connectivity between the districts and within them is
decided solely by the value of the interconnectedness ratio $r$, and each pair of districts has
the same probability of being connected. In other words, there is no spatial or geographical
structure in the network, except for the division into communities.
We specify the number of nodes $N$, the average degree $k$, and the number of communities $q$ -- each community is
of the same size with $N/q$ nodes (our framework also allows defining different sizes
for different communities). The total number of links is equal to $Nk/2$.

The network is generated slightly differently for simulations of real electoral systems.
After specifying the network size $N$ with the average degree $k$,
it is divided into communities corresponding to actual electoral districts. Therefore,
the proportions between sizes of topological communities are the same as
the proportions between voter populations in real electoral districts.
Moreover, for each community we specify geographical coordinates corresponding to the location
of the capital of the electoral district it represents. Every node in the network, in consequence,
has a geographical location (of the capital of its district). In the modified version of the SBM
that we use for countries, the probability of forming a connection between two nodes
decreases with the geographical distance (more precisely the great-circle distance) between them as:
\begin{equation}
    p(x) \propto \frac{1}{(x+c)^2} ,
\label{eqn:planar_affinity}
\end{equation}
where $x$ is the distance between the nodes and $c$ is a parameter we call the planar constant.
Probability $p$ is then normalized according to the specified average degree.
Note, that all nodes within a single electoral district will have the same location
and therefore the distance between them will be $x=0$, giving $p \sim 1/c^2$.
For a pair of nodes from distinct constituencies the distance $x$ is the distance between
the districts' capitals.
We choose [\ref{eqn:planar_affinity}] as the simplest relation which 
is finite in zero and can be normalized as probability
distribution (hence the second power, which gives a finite integral).
The planar constant $c$ is taken from the best fit
to empirical data on work commuting distance for a given country, since it was shown to be a good
proxy for the distance of interaction between voters \cite{fernandez2014voter}.
More precisely, for each country we take the data on the percentage of people
traveling to work up to 10 km, 20 km, 30 km etc, and this way we can approximate
the fraction of people who travel in certain distance bins (for example, what fraction of people
travel between 20 and 30 km). Then we use [\ref{eqn:planar_affinity}]
to compute the same fractions (for the same distance bins) as a function of parameter $c$.
Finally, we fit $c$ such that it minimizes the square distance between empirical and model bins.
Results of this procedure are shown in Figures~\ref{fig:pl_sejm-basic_info} 
(repeated in \ref{fig:pl_senate-basic_info}), \ref{fig:il_knesset-basic_info},
and \ref{fig:ind_house_otp-basic_info}.
In this manner we obtain a realistic data-driven connectivity between the districts.
You can also see our public code repository \cite{github_repo}
to check the details of network generation.


\subsection{Opinion dynamics -- the voter model}

After generating an appropriate network for the specific electoral setting the simulation of
electoral processes can be conducted.
The model we use to simulate opinion dynamics in a population of voters
is the noisy voter model (NVM)
\cite{granovsky1995noisy,carro2016noisy,redner2019reality,raducha2018coevolving,raducha2020emergence,jedrzejewski2020nonlinear}.
In its basic formulation,
there are two possible states of nodes, usually denoted by
$s_i \in \{ -1, 1 \}$. They represent two opposite opinions, which voters change
by interacting with others via links in the network or by independent decisions.
The opinion of voters
translates into votes, hence two possible states also represent two parties
or candidates running in the elections. Although binary choice
can be the case in some elections (e.g. the last round of majoritarian
presidential elections), the range of candidates is usually bigger
in the elections to national legislatures.
Therefore, we extend the standard voter model to accommodate
$n$ possible states $s_i \in \{a, b, c, ... \}$, representing $n$ possible
candidate choices (we use alphabetic characters to highlight that they only denote different
states or opinions and their numerical value does not matter).
Put simply, there are $n$ parties running in election and a node
in state $a$, for instance, will vote for party $a$.
This approach has been studied before under the name
of multi-state noisy voter model
\cite{herrerias2019consensus,nowak2021discontinuous,khalil2021zealots}.

\begin{figure}
\centering
\begin{subfigure}{\textwidth}
    \centering
    \includegraphics[scale=0.7]{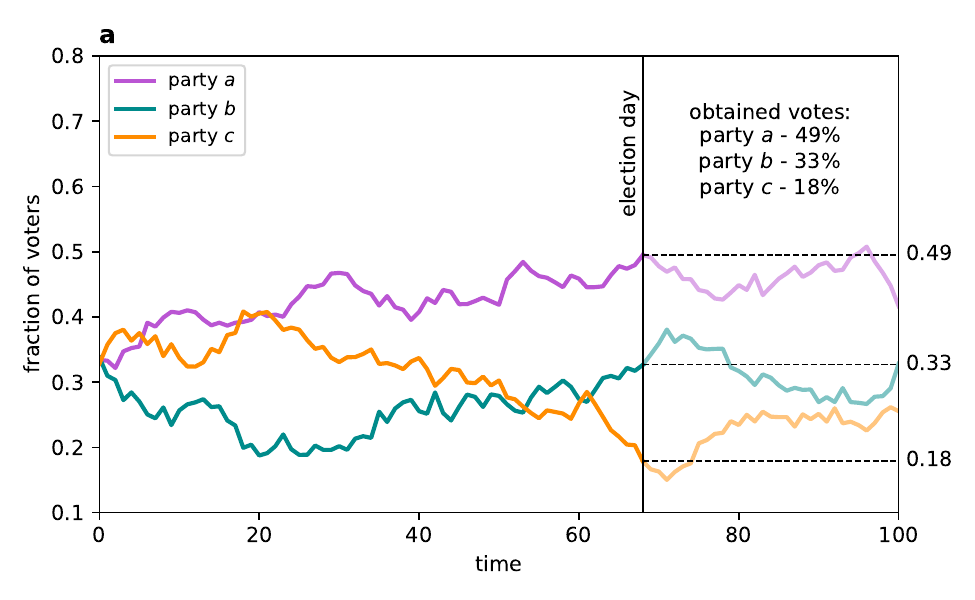}
\end{subfigure}\vspace{0.2cm}
\begin{subfigure}{\textwidth}
    \centering
    \includegraphics[scale=0.65]{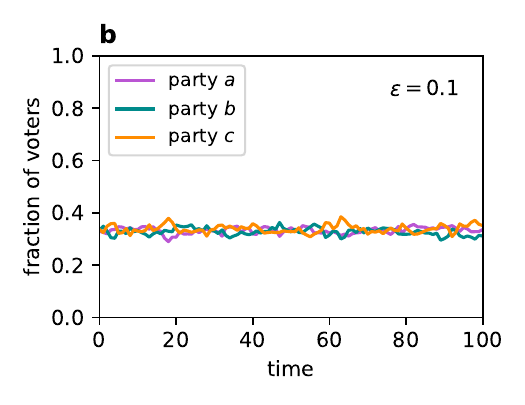}
    \includegraphics[scale=0.65]{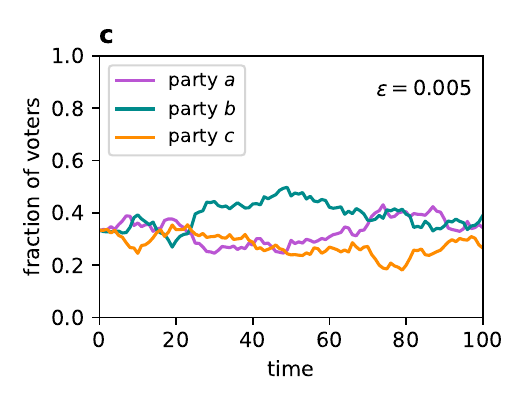}
    \includegraphics[scale=0.65]{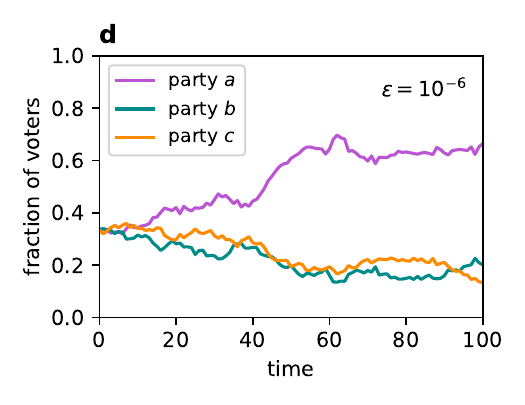}
\end{subfigure}
\caption[Support for parties over time in simulations]{\textbf{Support for parties over time in simulations.}
(a) Changes over time of the fraction of voters supporting each of 3 parties in the noisy voter model,
which directly translates
into the fraction of votes. An exemplary day of election is indicated together
with the fraction of votes each party would obtain on that day. We ran such simulations many times applying
different electoral systems to the obtained numbers of votes. This way we can compare ES over
large samples of election results. Here the simulation was ran for $\varepsilon=10^{-5}$,
$N=10^4$ divided into 100 equal communities in the SBM, which also form electoral districts.
The average degree $k=12$ and the ratio of probabilities of forming inter-links to intra-links is $r=0.002$.
Time is counted in 5 MC time steps (i.e. the simulation was ran for $5\times 10^6$ time steps).
(b, c, d) The same simulation ran for different values of the noise rate $\varepsilon$.
Note how the fluctuations change depending on noise. For very high noise all parties will maintain support of
approximately 1/3 of the population. For very low noise it is easy for one party to gain support of almost
the whole population and maintain it for a long time. We chose values of noise in between that allow changes
of the leading party over time.
}\label{fig:basic_trajectories}
\end{figure}

After generating the structure of the network,
every node is assigned one
of $n$ states at random with a uniform probability. Then, in every time step:
\begin{enumerate}
    \item we draw a random node (called the active node of this time step),
    \item we draw a random neighbor of the active node,
    \item the active node copies the state of the neighbor,
    \item with probability $\varepsilon$, regardless of what happened before,
    the active node adopts one of $n$ states at random (with $1/n$ probability for each state),
\end{enumerate}
where the noise rate $\varepsilon$ is a parameter of the simulation.
See Figure~\ref{fig:basic_trajectories} for exemplary trajectories of the model.
Noise in the model represents independent decisions of the voters, i.e. choosing the preferred
candidate while ignoring the influence of their peers. Conversely, state copying
represents social interactions and a decision made due to the pressure from peers.
If the active node and the selected neighbor have the same state, the action
no. 3 has no effect. If during the action no. 4 the active node draws the same state
as the one it had at the beginning, the time step effectively will cause no change.
Note also, that for $\varepsilon=1$ the state updates are
fully random, since the state copying is always overridden by the noise.
For $\varepsilon=0$ there is no noise in the system and voters
can change their states solely due to social interaction, what causes
the full consensus to be a frozen configuration.

There are two possible phases in the NVM: the disordered (or fully-mixing/coexistence)
phase is characterized by each of the $n$ parties
being supported by approximately $1/n$ voters, subject to fluctuations
(see e.g. Figure~\ref{fig:basic_trajectories}~(c)).
The distribution of vote share for each party is unimodal in this phase,
meaning that the support of $1/n$ of nodes is the most probable outcome.
In the second possible phase, the consensus phase,
one party gains control over the network.
In Figure~\ref{fig:basic_trajectories}~(d) we can see one of the parties
quickly gaining substantial fraction of votes.
This domination changes periodically with different
parties gaining control, hence the distribution of vote share for each party is
bimodal -- two most probable outcomes are gaining no votes and gaining all of them
(note, that we do not utilize the consensus phase and therefore we do not obtain a full
domination of one party in terms of votes).
It has been shown that an order-disorder phase transition is induced by changing
noise \cite{carro2016noisy,raducha2020emergence,herrerias2019consensus}.
For low values of $\varepsilon$ the network is in the consensus phase
and for high values in the fully-mixing phase, therefore
the preferred configuration can be obtained
by adjusting the noise rate $\varepsilon$ in the simulation.
We perform simulations only in the fully-mixing phase in order
to obtain symmetrical results for all parties. Otherwise it would be 
difficult to asses to what degree the superiority of one party is
caused by the zealots and media propaganda, and what is just a typical outcome
of the consensus phase.

The simulation is first ran for a fixed number of time steps (called thermalization time)
until a stationary state is obtained before collecting
election results. After thermalization, elections are performed a given number of times,
each election being separated
a given number of Monte Carlo (MC) time steps from the next one
(usually it's 50 MC time steps).
Note, that one time step consists of four actions given in the list above, while
one MC time step means performing the basic time step $N$ times.
Figure~\ref{fig:basic_trajectories}~(a) illustrates this process up to the first
election.
The other possibility is to initialize the states of nodes and
thermalize the system again before every election.
We did not observe any difference
in the results between the default mode and resetting, however
when studying zealot susceptibility we always reset the setup, 
so the zealots are in a different (and random) part of the network before each election.
A simulation ends after performing the desired number of elections;
this provides a statistics of votes and of results for each ES,
which are presented in Figure~\ref{fig:basic-es_histograms}.

As mentioned before, one of the studied external factors possibly influencing elections
are zealots. Zealots are people who do not change their opinion, neither due
to social interactions, nor due to individual choices. They represent political
agitators, aiming at spreading their political views.
The NVM with zealots has been studied and as expected
zealots were showed to skew the opinion distribution, even to extreme levels
\cite{Mobilia2007On,khalil2018zealots,Klamser2017Zealotry,Kowalski2020Role},
also in the multi-state version of the model
\cite{khalil2021zealots}.
Zealots are introduced in the network at the beginning
of the simulation after initializing state of the nodes. For example, if there are 10 zealots,
10 nodes are randomly selected, they are marked as zealots and they change the state
to the zealot state $s_b$, regardless of what state they had at the beginning.
Additionally, each time after performing election,
when the states are reset, zealots are randomly chosen again.
This way the effect of zealots is not
restricted to a particular district.
Zealots always have the same state $s_b=a$ and they never change
it during the simulation. This means that they always vote for party $a$ and
also agitate for this party.
When a zealot is chosen as the active node, the rest of the actions are
skipped and the procedure moves on to the next time step.
Nevertheless, they can and do influence others.

The second external influence we consider is propaganda spread by the media
\cite{sirbu2017opinion,gonzlezavella2007,crokidakis2012effects,rodriguez2010effects,moya2017agent,gonzalez2012model}.
This can be simulated by modifying the random state updates (action no. 4)
into biased state updates in favor
or against a given party.
Random state updates reflect free choices of
individual voters, therefore existence of bias can be interpreted as
the influence of unbalanced media campaigns or straightforward propaganda.
In the non-biased case every party is equally likely to be chosen by
the active node (i.e. the probability is $1/n$ for every party).
Media propaganda is introduced by fixing the probability $p_m \in [0, 1]$ of choosing
party $a$. If $p_m<1/n$ the propaganda is directed against party $a$,
if $p_m>1/n$ the propaganda works in favor of party $a$.
For simplicity, we always choose party $a$ as the one with the probability $p_m$.
All other parties are equally likely
to be chosen, and the probability for each of them is equal to $(1-p_m)/(n-1)$.
We measure the strength of media bias as $p_m-1/n \in [-1/n, (n-1)/n]$.
This range depends on the number of parties $n$, however a negative or positive value
always indicates a propaganda directed against or in favor of party $a$ respectively,
and a value of 0 corresponds to no bias with all parties having the same probability.

Our framework offers additional possibility of performing simulations
with a different kind of social influence,
namely the majority rule \cite{redner2019reality,boyd2005origin,grofman1983thirteen,conradt2009group}.
In this case the active node copies the state which is
shared by the majority of her neighbors. If there is a draw,
a random state is copied (from among the major ones, if we have
more than 2 states). Majority rule represents non-linear social pressure,
i.e. influence of the whole neighborhood instead of one chosen agent.
Nevertheless, our initial simulations showed qualitatively consistent
results for the standard voter model and the majority rule.


\subsection{Applying electoral systems}

A crucial part of our simulations is performing elections using different electoral systems.
Each electoral system may have a distinct division into constituencies.
We determine the affiliation of voters with electoral districts at the beginning of the simulation, and
this affiliation does not change later. The basic configuration corresponds to electoral districts
overlapping with topological communities of the SBM. In other words, voters of each electoral district
are more densely connected with each other than with voters of separate districts. Nevertheless,
our framework allows merging multiple topological communities into a single constituency, up
to the level of one global electoral district. In such case constituencies are not homogeneous
inside and have distinct internal communities. Typically, borders of electoral districts are
defined based on urban agglomeration, population distribution, topography, and perhaps historical reasons,
therefore voters of the same constituency are more likely to interact than those of separate constituencies.
In some cases, however, electoral districts are large and have diverse internal structure, like for Israel's
Knesset where a single constituency covers the whole country. Merging different topological communities into
one electoral district enables covering such cases. Importantly, we can test various divisions
into districts during the same simulation, making sure that observed differences are not merely an effect of
a particular trajectory of the simulation.

In addition, we can define an arbitrary number of seats available in each district and, consequently,
the total number of seats in the elections. The number of seats may vary between districts.
This flexibility makes it possible to simulate plurality voting ES with single-member districts,
i.e. first-past-the-post voting, as well as party block voting with more seats, covering most of the
real-world scenarios for PV.
If the number of seats is the same in each district, which is the case for some simulations,
we denote it by $qs$.
In order to simulate a wide range of proportional representation ES
we also accommodate the possibility of applying an electoral entry threshold and choosing a specific
seat assignment method. The entry threshold can be set at any percentage of votes and is nation-wide,
what means that votes cast for parties that obtain fewer votes in total are ignored in the final calculation.
Our model is capable of applying several of the most popular seat assignment methods, most importantly
the Jefferson method and the Webster method.

The first step of performing elections is to count nodes in every state $s_i$,
which directly translates into the number of votes cast for each party.
The entry threshold is applied, if there is one, to exclude voters of parties that fall below it.
Then, seats available in each district are assigned to parties according to the specified method
(FPTP and PBV can be seen as a seat assignment method giving all seats to the party with the best result).
Finally, seats won in each district are aggregated providing the final number of seats obtained by each party.
The vote share and the seat share of all parties under all considered ES are saved,
and if needed the simulation can continue.

When analyzing general properties of ES we consider four abstract ES: a PR system with a division into
many constituencies and the Jefferson method (indicated just by ,,PR'' in figures' legends),
an identical to the first one PR system but with the Webster method (indicated by ,,PR-W''),
a PR system with one constituency and the Jefferson method (indicated by ,,PR-G''),
and a PV system with a division into many constituencies which is either FPTP or PBV depending 
on the number of seats per district (indicated by ,,PV''). By default, the systems listed above do not have
a minimal electoral entry threshold. Nevertheless, when studying real ES we simulate the original system
with all its features (districts, seats, threshold, seat assignment etc.) and additionally
several alternative PR and PV systems, which change selected aspects of the original one.


\section{Binomial approximation}

The simplest version of the model is obtained for $\varepsilon=1$, where the
structure of the network is irrelevant, because voters always choose a random
state at the end of a time step. Although social interactions are a crucial
part of opinion dynamics, the case of their absence is theoretically interesting, as 
it can be fully described analytically.
The NVM with $\varepsilon=1$ is equivalent to a simple model where each node takes opinion $a$ or $b$
at any given moment at random, i.e. with probability $\frac{1}{2}$.
Therefore, the distribution of votes obtained by one
of two political parties has a fully stochastic character and is described by a binomial distribution:
\begin{equation}
    P(X=V) = \genfrac(){0pt}{0}{N}{V} p^V (1 - p)^{N-V} \equiv B(V; N, p),
    \label{eqn:binom_dist}
\end{equation}
where $V$ represents the number of gained votes, $N$ is the number of voters,
and the probability of voting for a given party is equal to $p=\frac{1}{2}$.
For a PR electoral system with one nation-wide district and without an entry threshold
Eq.~[\ref{eqn:binom_dist}] is enough to describe the seat share distribution -- 
it is simply the same as the vote share distribution up to one seat per party, which can be rounded differently
depending on the method.
However, FPTP voting requires aggregation on a district level.
In this case $N$ becomes $q$, which is the number of districts, and $p$ becomes $p_d$, which represents the
probability of winning in a single district. We assume that the districts have the same size
$N_d = N/q$, which is the case in our simulations for abstract ES.
In the binomial approximation setting we get:
\begin{equation}
    p_d = \left\{ \begin{array}{ll}
    \sum_{V > \frac{N_d}{2}} B(V; N_d, p) & \,\textrm{if}\,\, N_d \,\,\textrm{is \textit{odd}} \\
    \frac{1}{2} B(\frac{N_d}{2}; N_d, p) + \sum_{V > \frac{N_d}{2}} B(V; N_d, p) & \,\textrm{if}\,\, N_d \,\,\textrm{is \textit{even}}
    \end{array} \right.
    \label{eqn:binom_dist_fptp}
\end{equation}

This setting also allows us to take into account mass media and zealots.
The former can easily be added to both electoral systems by changing $p=\frac{1}{2}$
into $p_m\in [0, 1]$ for party $a$ and $1-p_m$ for party $b$.
Incorporating $N_z$ zealots into the proportional representation can be done by updating
Eq.~[\ref{eqn:binom_dist}] into:
\begin{equation}
    P(X=V) = \left\{ \begin{array}{ll}
    B(V-N_z; N-N_z, p) & \,\textrm{if}\,\, V \geq N_z \\
    0 & \,\textrm{if}\,\, V < N_z
    \end{array} \right.
\end{equation}
For FPTP voting, instead of summing over all possible combinations of zealots,
we assume that each district gets an average (per district) number of zealots $\hat{N}_z$.
This approximation allows us to rewrite the probability of winning in a district as:
\begin{equation}
    p_d = \left\{ \begin{array}{ll}
    \sum_{V > \frac{N_d}{2}-\hat{N}_z} B(V; N_d-\hat{N}_z, p) & \,\textrm{if}\,\, N_d \,\,\textrm{is \textit{odd}} \\
    \frac{1}{2} B(\frac{N_d}{2}-\hat{N}_z; N_d-\hat{N}_z, p) + \sum_{V > \frac{N_d}{2}-\hat{N}_z} B(V; N_d-\hat{N}_z, p) & \,\textrm{if}\,\, N_d \,\,\textrm{is \textit{even}}
    \end{array} \right.
    \label{eqn:binom_dist_fptp_zealots}
\end{equation}
Note, that we round the average number of zealots up, increasing their effect on the results.
Eq.~[\ref{eqn:binom_dist_fptp_zealots}] describes only the seat share of the party supported by zealots, but
the collective result of other parties is just a symmetrical reflection of the zealot party
(seat fractions off all parties must add up to 1).

This approximation can easily be generalized to an arbitrary number of parties,
however, the numerical burden will increase accordingly.
The general distribution for $n$ parties is as follows:
\begin{equation}
    P(X_1=V_1,X_2=V_2,\dots,X_n=V_n) = \genfrac(){0pt}{0}{N}{V_1,\dots,V_n} p_1^{V_1}\cdot\ldots\cdot p_n^{V_n} \equiv M\left(\{V_i\}; N, \{p_i\}\right),
    \label{eqn:multinom_dist}
\end{equation}
where $\sum_i V_i = N$ and $\sum_i p_i = 1$.
Moreover, in the neutral case with no zealots or media $\forall_i \,\, p_i =1/n$.
If one is only interested in a distribution of votes for a single party in PR, it is sufficient to use the previous approximation derived for two parties but with $p=1/n$ or $p$ accordingly modified by media.
This also includes the case with zealots.
Adapting the approximation for PV, however, is much more complex for 3 or more parties and it requires computing:
\begin{equation}
    p_d = \sum_{\mathclap{V_1 > \frac{N_d}{n}}} \,\,\,\,\,\,\, \sum_{\mathclap{V_2 < V_1}} \dots \sum_{\mathclap{V_n < V_1}} \, \mathds{1}_{\left(\sum_i V_i = N_d\right)} M\left(\{V_i\}; N_d, \{p_i\}\right).
    \label{eqn:multinom_dist_fptp}
\end{equation}
The indicator is added only to highlight that the sums make sense only for particular cases of the set $\{V_i\}$.
In practice, this is already accounted for in $M\left(\{V_i\}; N_d, \{p_i\}\right)$,
which will be equal to zero for impossible configurations of $\{V_i\}$.
For simplicity, we do not take into account the special case of a draw, but this is analogous to the $n=2$ case.
Adding media bias is analogous to the case of two parties, one only needs to modify the set $\{p_i\}$ accordingly.
Zealots can also be added similarly as before:
\begin{equation}
    p_d = \sum_{\mathclap{V_1 > \frac{N_d}{n}-\hat{N}_z}} \,\,\,\,\,\,\,\,\,\,\,\,\,\,\,\,\, \sum_{\mathclap{V_2 < V_1+\hat{N}_z}} \,\,\,\,\, \dots \,\,\,\,\, \sum_{\mathclap{V_n < V_1+\hat{N}_z}} \,\,\, \mathds{1}_{\left(\sum_i V_i = N_d-\hat{N}_z\right)} M\left(\{V_i\}; N_d-\hat{N}_z, \{p_i\}\right),
    \label{eqn:multinom_dist_fptp_zealots}
\end{equation}
using the same approximation about the number of zealots in each district.
Exact equations and practical implementation for $n=3$ is available in our online code repository \cite{github_repo}.

As expected and shown in Figure~\ref{fig:basic-limit_cases}~(d) and~(p), above equations
are in good agreement with results of simulations for $\varepsilon=1$.
Nevertheless, the developed approximation, while theoretically interesting and valuable 
for verifying simulations, is limited to describing only a specific case of the model.
It can not capture the intricacies of real-world electoral processes,
where social interactions play a crucial role and can have a significant influence on
the stability of electoral systems (see Figure~\ref{fig:basic-limit_cases}).
Computer simulations, on the other hand, offer a more comprehensive 
understanding of the system, accounting for the interdependencies 
at play. Therefore, the primary approach to studying the model 
relies on simulations, as they provide a more accurate representation of the complex
dynamics.


\section{Disproportionality and political fragmentation indexes}

In addition to normal statistical measures and our novel measures of zealot and media susceptibility,
we use three indexes commonly applied in political sciences to evaluate electoral systems.
The Gallagher index \cite{gallagher1991proportionality,gallagher1992comparing,benoit2000electoral}
and the Loosemore-Hanby index \cite{loosemore1971theoretical,fry1991note}
quantify the amount of disproportionality between the votes gained and seats assigned to parties.
The effective number of parties \cite{laakso1979effective}
measures political fragmentation by counting parties based on their relative strength.
Let $n$ be the number of parties running in the elections, the fraction of votes obtained
by party $i$ be $v_i\in[0,1]$, and the fraction of seats assigned for the party be $t_i\in[0,1]$.
Then, the Gallagher index can be expressed as:
\begin{equation}
    G = \sqrt{\frac{1}{2} \sum_{i=1}^n (v_i - t_i)^2 } ,
    \label{eqn:gallagher}
\end{equation}
and the Loosemore-Hanby index as:
\begin{equation}
    LH = \frac{1}{2} \sum_{i=1}^n |v_i - t_i|  .
    \label{eqn:loosemore-hanby}
\end{equation}
Note, that these indexes can also utilize percentages of votes and seats, instead of fractions, and so
give a result which is a percentage as well. We use fractions of votes and seats, therefore both indexes
take values from 0 to 1, with higher value indicating more disproportionality.
The effective number of parties is given by:
\begin{equation}
    E = \frac{1}{\sum_{i=1}^n t_i^2 } ,
    \label{eqn:eff_num_parties}
\end{equation}
and it can take values from 1 up the the real number of parties $n$, where higher values indicate more
political fragmentation.

We compute all above indexes in most of our numerical experiments. In Figure~\ref{fig:basic-indexes}
we present basic differences between statistics of the three indexes for abstract PR and PV electoral systems.
Typically, PR systems minimize the Gallagher index and the Loosemore-Hanby index,
while PV systems minimize the effective number of parties.


\section{Results for abstract electoral systems}

Before applying our framework to real-world electoral systems, we conducted
simulations for several theoretical systems. This approach serves multiple 
purposes. Firstly, it allows us to elucidate the fundamental differences
between various electoral systems and how these differences manifest within 
the simulations. Secondly, we aim to explore the significance of key parameters.
Our objective is to obtain results that are qualitatively consistent with established
findings from the field of political sciences, 
concerning different aspects of electoral systems, such as district
magnitude, number of districts, and seat assignment methods. Subsequently,
we investigate how parameters governing network structure and opinion dynamics, such as network size,
degree, and noise rate, influence these outcomes. Lastly, we examine the impact 
of zealots and media propaganda within this controlled environment,
what provides a better understanding before the exploration of empirical cases.

We consider four abstract ES which are indicated in the figures by the following abbreviations:
\begin{itemize}
    \item PR - a PR system with a division into many constituencies and the Jefferson seat assignment method,
    \item PR-W - a PR system with a division into many constituencies and the Webster seat assignment method,
    \item PR-G - a PR system with one nation-wide constituency and the Jefferson seat assignment method, 
    \item PV - a PV system with a division into many constituencies 
    either FPTP or PBV depending on the number of seats per district.
\end{itemize}
Systems PR, PR-W, and PV always have the same division into districts and always have the same number of seats
available in each district, while PR-G has the same total number of seats as others,
but available in a single district. None of the systems have an entry threshold.
In the initial phase of our work we also performed some of the simulations for a mixed ES called
parallel voting. However, all the results were simply an average of results for PR and PV.


\subsection{General election statistics}

\begin{figure}
\centering
\begin{subfigure}{\textwidth}
    \centering
    \includegraphics[scale=0.65]{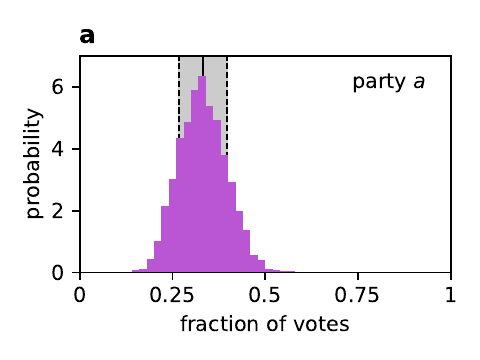}
    \includegraphics[scale=0.65]{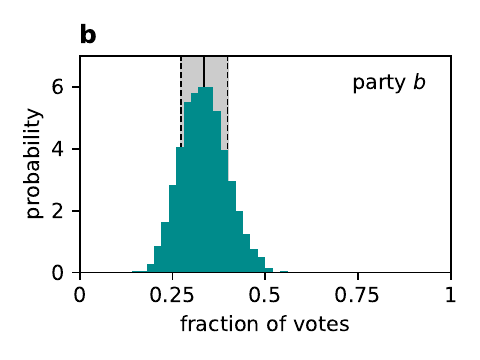}
    \includegraphics[scale=0.65]{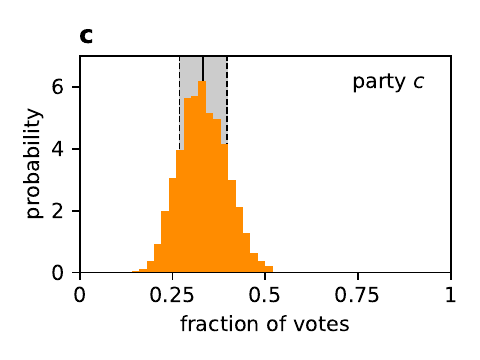}
\end{subfigure}
\begin{subfigure}{\textwidth}
    \centering
    \includegraphics[scale=0.65]{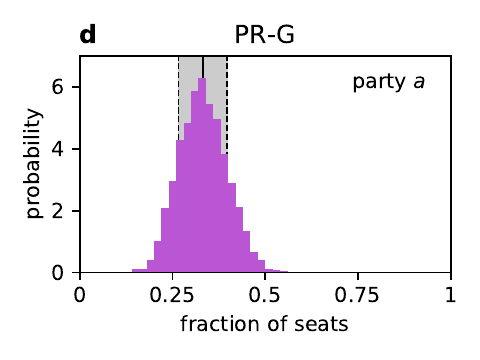}
    \includegraphics[scale=0.65]{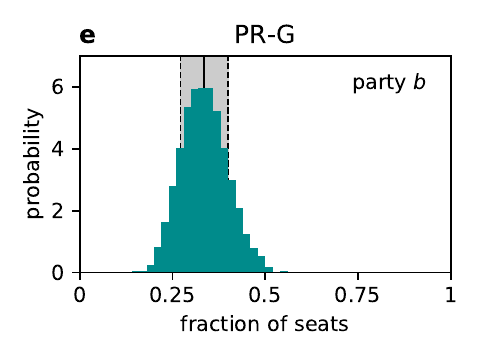}
    \includegraphics[scale=0.65]{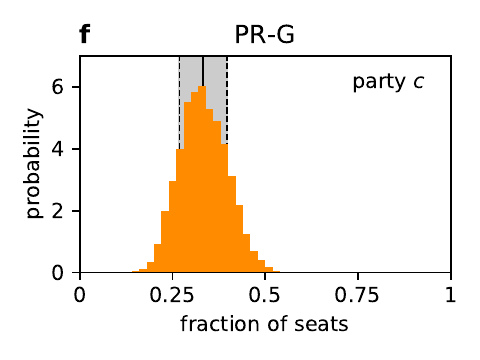}
\end{subfigure}
\begin{subfigure}{\textwidth}
    \centering
    \includegraphics[scale=0.65]{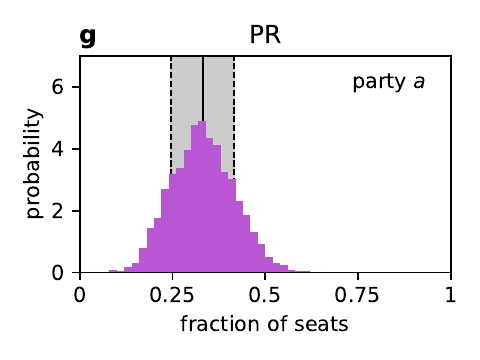}
    \includegraphics[scale=0.65]{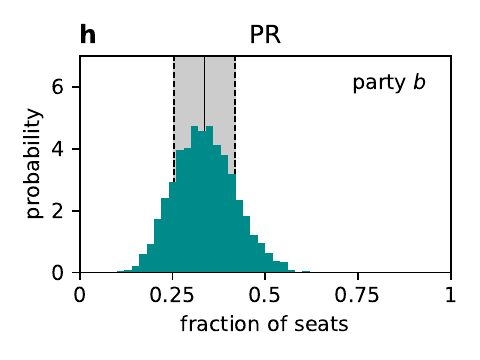}
    \includegraphics[scale=0.65]{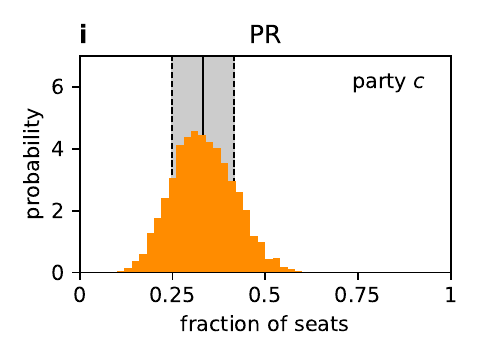}
\end{subfigure}
\begin{subfigure}{\textwidth}
    \centering
    \includegraphics[scale=0.65]{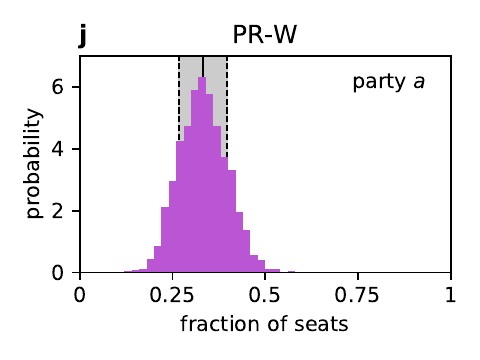}
    \includegraphics[scale=0.65]{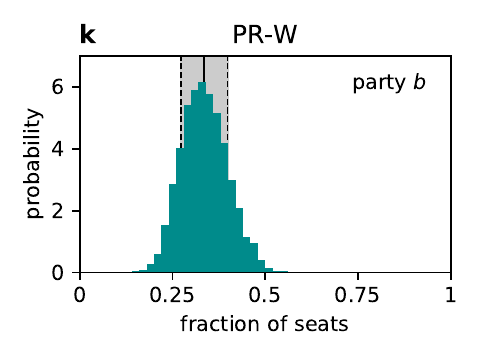}
    \includegraphics[scale=0.65]{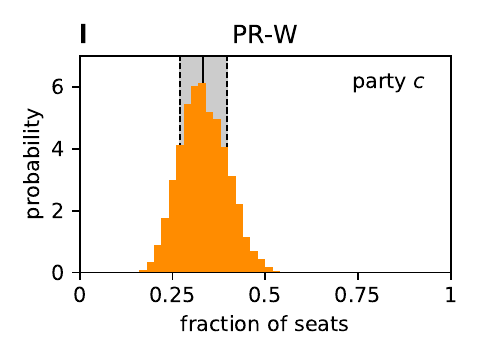}
\end{subfigure}
\begin{subfigure}{\textwidth}
    \centering
    \includegraphics[scale=0.65]{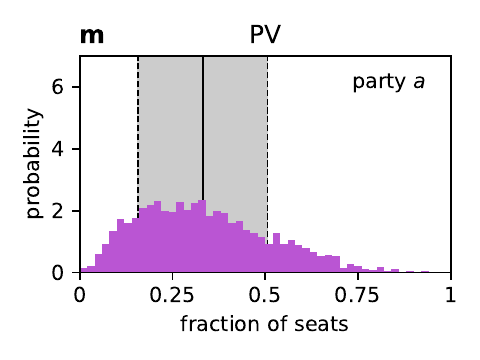}
    \includegraphics[scale=0.65]{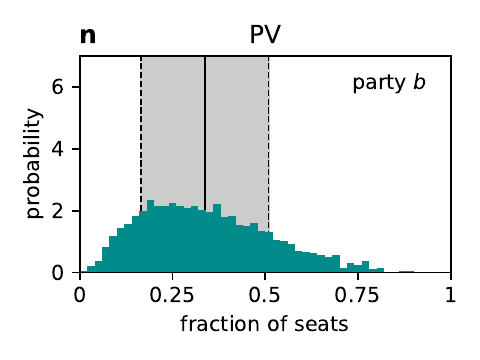}
    \includegraphics[scale=0.65]{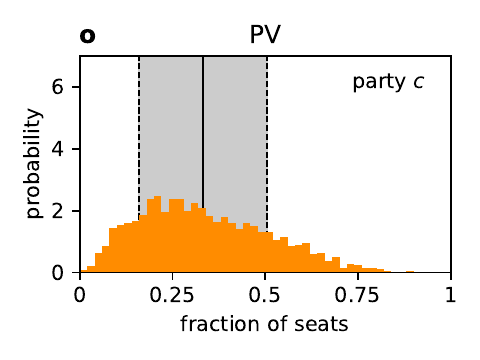}
\end{subfigure}
\caption[Election results under different electoral systems]{\textbf{Election results under different electoral systems.} The histograms represent
vote share and seat share distributions
(i.e. the probability distribution of the fraction of votes/seats) for $n=3$ parties:
party $a$ in purple, party $b$ in green, party $c$ in orange, obtained over
4000 simulated elections. Solid vertical line marks the average value and the gray area shows
results within one standard deviation from the average. 
(a, b, c) Vote share distribution obtained directly by counting the nodes in
a given state, as in Figure~\ref{fig:basic_trajectories}.
(d, e, f) Seat share distribution under proportional representation with one global electoral district
and Jefferson method (PR-G).
(g, h, i) Seat share distribution under proportional representation with 100 electoral districts
corresponding to topological communities and Jefferson method (PR).
(j, k, l) Seat share distribution under proportional representation with 100 electoral districts
corresponding to topological communities and Webster method (PR-W).
(m, n, o) Seat share distribution under plurality voting with 100 electoral districts
corresponding to topological communities (PV).
Note, that distributions are identical for each party, except from small fluctuations.
This is to be expected from a fully symmetrical setup in respect to parties and for that reason
we show results only for one of them in the following figures.
Simulations were ran for $\varepsilon=0.005$, $k=12$, $r=0.002$,
$N=10^4$ divided into 100 equal communities in the SBM.
Each electoral district corresponding to a topological community
has $qs=5$ seats assigned, giving 500 seats in total to win in elections.
}\label{fig:basic-es_histograms}
\end{figure}

\begin{figure}
\centering
\begin{subfigure}{\textwidth}
    \centering
    \includegraphics[scale=0.65]{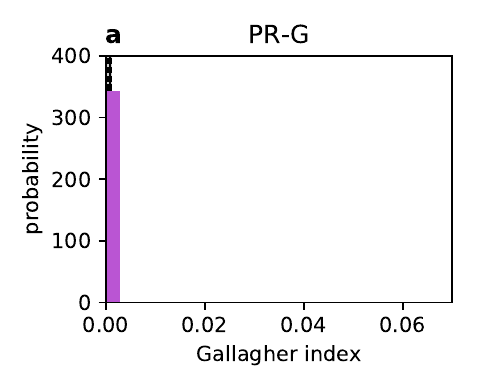}
    \includegraphics[scale=0.65]{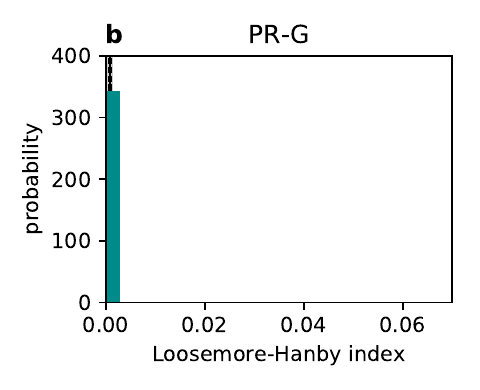}
    \includegraphics[scale=0.65]{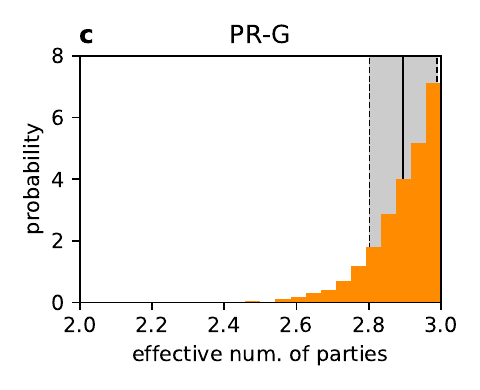}
\end{subfigure}
\begin{subfigure}{\textwidth}
    \centering
    \includegraphics[scale=0.65]{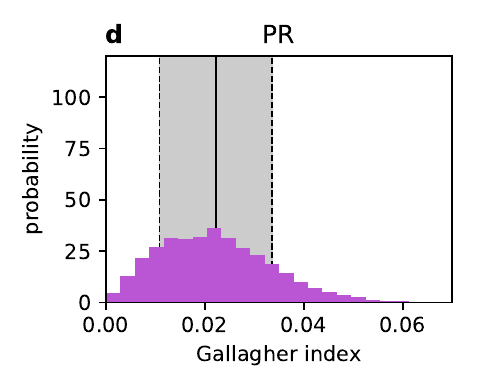}
    \includegraphics[scale=0.65]{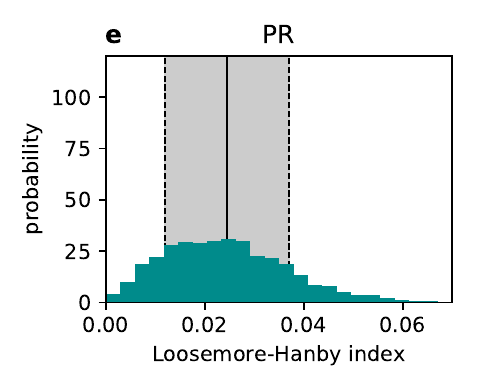}
    \includegraphics[scale=0.65]{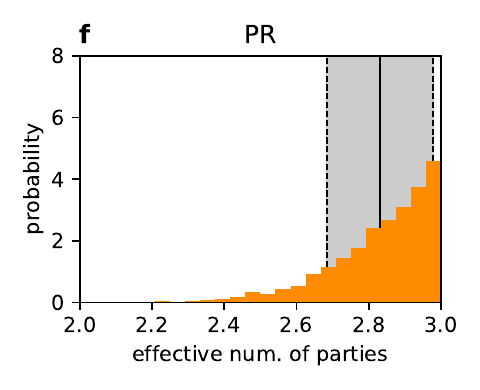}
\end{subfigure}
\begin{subfigure}{\textwidth}
    \centering
    \includegraphics[scale=0.65]{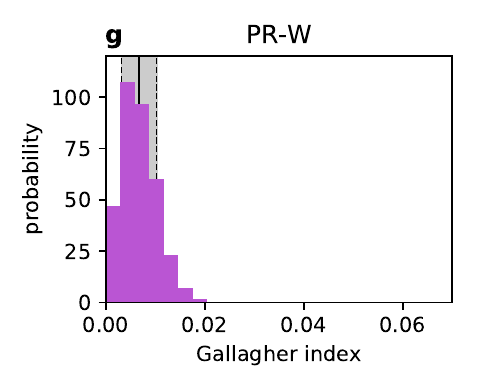}
    \includegraphics[scale=0.65]{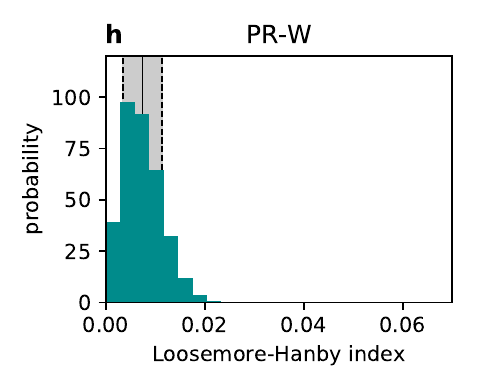}
    \includegraphics[scale=0.65]{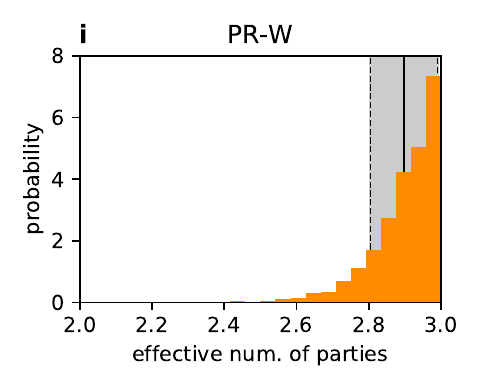}
\end{subfigure}
\begin{subfigure}{\textwidth}
    \centering
    \includegraphics[scale=0.65]{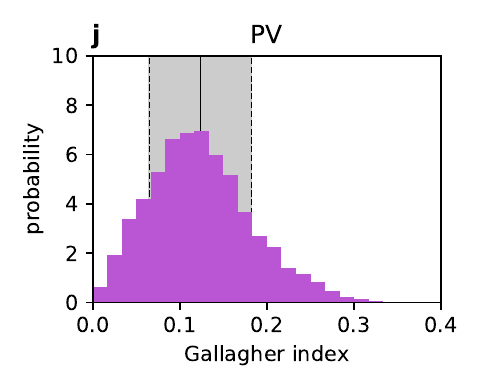}
    \includegraphics[scale=0.65]{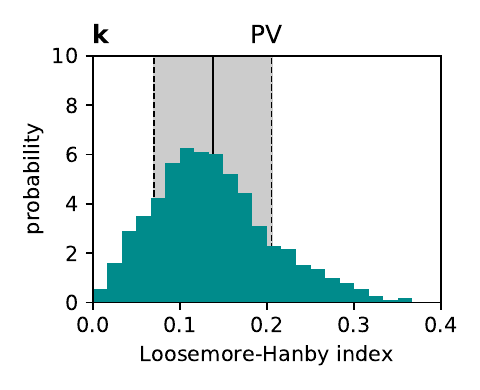}
    \includegraphics[scale=0.65]{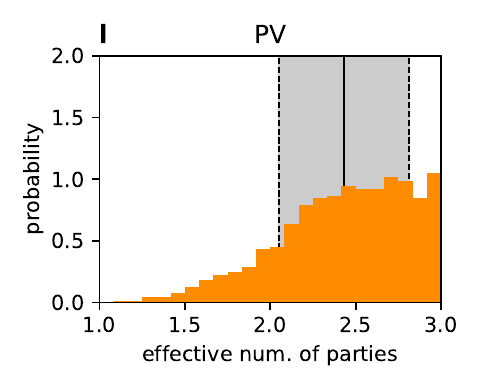}
\end{subfigure}
\caption[Disproportionality and political fragmentation in simulated elections]{\textbf{Disproportionality and political fragmentation in simulated elections.}
For each election in Figure~\ref{fig:basic-es_histograms} we compute three indexes:
the Gallagher index in purple, the Loosemore–Hanby index in green, and the effective number of parties in orange.
Solid vertical line marks the average value and the gray area shows
results within one standard deviation from the average.
Obtained values depend on the applied electoral system:
(a, b, c) proportional representation with one global electoral district
and Jefferson method (PR-G).
(d, e, f) proportional representation with 100 electoral districts
corresponding to topological communities and Jefferson method (PR).
(g, h, i) proportional representation with 100 electoral districts
corresponding to topological communities and Webster method (PR-W).
(j, k, l) plurality voting with 100 electoral districts
corresponding to topological communities (PV).
Results are obtained for the same parameters as in Figure~\ref{fig:basic-es_histograms}:
histograms are obtained over 4000 elections with 3 parties,
$\varepsilon=0.005$, $k=12$, $r=0.002$,
$N=10^4$ divided into 100 equal communities in the SBM.
Each electoral district corresponding to a topological community
has 5 seats assigned, giving 500 seats in total to win in elections.
}
\label{fig:basic-indexes}
\end{figure}

\begin{figure}
\centering
\begin{subfigure}{\textwidth}
    \centering
    \includegraphics[scale=0.61]{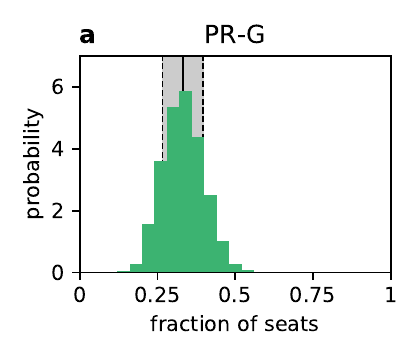}
    \includegraphics[scale=0.61]{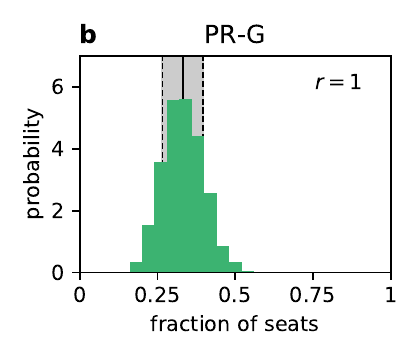}
    \includegraphics[scale=0.61]{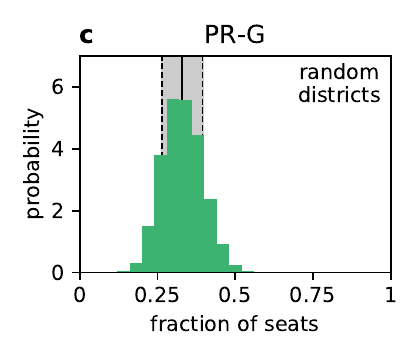}
    \includegraphics[scale=0.61]{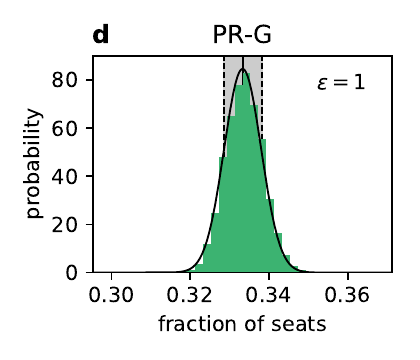}
\end{subfigure}
\begin{subfigure}{\textwidth}
    \centering
    \includegraphics[scale=0.61]{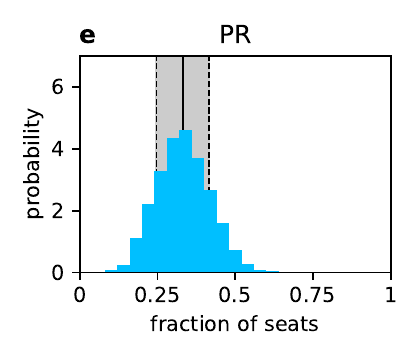}
    \includegraphics[scale=0.61]{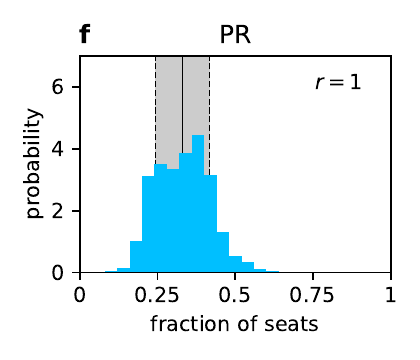}
    \includegraphics[scale=0.61]{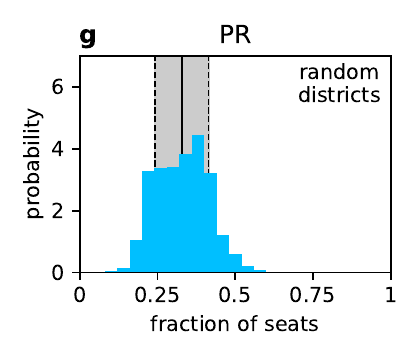}
    \includegraphics[scale=0.61]{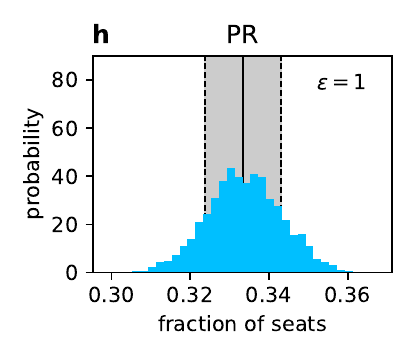}
\end{subfigure}
\begin{subfigure}{\textwidth}
    \centering
    \includegraphics[scale=0.61]{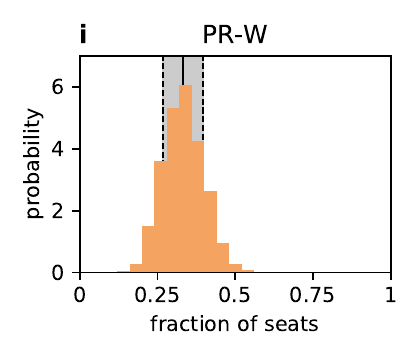}
    \includegraphics[scale=0.61]{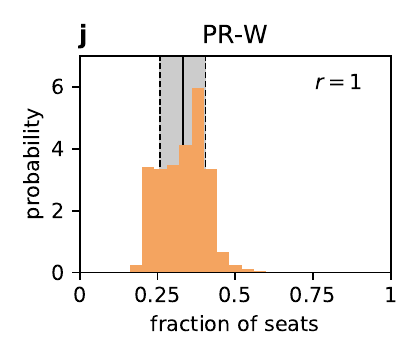}
    \includegraphics[scale=0.61]{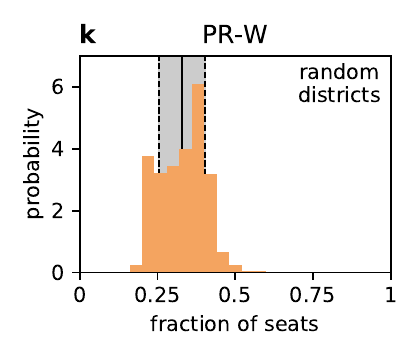}
    \includegraphics[scale=0.61]{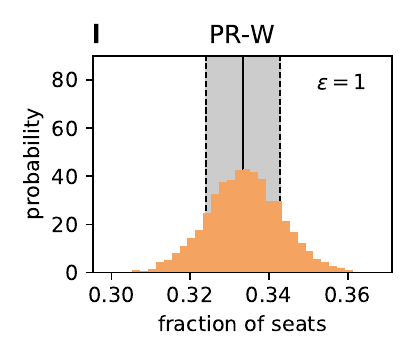}
\end{subfigure}
\begin{subfigure}{\textwidth}
    \centering
    \includegraphics[scale=0.61]{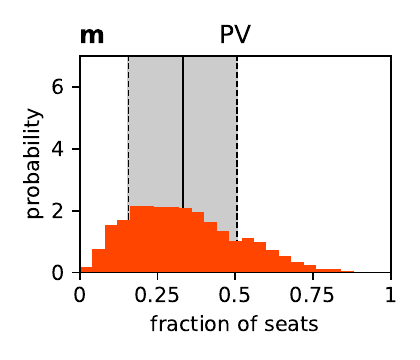}
    \includegraphics[scale=0.61]{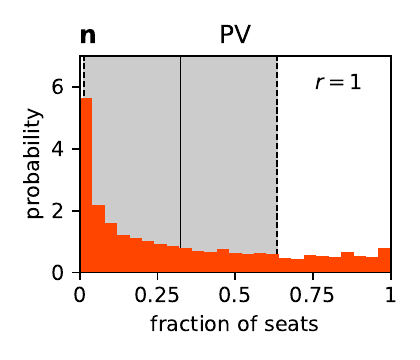}
    \includegraphics[scale=0.61]{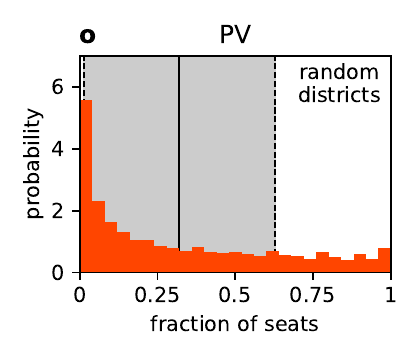}
    \includegraphics[scale=0.61]{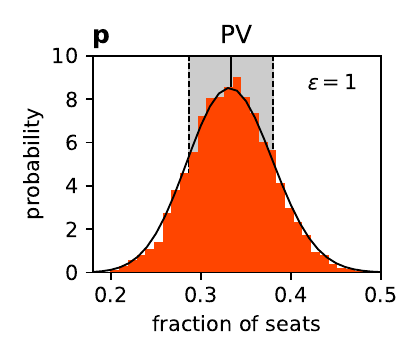}
\end{subfigure}
\caption[Election results in limit cases]{\textbf{Election results in limit cases.} (a, e, i, m) Histograms in the first column
represent a seat share distribution for one of the three parties obtained over
4000 elections for four different electoral systems: PR-G, PR, PR-W, and PV.
All details of the simulation are identical as for Figure~\ref{fig:basic-es_histograms}.
In every other column one of the parameters is changed to an extreme value.
(b, f, j, n) In the second column the ratio parameter is set to one,
$r=1$, which results in no topological communities
and effectively a random graph.
(c, g, k, o) In the third column the nodes are randomly grouped into districts, without taking 
into account community structure. Note, that it has the same effect as setting $r=1$.
(d, h, l, p) In the fourth column the noise rate $\varepsilon=1$ and effectively
there are no social interactions. The solid curves in the panel (d) and (p) represent
the binomial approximation of that case given in
Eq.~[\ref{eqn:binom_dist}] and~[\ref{eqn:multinom_dist_fptp}] respectively.
Solid vertical line marks the average value and the gray area shows
results within one standard deviation from the average.
}
\label{fig:basic-limit_cases}
\end{figure}

We performed elections up to 4000 times for each abstract electoral system to ensure sufficient 
statistical relevance. See Figure~\ref{fig:basic_trajectories} for the trajectories of opinion 
dynamics, with a visualization of the elections. After aggregating all the results, we construct 
probability distributions for each political party, which are presented in
Figure~\ref{fig:basic-es_histograms}. The results in every column of the figure are essentially
identical. This outcome is to be expected -- the design of the simulations is fully symmetrical
with respect to the parties; therefore, statistically, they should obtain the same results. 
For that reason, in further analysis, we present only the results for one of the parties.

Comparing the distribution of votes with those of seats in Figure~\ref{fig:basic-es_histograms},
we can see that the PR-G and PR-W systems produce seat distributions with the highest similarity
to the vote distribution. Slightly behind, with a bigger variance of results, is PR, and
the last one is PV, which produces much more volatile results. Even in the simplest case without
external influences, the contrast between plurality voting and proportional representation is 
clearly visible. The PR systems exhibit a distribution with a peak at 1/3.
Consequently, the most probable election outcome in this configuration 
is an equal number of seats for all parties on average. In contrast, the distribution 
for PV displays a peak below 1/3 and gives a non-zero probability of winning almost all seats 
and winning zero seats. This implies that the most probable outcome is a domination of one 
party (obviously, the average outcome is still 1/3 for each party). This result is conceptually 
straightforward: in first-past-the-post or party block voting, persuading even 34\% of voters
in each district would suffice to secure 100\% of the seats. In this sense, 
one can say that PV electoral systems are less stable.

Differences in the statistics of disproportionality and political fragmentation are 
also pronounced, as we can see in Figure~\ref{fig:basic-indexes}. Analyzing 
the Gallagher index and the Loosemore-Hanby index, we can see that PR-G is by far 
the most proportional electoral system, followed by PR-W and PR. PV produces much less
proportional results with indexes higher by an order of magnitude (note the scale on the x-axis). 
Such a result is to be expected -- with only one constituency in PR-G, the error of rounding
the vote fraction to fit the available seats is not propagating over 100 districts.
Additionally, the Webster method of seat assignment is known to be more proportional than 
the Jefferson method, with the latter working in favor of bigger parties. Finally, plurality voting 
systems are recognized in political sciences as much less proportional, as described
in the first section. Nevertheless, PV displays the least political fragmentation with
the lowest values of the effective number of parties, followed by PR, PR-W, and PR-G.

It is evident that the PR system with the Jefferson method is the one from among 
the PR systems that produces results the least dissimilar to PV. For this reason,
we choose it for simulations comparing plurality voting with proportional representation --
if there is a significant difference when using the Jefferson method, 
it will be even more pronounced for the two other PR systems.


\subsection{Exploration of the parameter space}

We start the exploration of the parameter space by studying the extreme cases presented in 
Figure~\ref{fig:basic-limit_cases}. The first column of the figure contains histograms of
the same election results under four ES as in Figure~\ref{fig:basic-es_histograms}, but 
just for one party. The other presented cases contain a modification of social interactions
between the voters. The second and third columns show results for a setting that effectively 
results in a random pattern of interaction. More precisely, in these both cases, there is no difference 
between a pair of voters from distinct districts and a pair from the same district -- both have the 
same probability of interacting. In the second column, it is achieved by setting the ratio $r=1$,
resulting in a random graph with no topological communities. In the third column, it is 
accomplished using random district allocation, i.e. nodes for each district are chosen randomly 
from the whole network, not based on the topological communities. 

This modification barely changes the results for the PR-G system, as district division is omitted 
here, and the whole network is considered as one constituency. However, it does influence the PR 
and PR-W systems by slightly increasing the standard deviation and shifting the peak of the 
distribution above 1/3. Nevertheless, the biggest impact can be seen in the seat share 
distribution of the PV system. Not only does the standard deviation become much bigger, causing
higher unpredictability of the results, but also the most probable outcome is now 0 -- not gaining
a single seat. The second peak of the distribution is at 1, i.e. winning all seats. In other 
words, landslide victories of one party become much more common. This is an important result -- higher
isolation of individual districts gives smaller parties a chance of winning seats under PV,
because the same number of votes that is enough to win in a district can be a tiny minority
on the national level. On the other hand, if all the districts are well-connected,
which can be the case for small countries, it provides a better opportunity 
for bigger parties to dominate the parliament, because the opinion of the majority 
can easily spread across the constituencies.

The last column of Figure~\ref{fig:basic-limit_cases} shows the results for $\varepsilon=1$. 
In this scenario, there is no social interaction at all because whatever happens between
the voters is always overridden by a random choice of the candidate at the end of the time step.
There is no spatial correlation of opinions, and the vote share can be described by the binomial
distribution (see section \textit{Binomial approximation}). Notably, the variance in the results decreases by 
an order of magnitude for all ES (note the scale on the x-axis), maintaining relative differences 
between them. Since there are no social interactions, an opinion cannot propagate through the 
population, and an advantage of any party can be gained only via random fluctuations among $10^4$
voters. Clearly, the character of social interactions can be crucial for the stability of electoral systems;
therefore, taking into account the actual geography of a country, the mobility of its voters, and their
implications on the social interactions as captured in the networked voter model, 
is essential when studying real-world ES.

In the following figures, we study in more detail the impact of several parameters on the election
results for PR and PV. We present the basic histogram of the seat share distribution for each 
parameter value under each ES (PR is in blue, PV is in red in the histograms). This provides 
the basic idea about the shape and variance of the distribution. However, to provide a better 
comparison between the systems and between different values of the parameters, we plot the main
characteristics of the distributions in a box plot, together with all indexes of disproportionality
and political fragmentation. A box plot contains all essential features of a distribution: 
the average value is indicated by an $x$, the median by an orange line, the width of the box 
is given by the first and third quartiles, the whiskers indicate values within 1.5 times 
the inter quartile range (IQR), and the outliers are marked by a circle.

\subsubsection{Number of parties}

\begin{figure}
\centering
\begin{subfigure}{0.66\textwidth}
\centering
    \includegraphics[scale=0.67]{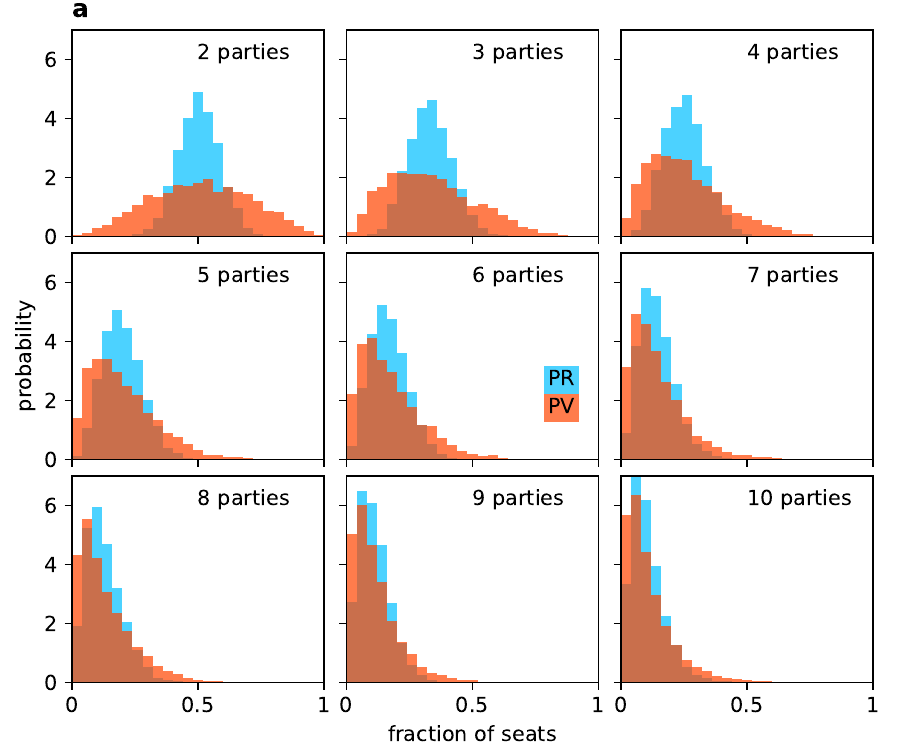}\hfill
\end{subfigure}\hfill
\begin{subfigure}{0.33\textwidth}
    \includegraphics[scale=0.65]{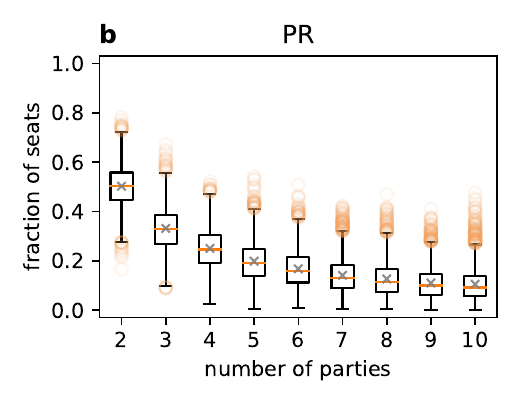}
    \includegraphics[scale=0.65]{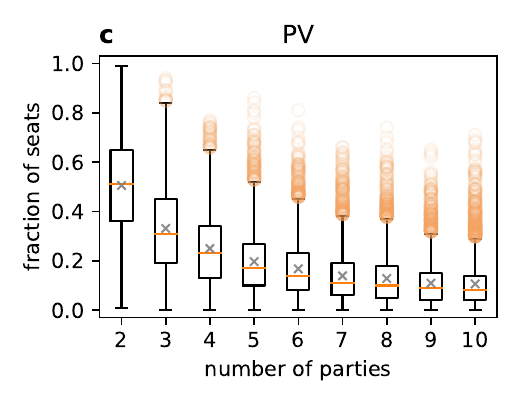}
\end{subfigure}\hfill
\begin{subfigure}{1\textwidth}
\centering
    \includegraphics[scale=0.65]{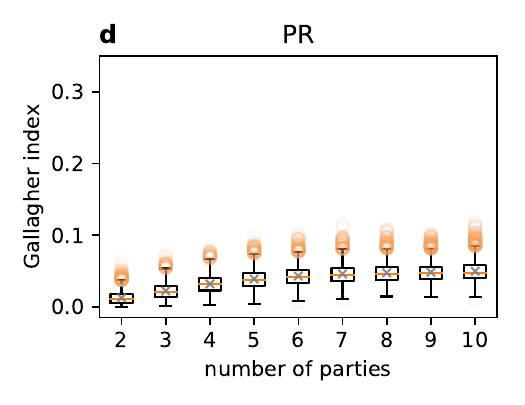}
    \includegraphics[scale=0.65]{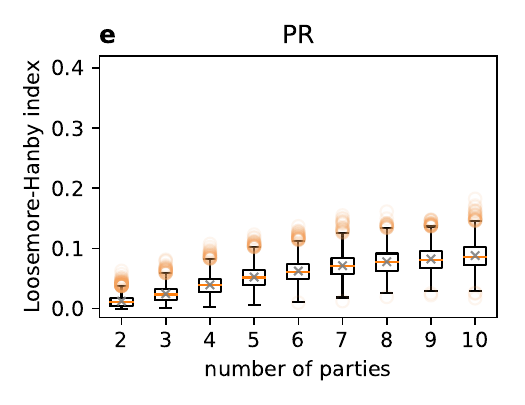}
    \includegraphics[scale=0.65]{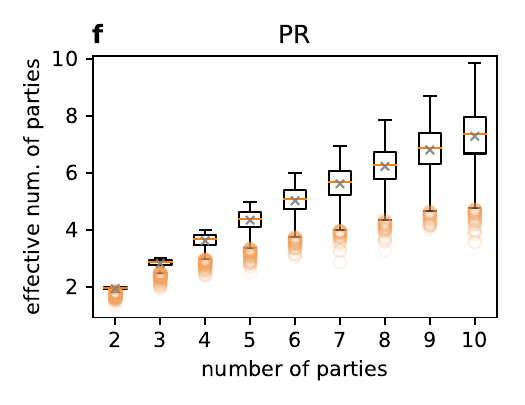}
    \includegraphics[scale=0.65]{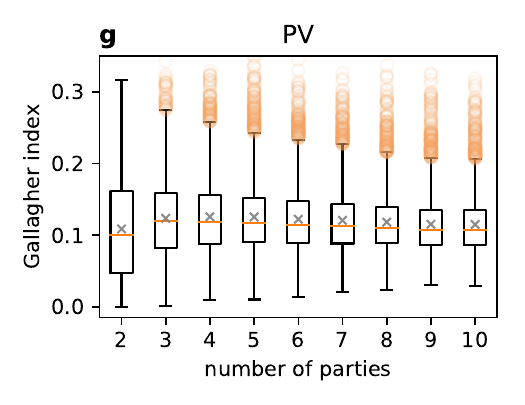}
    \includegraphics[scale=0.65]{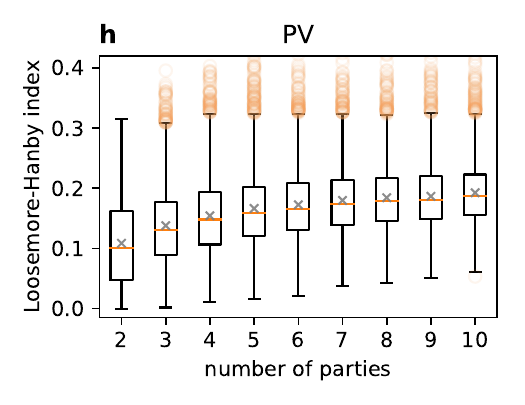}
    \includegraphics[scale=0.65]{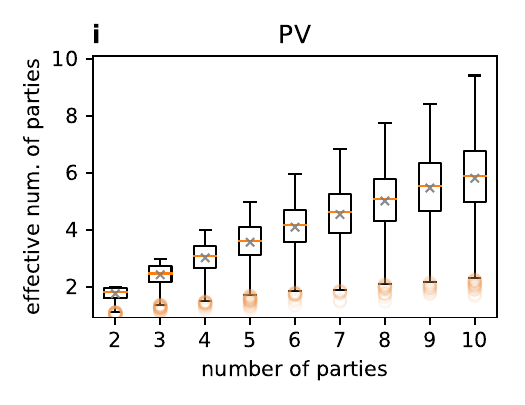}
\end{subfigure}
\caption[Comparison between PR and PV for different numbers of parties]{\textbf{Comparison between PR and PV for different numbers of parties.}
(a) Histograms of the seat share distribution of one of the parties when using
proportional representation (PR, blue) and plurality voting (PV, red) for varying total number of parties.
Note how the difference in variance of the distributions decreases with increasing number of parties.
(b, c) Box plots of the same results highlighting the differences in the distribution.
(d, e, f, g, h, i) Box plots of disproportionality and political fragmentation indexes for both ES.
The average value in box plots is indicated by an \textit{x}, the median by an orange line,
the width of the box is given by the first and third quartile,
the whiskers indicate values within 1.5 times the IQR,
and outliers are marked by a circle.
Other simulation parameters are the same as in Figure~\ref{fig:basic-es_histograms}:
$N=10^4$, $q=100$, $qs=5$, $k=12$, $r=0.002$, $\varepsilon=0.005$, results obtained over 4000 elections.
}
\label{fig:hist-box_parties}
\end{figure}

In Figure~\ref{fig:hist-box_parties}, we show the dependence of election results on the number
$n$ of parties running. Looking at the histograms in panel (a), we can see that the difference 
between PR and PV systems is much more pronounced for fewer parties. With an increasing number
of parties, the standard deviation of the seat share decreases for both systems. Note that this 
is a symmetrical case with no external influence, and every party has the same support on 
average. In the box plots of panels (b) and (c), we can see that the discrepancy between PR and PV,
although smaller,
is still present for higher numbers of parties -- the distribution for PV is wider, has more 
extreme values, and is more skewed, what can be seen from the difference between the average
value and the median.

Interestingly, the difference between PR and PV systems persists even for more parties when
comparing the indexes of disproportionality and fragmentation in 
Figure~\ref{fig:hist-box_parties} panes (d) to (i).
The values of the Gallagher index and the Loosemore-Hanby 
index are much lower for PR, in some cases by an order of magnitude. The Loosemore-Hanby index 
increases for both systems with every additional party, but the Gallagher index reaches a 
plateau at around 7 parties for PR and surprisingly slightly decreases after 3 parties for PV.
This contrasting behavior of the two indexes comes from the fact that one uses quadratic 
difference (Gallagher) and the other uses absolute difference (Loosemore-Hanby) between a vote 
share and a seat share (see section \textit{Disproportionality and political fragmentation 
indexes}). For more parties, the error of rounding the vote share into available seats on average
is smaller for each party, just because the vote fraction obtained by each party is smaller,
but there are more parties to sum the error over. The quadratic function is more sensitive to
big errors or outliers but less sensitive to small errors than a linear function. Hence, if we 
sum small differences over many parties, the Loosemore-Hanby index should produce higher 
values than the Gallagher index.

This divergence between two indexes can be visualized with a simple example of the absolute
difference being equal to~1. If we have two parties, both with a vote share of 0.5, and one 
obtaining all seats while the other none, the absolute difference between the vote share and 
seat share will be $|0.5|+|-0.5|=1$ (so $LH=0.5$) and the quadratic difference will be
$0.5^2+(-0.5)^2=0.5$ (so $G=0.5$ as well). Now let's assume we have the same absolute difference spread
across 10 parties with a vote share of 0.1 each, due to half of them obtaining 0.2 of seats 
and the other half zero seats. The absolute difference stays the same because for each party
the difference is either $|0.1|=0.1$ or $|-0.1|=0.1$, times 10 parties gives 1 
(so $LH=0.5$ again). The quadratic error, however, will be either $0.1^2=0.01$ or 
$(-0.1)^2=0.01$, which after multiplying by 10 parties gives 0.1 (so this time $G\approx0.22$).

The effective number of parties increases with the real number of parties for both systems,
however slower than
linearly. It is always bigger for PR than PV, which indicates higher political fragmentation
or lower concentration of power in the former case. This difference becomes even more 
pronounced for a higher number of parties.

\subsubsection{Number of districts}

\begin{figure}
\centering
\begin{subfigure}{0.66\textwidth}
\centering
    \includegraphics[scale=0.67]{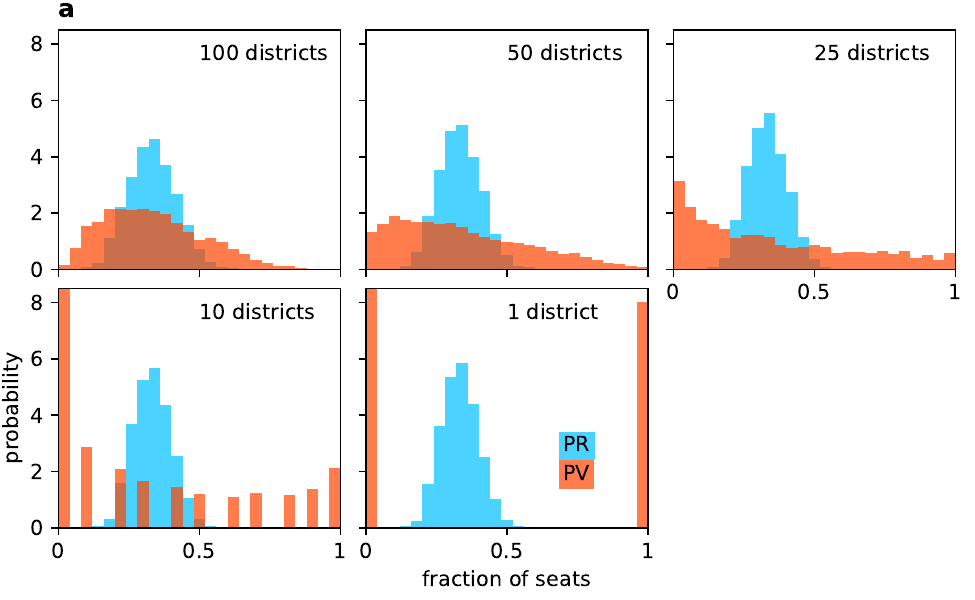}\hfill
\end{subfigure}\hfill
\begin{subfigure}{0.33\textwidth}
    \includegraphics[scale=0.65]{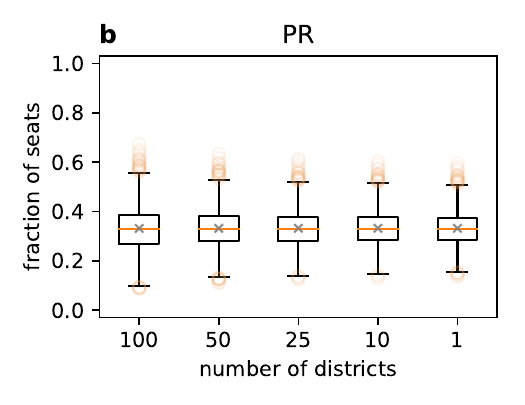}
    \includegraphics[scale=0.65]{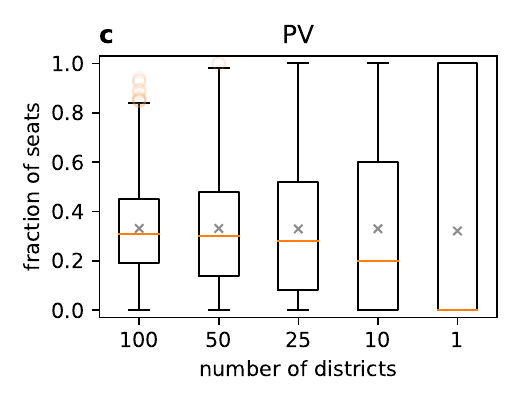}
\end{subfigure}\hfill
\begin{subfigure}{1\textwidth}
\centering
    \includegraphics[scale=0.65]{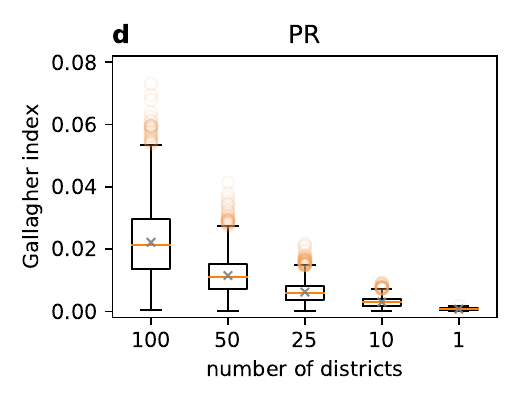}
    \includegraphics[scale=0.65]{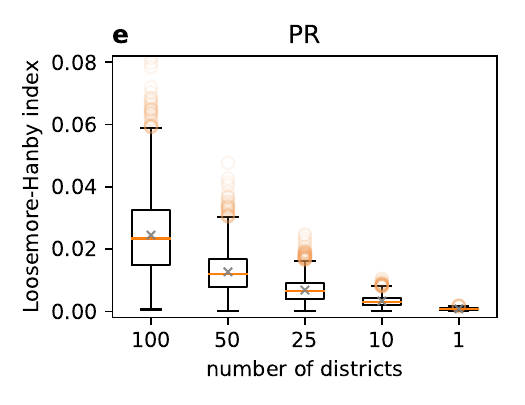}
    \includegraphics[scale=0.65]{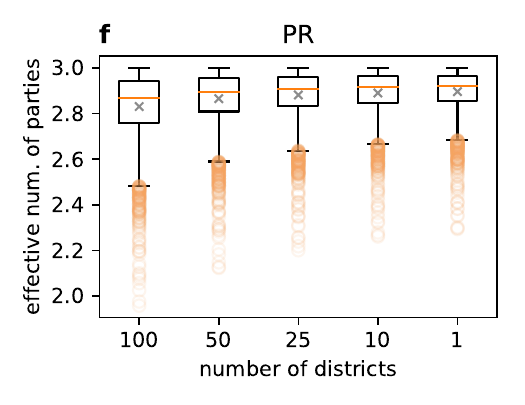}
    \includegraphics[scale=0.65]{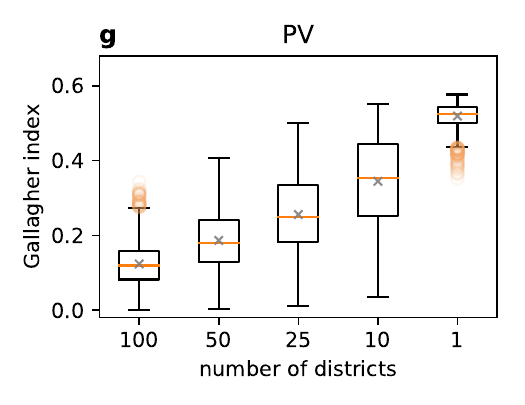}
    \includegraphics[scale=0.65]{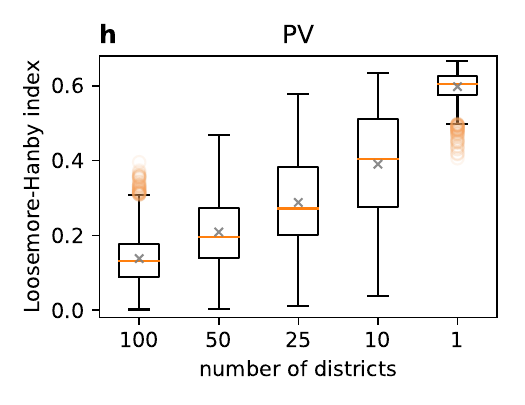}
    \includegraphics[scale=0.65]{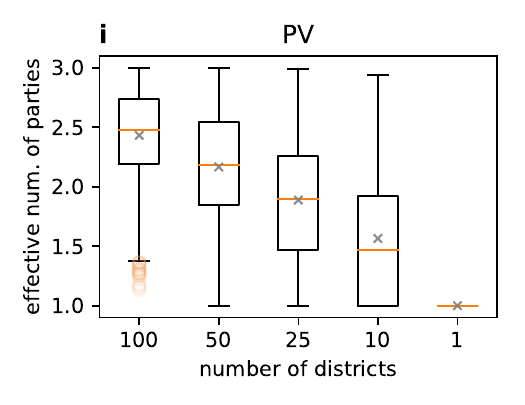}
\end{subfigure}
\caption[Comparison between PR and PV for different numbers of districts]{\textbf{Comparison between PR and PV for different numbers of districts.}
(a) Histograms of the seat share distribution of one of three parties when using
proportional representation (PR, blue) and plurality voting (PV, red) for varying number of districts.
The total number of seats is kept constant at 500 - when districts are merged their seats are added
to create a bigger district with more seats.
Note how decreasing number of districts has opposite effect on the variance of the distribution
for the two ES.
For one district the distribution for PV becomes bimodal meaning
that the most probable outcome for a party is to win
all seats or to not win a single one.
(b, c) Box plots of the same results highlighting the differences in the distribution.
(d, e, f, g, h, i) Box plots of disproportionality and political fragmentation indexes for both ES.
The average value in box plots is indicated by an \textit{x}, the median by an orange line,
the width of the box is given by the first and third quartile,
the whiskers indicate values within 1.5 times the IQR,
and outliers are marked by a circle.
Other simulation parameters are the same as in Figure~\ref{fig:basic-es_histograms}:
$N=10^4$, $k=12$, $r=0.002$, $\varepsilon=0.005$, total seats 500, and $n=3$ parties, results obtained over 4000 elections.
}
\label{fig:hist-box_districts}
\end{figure}

In Figure~\ref{fig:hist-box_districts}, we show the dependence of election results on the number
of electoral districts the country is divided into. The network always has the same number of
topological communities $q = 100$, but the number of districts changes from 100 to just 1. This
is achieved by treating one or multiple topological communities as a single constituency.
From the histograms in panel~(a) and the box plots in panels~(b) and~(c),
we can see that the influence it has on the PR system is opposite to the
influence on the PV system. With fewer districts, votes have to be rounded to fit available seats fewer times, hence
the variance of seat share under PR is lower. However, under PV, a party can win only all or none
of the seats in each district. Therefore, with fewer districts, results become more volatile. For 10 districts,
there are only 11 possible election outcomes for each party under PV, resulting in the same number of
separate histogram bars. By comparison, in PR, there are always 501 possible outcomes (from 0 to 500 seats) because
every seat is assigned separately. In the extreme case of one global constituency, the difference
between the two systems becomes colossal, since in PV, only one party can win all seats.
Such an electoral system is, of course, unlikely, and party block voting always involves division into
multiple districts. Nevertheless, these results show the general impact of the number of districts
on the election results under PR and PV, and the difference between them.

As we can see in panels~(d) to~(i) of Figure~\ref{fig:hist-box_districts}, PR is always more proportional
and more fragmented than PV. The average value of the Gallagher index and the Loosemore-Hanby index
quickly decrease with decreasing number of districts for PR, but the effective number of parties increases slowly.
It suggests that the balance between low disproportionality and low political fragmentation
is best achieved for smaller numbers of districts in PR. The effects are exactly opposite for PV.

\subsubsection{District magnitude}

Figure~\ref{fig:hist-box_seats_per_dist} illustrates 
the relationship between election results and the number of seats available
in each electoral district, also known as district magnitude.
The number of districts and other parameters are kept constant,
therefore the total number of seats in the parliament changes 
together with district magnitude. For one seat assigned
to each district, both electoral systems display the same results because both effectively use FPTP voting.
With an increasing number of seats, the variance of seat share under PR decreases, while the distribution
for PV remains unchanged. This is because under PV, 
a party can win only all (or none) of the seats in a district, and consequently, it does not
matter how many seats are available in one district since it is always the same fraction of all seats.

An observation evident from panels~(d) to~(i) of Figure~\ref{fig:hist-box_seats_per_dist}
is that disproportionality and political fragmentation
also remain the same for PV across different district magnitudes
and are equal to those observed under PR only when there 
is one seat per district. In the PR system, on the other hand,
the Gallagher index and the Loosemore-Hanby index decrease quickly when increasing the number of seats
from 1 to 5 and then only slightly when increasing it further to 9. Similarly, the effective number of parties
is more influenced by the initial increase in the number of seats and does not change much from 5 seats onwards.
These results are consistent with the widely recognized fact that district magnitude
can be used as a mechanism for altering the proportionality of elections under PR. Note, however, that
it is impossible to obtain low disproportionality and low political fragmentation at the same time.

\subsubsection{Noise rate}

\begin{figure}
\centering
\begin{subfigure}{0.66\textwidth}
\centering
    \includegraphics[scale=0.67]{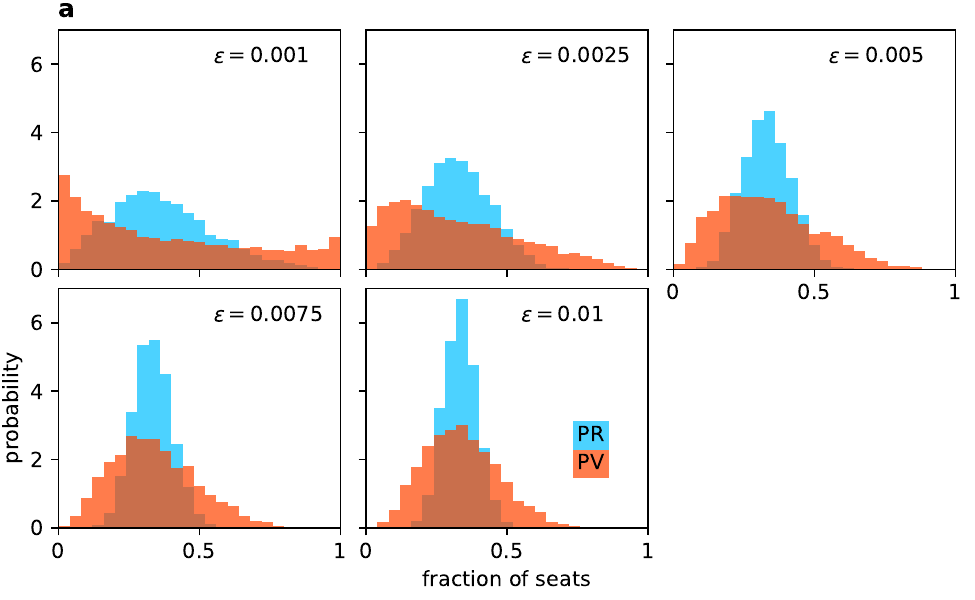}\hfill
\end{subfigure}\hfill
\begin{subfigure}{0.33\textwidth}
    \includegraphics[scale=0.65]{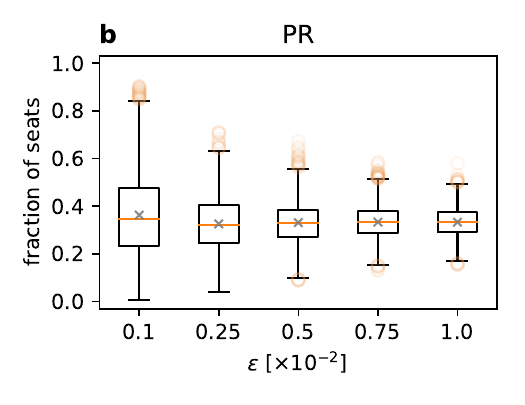}
    \includegraphics[scale=0.65]{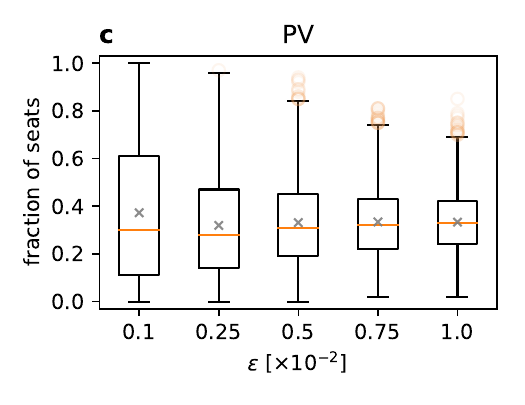}
\end{subfigure}\hfill
\begin{subfigure}{1\textwidth}
\centering
    \includegraphics[scale=0.65]{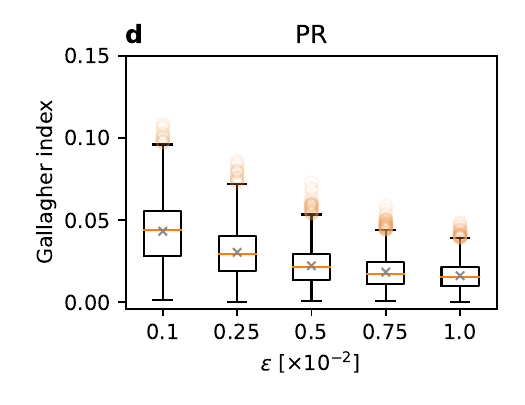}
    \includegraphics[scale=0.65]{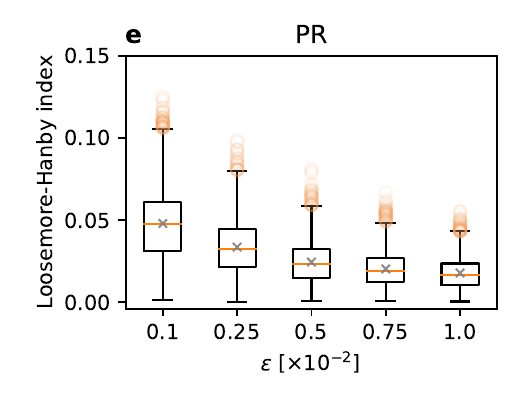}
    \includegraphics[scale=0.65]{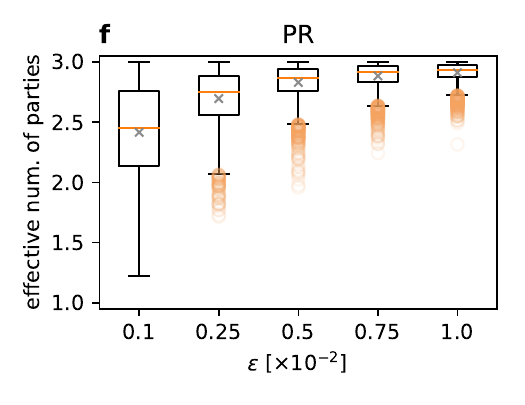}
    \includegraphics[scale=0.65]{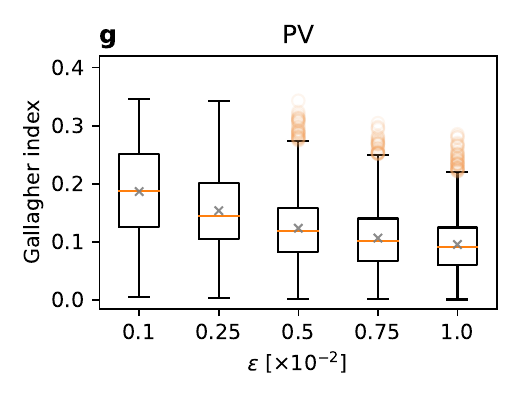}
    \includegraphics[scale=0.65]{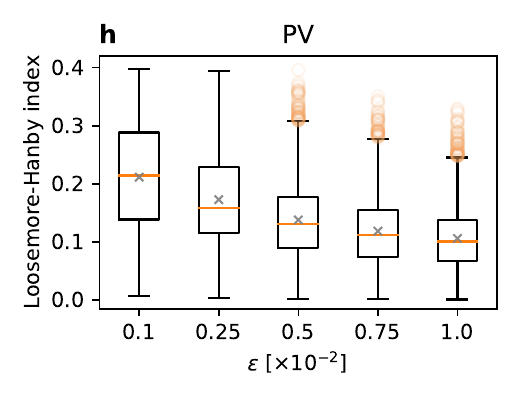}
    \includegraphics[scale=0.65]{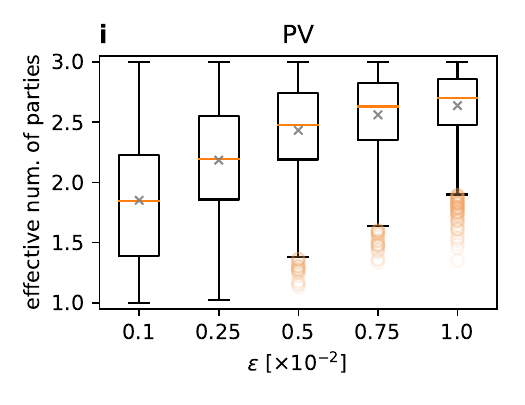}
\end{subfigure}
\caption[Comparison between PR and PV for different noise rates]{\textbf{Comparison between PR and PV for different noise rates.}
(a) Histograms of the seat share distribution of one of three parties when using
proportional representation (PR, blue) and plurality voting (PV, red) for varying noise rate $\varepsilon$.
Note how the the variance of the distribution decreases with increasing noise.
For $\varepsilon=0.001$ the distribution becomes bimodal in case of PV,
i.e. the most probable outcome is one party taking all seats and two parties obtaining none.
See Figure~\ref{fig:basic-limit_cases} for the limit case of $\varepsilon=1$.
(b, c) Box plots of the same results highlighting the differences in the distribution.
(d, e, f, g, h, i) Box plots of disproportionality and political fragmentation indexes for both ES.
The average value in box plots is indicated by an \textit{x}, the median by an orange line,
the width of the box is given by the first and third quartile,
the whiskers indicate values within 1.5 times the IQR,
and outliers are marked by a circle.
Other simulation parameters are the same as in Figure~\ref{fig:basic-es_histograms}:
$N=10^4$, $q=100$, $qs=5$, $k=12$, $r=0.002$, and $n=3$ parties, results obtained over 4000 elections.
}
\label{fig:hist-box_epsilon}
\end{figure}

Figure~\ref{fig:hist-box_epsilon} demonstrates how election results
are influenced by the noise rate $\varepsilon$.
From the histograms and box plots in panels (a) to (c), we can see that the variance of seat share decreases
with increasing noise for both systems, PR and PV. As expected, for any given value of $\varepsilon$,
results under PR are more stable, i.e. the standard deviation is smaller than in PV. Remarkably,
the seat share distribution under PV becomes bimodal for low values of the noise rate, indicating that
landslide victories of one party become highly probable. As discussed in the subsection
\textit{Opinion dynamics – the voter model}, for smaller 
values of $\varepsilon$, single parties more easily and more
often gain a substantial fraction of votes, and the pendulum of the political climate becomes more shaky.
The PV electoral system can amplify the dominance of a party 
when translating it into seats, up to the point of winning all seats
if the advantage was substantial, hence we observe a bimodal distribution.
See also Figure~\ref{fig:basic-limit_cases} for results with the maximal noise rate $\varepsilon=1$.

For all the considered values of the noise rate, PR is more proportional, and PV is less fragmented,
as illustrated in panels (d) to (i) of Figure~\ref{fig:hist-box_epsilon}. For both systems,
the Gallagher index and the Loosemore-Hanby index decrease, while the effective
number of parties increases with increasing noise. This can be explained by the fact
that for lower $\varepsilon$, one party can more easily dominate the election, causing other parties
to lose all seats despite having non-zero support under PV or causing the bias of the Jefferson method
to propagate across the districts in PR.

\subsubsection{Size of the population}

\begin{figure}
\centering
\begin{subfigure}{0.66\textwidth}
\centering
    \includegraphics[scale=0.67]{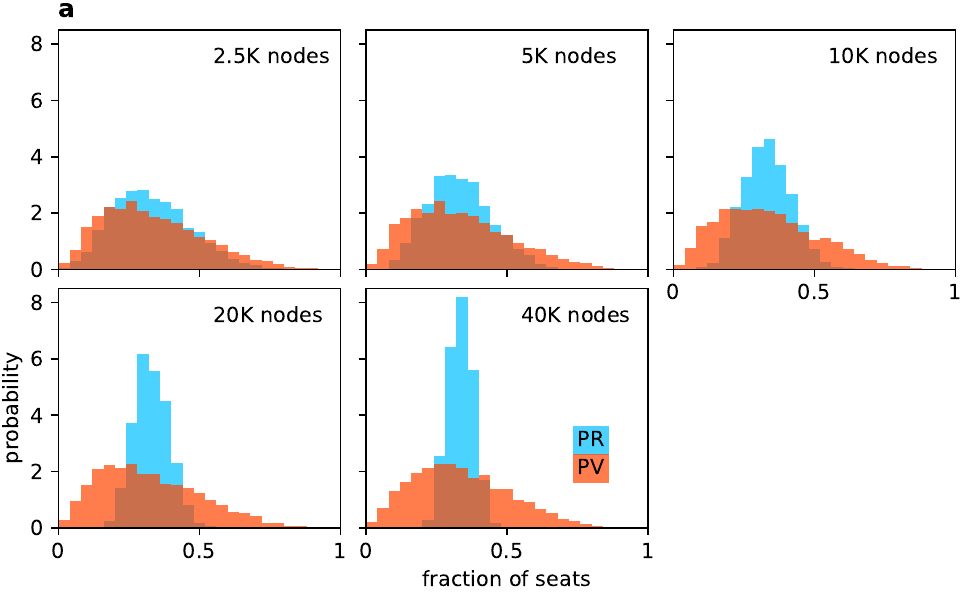}\hfill
\end{subfigure}\hfill
\begin{subfigure}{0.33\textwidth}
    \includegraphics[scale=0.65]{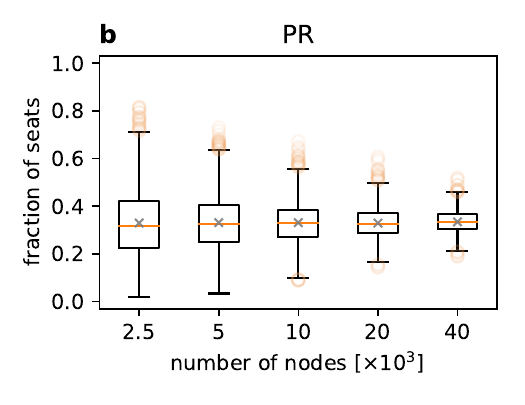}
    \includegraphics[scale=0.65]{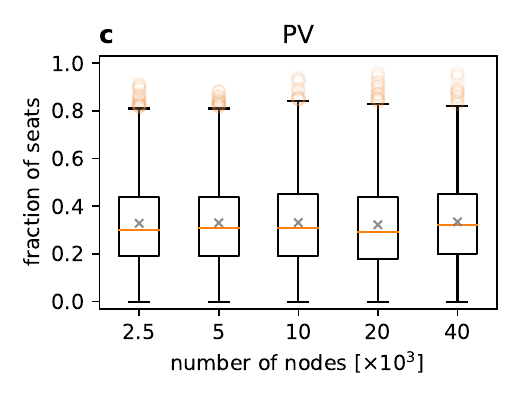}
\end{subfigure}\hfill
\begin{subfigure}{1\textwidth}
\centering
    \includegraphics[scale=0.65]{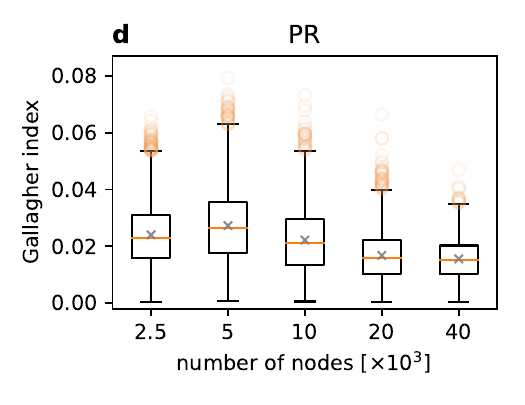}
    \includegraphics[scale=0.65]{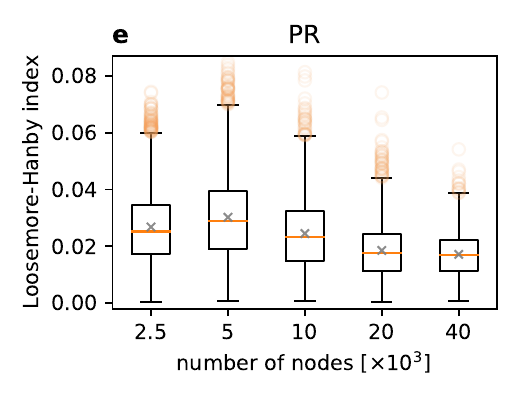}
    \includegraphics[scale=0.65]{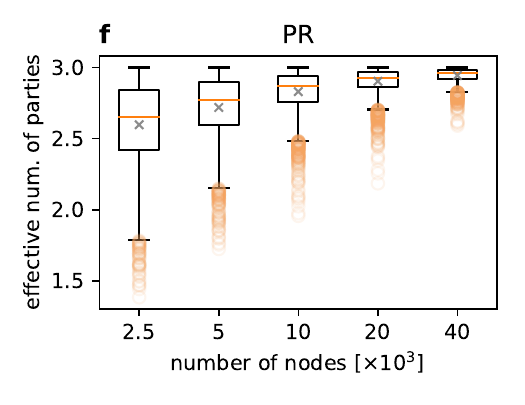}
    \includegraphics[scale=0.65]{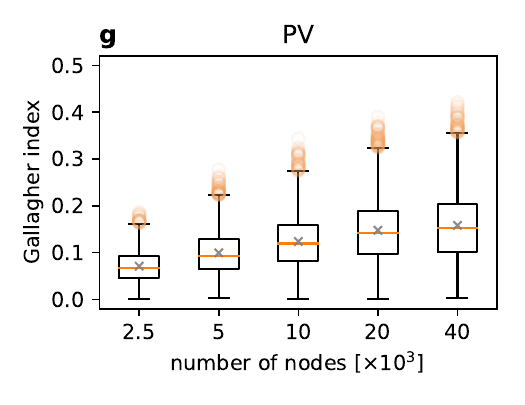}
    \includegraphics[scale=0.65]{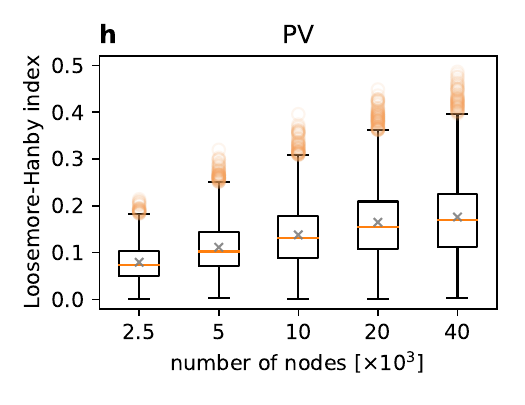}
    \includegraphics[scale=0.65]{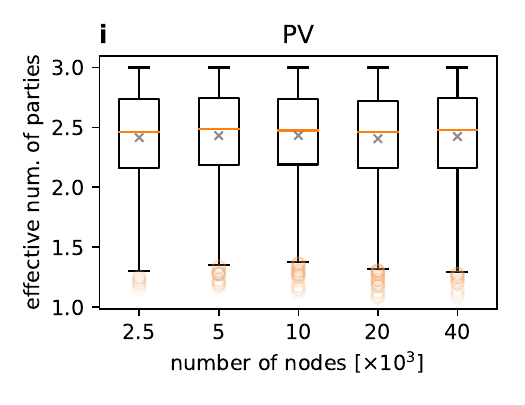}
\end{subfigure}
\caption[Comparison between PR and PV for different sizes of the network]{\textbf{Comparison between PR and PV for different sizes of the network.}
(a) Histograms of the seat share distribution of one of three parties when using
proportional representation (PR, blue) and plurality voting (PV, red) for varying number of nodes $N$.
Note how the variance of the distribution decreases with increasing network size for PR,
but does not change for PV.
(b, c) Box plots of the same results highlighting the differences in the distribution.
(d, e, f, g, h, i) Box plots of disproportionality and political fragmentation indexes for both ES.
The average value in box plots is indicated by an \textit{x}, the median by an orange line,
the width of the box is given by the first and third quartile,
the whiskers indicate values within 1.5 times the IQR,
and outliers are marked by a circle.
Other simulation parameters are the same as in Figure~\ref{fig:basic-es_histograms}:
$q=100$, $qs=5$, $k=12$, $r=0.002$, $\varepsilon=0.005$, and $n=3$ parties, results obtained over 4000 elections.
}
\label{fig:hist-box_net_size}
\end{figure}

In Figure~\ref{fig:hist-box_net_size}, we present the dependence
of election results on the network size $N$.
In panel~(a), we can see that for a small network with $N=2500$ nodes, 
we obtain a very similar distribution for PR and PV.
With an increasing size of the population, the variance of the seat
share significantly decreases for PR but stays unchanged for PV,
as visible in panels~(b) and~(c). This intriguing result can be
understood by looking at each of the 100 districts
as a small network in itself. It is well known that the threshold 
levels of noise in the voter model depend on the network size,
or in other words, the same value of the noise rate will have 
a different effect in networks of different sizes \cite{raducha2020emergence}.
In smaller networks, it is easier to obtain the consensus phase 
(the critical value of the noise rate separating the consensus phase from the coexistence phase is higher),
and for the same noise rate $\varepsilon$, the support for 
a given party will more often grow to considerable levels, providing an advantage over other parties.
In a small district, a party will get close to 5 seats more often than in a big district.
But since the simulation is symmetrical with respect to parties,
it will also get 0 seats more often than in a big district.
In PR systems, the bigger the advantage of one party over the others,
the higher the variance of the distribution because each party can gain any number of the available seats.
This explains why the variance of the seat share decreases with 
growing network size for PR -- the fluctuations of the vote share 
and consequently of the seat share are smaller in bigger networks.
There is also a rounding effect -- when $N=2500$, one district has 
25 voters and 5 seats, so the fraction of seats can change by not less than
$5/25=0.2$ (if only one vote changes);
for $N=4\times 10^4$, it can change more smoothly with the smallest
increment of $(5/4) \times 10^{-4} = 0.000125$.

In PV, however, a party can win only 0 or 5 seats in a district.
How big is the support for a given party in a given constituency
does not matter as long as it has the plurality.
For example, in a district with three parties, gaining 34\% of votes,
while the two political opponents win 33\% each, will have the same outcome as gaining 90\%
and leaving 10\% to split between the opponents. In both cases the party with plurality
will obtain all 5 seats.
Therefore, it is not important whether the fluctuations of support are
small or big -- every party will obtain plurality the same number of times 
on average because the simulation is symmetrical with respect to parties.
Consequently, there is no influence of the district size, dictated by 
the network size, on the probability distribution of the seat share.

In panels (d) to (i) of Figure~\ref{fig:hist-box_net_size},
we can see that indexes of disproportionality take much lower 
values for PR, even in the case of $N=2500$,
while political fragmentation is higher in this system than in PV.
The effective number of parties grows on average with increasing 
network size in PR and stays approximately constant in PV.
There is an interesting effect in the Gallagher index and 
the Loosemore-Hanby index -- for PV, they increase with $N$, but for PR,
they take a maximal value at $N=5 \times 10^3$.
There are two competing effects influencing proportionality in the PR system.
One is the variance of the vote share -- as discussed previously,
for higher fluctuations, proportionality decreases due to repeated seat assignment
bias towards the biggest party in each district.
The second is rounding error, which is inherent in any electoral system,
which has a different number of voters and seats in districts (in practice this is the case for every ES).
After decreasing the number of nodes to an extreme case of 5 voters per district, as well as 5 seats per district,
any reasonable seat assignment method would obtain perfect proportionality. The case of 25 voters per district,
i.e. $N=2500$, is close enough to this extreme to observe a drop in disproportional indexes. Nevertheless,
it should be treated as a finite size effect, since the number of voters is vastly greater than the number of seats
in real ES. In PV the disproportionality increases with the network size, because the variance of the vote share
decreases but the variance of the seat share does not. If a party wins,
it wins all the seats in a district, but in bigger networks
it happens with a smaller advantage over others. Hence, the discrepancy between
the votes obtained and the seats assigned gets bigger.

\subsubsection{Average connectivity}

\begin{figure}
\centering
\begin{subfigure}{0.66\textwidth}
\centering
    \includegraphics[scale=0.67]{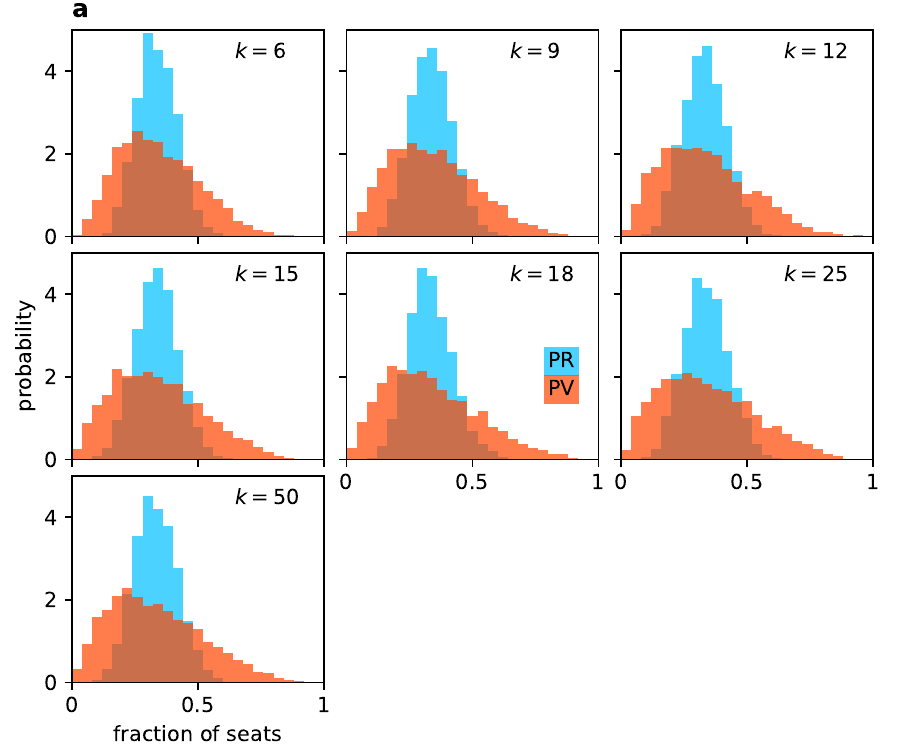}\hfill
\end{subfigure}\hfill
\begin{subfigure}{0.33\textwidth}
    \includegraphics[scale=0.65]{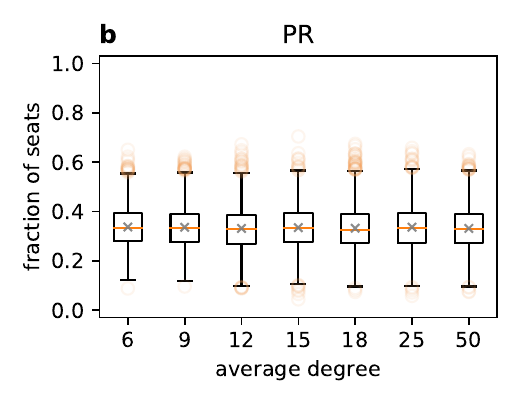}
    \includegraphics[scale=0.65]{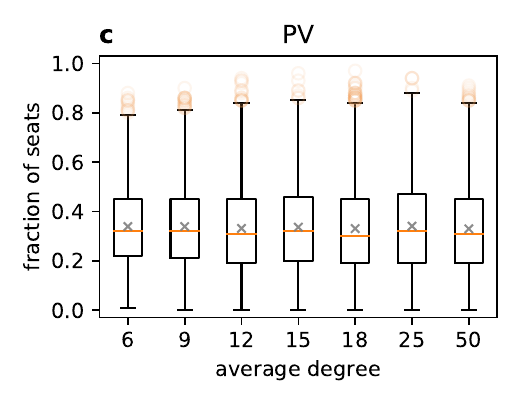}
\end{subfigure}\hfill
\begin{subfigure}{1\textwidth}
\centering
    \includegraphics[scale=0.65]{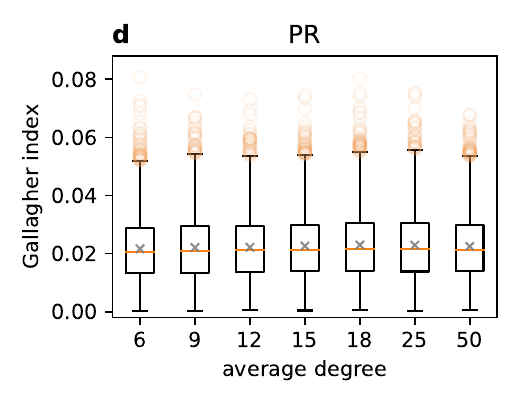}
    \includegraphics[scale=0.65]{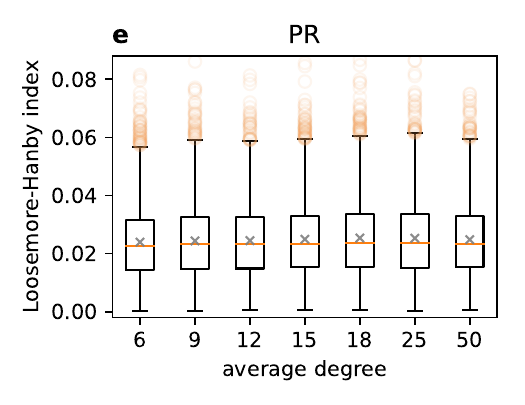}
    \includegraphics[scale=0.65]{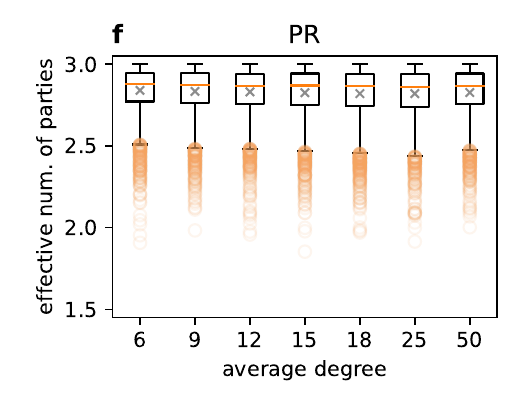}
    \includegraphics[scale=0.65]{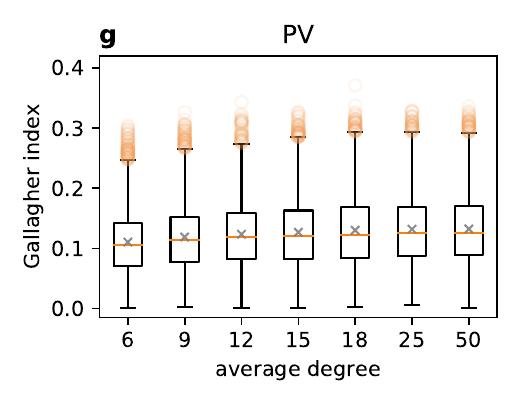}
    \includegraphics[scale=0.65]{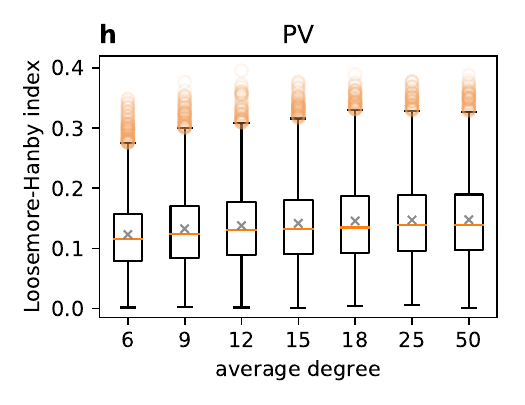}
    \includegraphics[scale=0.65]{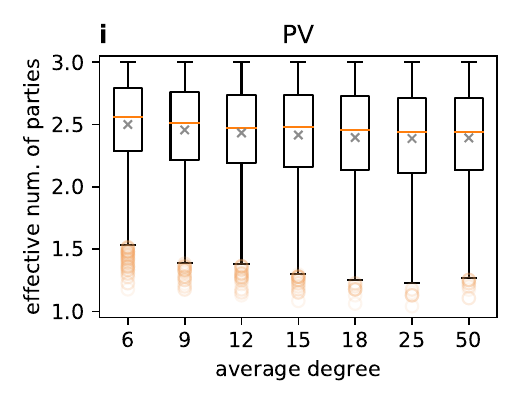}
\end{subfigure}
\caption[Comparison between PR and PV for different average degrees]{\textbf{Comparison between PR and PV for different average degrees.}
(a) Histograms of the seat share distribution of one of three parties when using
proportional representation (PR, blue) and plurality voting (PV, red) for varying average degree $k$.
Note, that it does not have any significant effect on the results for either of the ES. However,
it does slightly influence disproportionality and political fragmentation.
(b, c) Box plots of the same results highlighting the differences in the distribution.
(d, e, f, g, h, i) Box plots of disproportionality and political fragmentation indexes for both ES.
The average value in box plots is indicated by an \textit{x}, the median by an orange line,
the width of the box is given by the first and third quartile,
the whiskers indicate values within 1.5 times the IQR,
and outliers are marked by a circle.
Other simulation parameters are the same as in Figure~\ref{fig:basic-es_histograms}:
$N=10^4$, $q=100$, $qs=5$, $r=0.002$, $\varepsilon=0.005$, and $n=3$ parties, results obtained over 4000 elections.
}
\label{fig:hist-box_aveg_deg}
\end{figure}

Figure~\ref{fig:hist-box_aveg_deg} displays the effect of
varying the average node degree $k$ on election results. Its influence on the
seat share distribution is minimal, as we can see from
panels~(a) to~(c). The shape of the histograms seems not to
change, however, from the box plots, we can observe that
the variance of the seat share increases very slightly with
the growing average degree. This might be due to a higher
range of social interaction in a well-connected network,
which gives it an advantage over independent (random)
state updates. Nevertheless, the effect is insignificant.

Similarly, in PR the indexes of disproportionality and fragmentation in panels~(d) to~(f) remain approximately
constant with increasing connectivity. Under PV the Gallagher index and the Loosemore-Hanby index slightly
increase, and the effective number of parties slightly decreases on average for bigger $k$, as can be seen in
panels~(g) to~(i). This might be caused by the fact that districts become better connected between each other,
making it marginally easier to win more districts for a
given party (see \textit{Interconnectedness ratio} below). Note, however, that the relative connectivity
between the districts in respect to the connectivity inside
them does not change.

\subsubsection{Interconnectedness ratio}

\begin{figure}
\centering
\begin{subfigure}{0.66\textwidth}
\centering
    \includegraphics[scale=0.67]{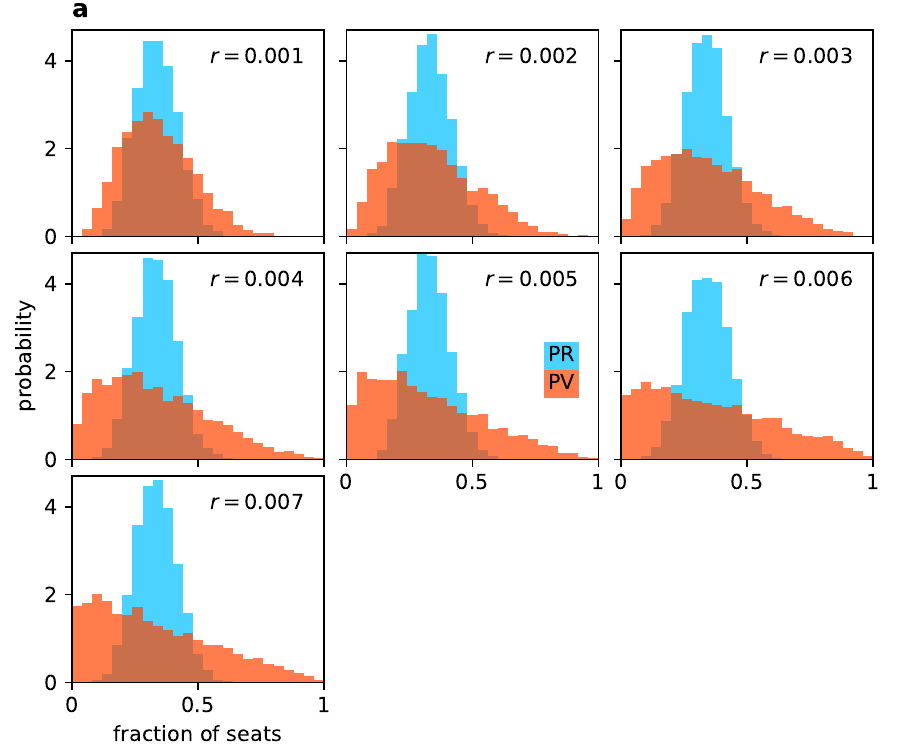}\hfill
\end{subfigure}\hfill
\begin{subfigure}{0.33\textwidth}
    \includegraphics[scale=0.65]{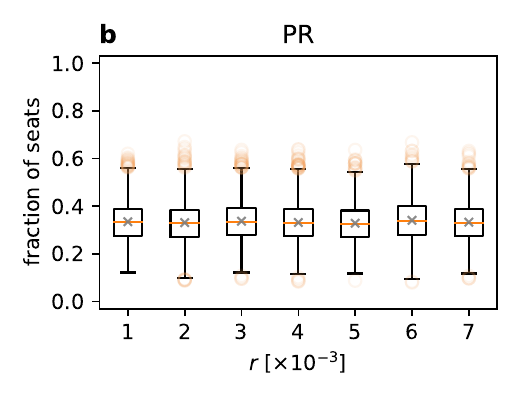}
    \includegraphics[scale=0.65]{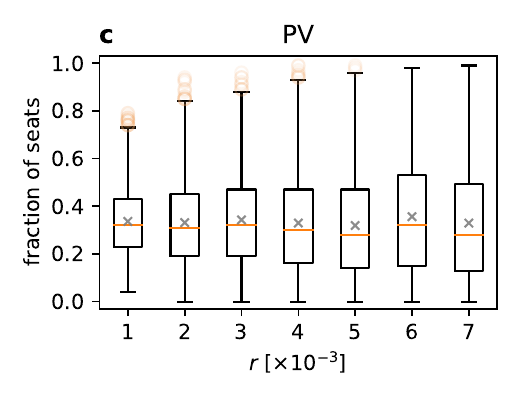}
\end{subfigure}\hfill
\begin{subfigure}{1\textwidth}
\centering
    \includegraphics[scale=0.65]{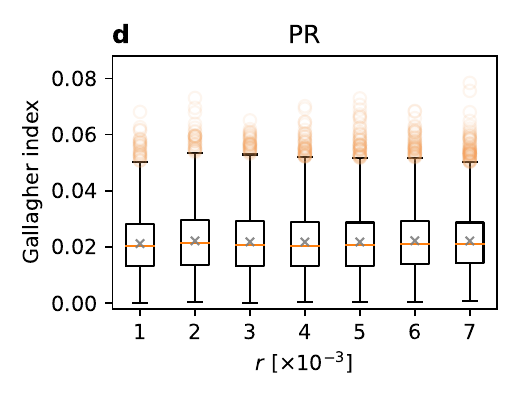}
    \includegraphics[scale=0.65]{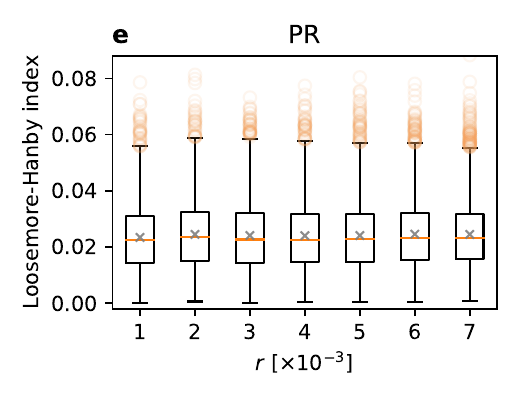}
    \includegraphics[scale=0.65]{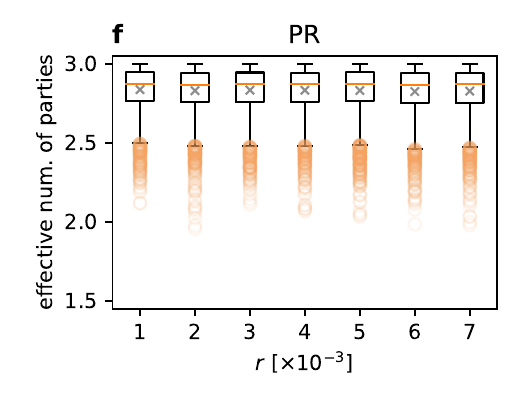}
    \includegraphics[scale=0.65]{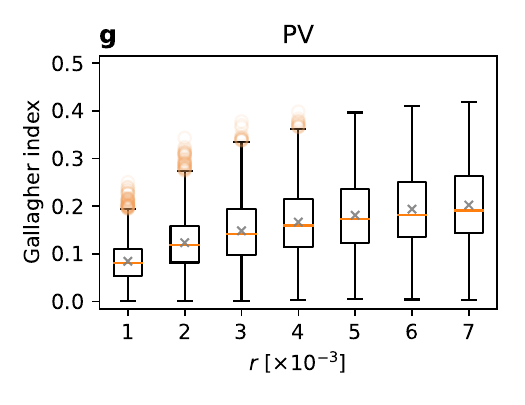}
    \includegraphics[scale=0.65]{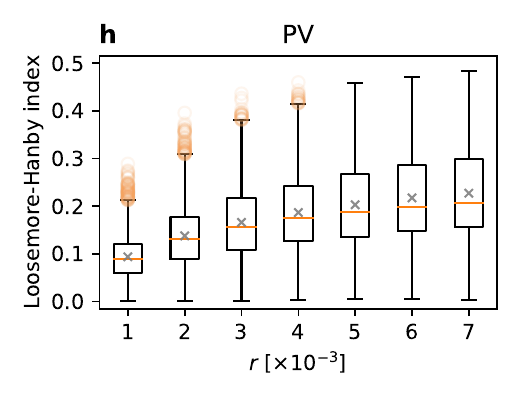}
    \includegraphics[scale=0.65]{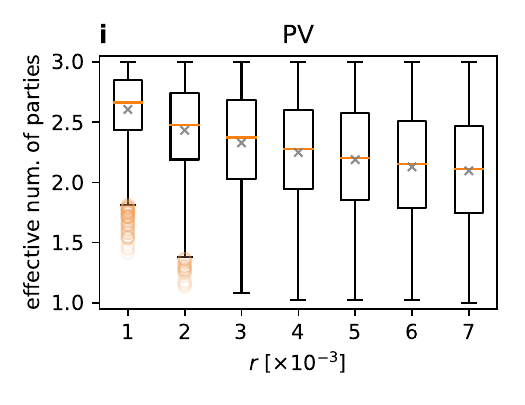}
\end{subfigure}
\caption[Comparison between PR and PV for different interconnectedness ratios]{\textbf{Comparison between PR and PV for different interconnectedness ratios.}
(a) Histograms of the seat share distribution of one of three parties when using
proportional representation (PR, blue) and plurality voting (PV, red) for varying ratio $r$.
Note how increasing ratio, which means loosing topological communities and approaching random graph,
increases the variance for PV, but does not have a clear influence on the results for PR.
See Figure~\ref{fig:basic-limit_cases} for the limit case of $r=1$.
(b, c) Box plots of the same results highlighting the differences in the distribution.
(d, e, f, g, h, i) Box plots of disproportionality and political fragmentation indexes for both ES.
The average value in box plots is indicated by an \textit{x}, the median by an orange line,
the width of the box is given by the first and third quartile,
the whiskers indicate values within 1.5 times the IQR,
and outliers are marked by a circle.
Other simulation parameters are the same as in Figure~\ref{fig:basic-es_histograms}:
$N=10^4$, $q=100$, $qs=5$, $k=12$, $\varepsilon=0.005$, and $n=3$ parties, results obtained over 4000 elections.
}
\label{fig:hist-box_ratio}
\end{figure}

In Figure~\ref{fig:hist-box_ratio}, we show the dependence of
election results on the interconnectedness ratio $r$. In
panels~(a) to~(c), we can see that the ratio has a very strong
influence on the outcome of PV but has no meaningful consequences
under PR. In the latter system, it does not matter if the support
for a party is concentrated in a few districts or spread over the
whole population, as long as it involves the same number of
supporting voters. In PV this distinction is crucial because even
the smallest plurality equally distributed across all districts will
result in a landslide victory, whereas the same support
concentrated in a few constituencies may result in losing the
leading position in the parliament. This fact is well known and
utilized in the manipulative practice of gerrymandering
\cite{stephanopoulos2015partisan, stewart2019information,kenny2023widespread}. It also
means that more separated districts provide less opportunity for
the dominance of a single party in PV since its support cannot
easily spread across the districts. That is why we observe a much
lower variance of the seat share for lower values of the
interconnectedness ratio $r$. Conversely, for high values of $r$,
the seat share in PV becomes more unstable, up to the point where
the distribution becomes bimodal, as visible in
Figure~\ref{fig:basic-limit_cases} for $r=1$.

In panels (d) to (i) of Figure~\ref{fig:hist-box_ratio}, we can
see that results under PR are more proportional and more fragmented
than under PV for any value of $r$. All the indexes stay
statistically unchanged in PR across different values of the
interconnectedness ratio. In PV, on the other hand,
disproportionality increases and political fragmentation decreases
with a growing ratio $r$. This can be directly linked to
increasing variance -- when one party dominates the landscape, the
proportionality of apportionment drops because smaller parties can
obtain no seats despite having non-zero support from voters.
Similarly, the effective number of parties decreases since fewer
parties enter the parliament or they enter with only a few seats.


\subsection{Zealot susceptibility}

The first category of external influences on the voting process
that we study are zealots. Zealots are individuals who exhibit
unwavering opinions and solely aim to sway others to vote for
the same party or candidate. They can operate in both the real
(political agitators) and online realms, with concerns
increasingly emerging regarding the impact of online bots on
democratic processes. In our simulations, all zealots assume
the same state $a$ and they are randomly distributed throughout
the entire population.

\begin{figure}
\centering
\begin{subfigure}{\textwidth}
    \includegraphics[scale=0.65]{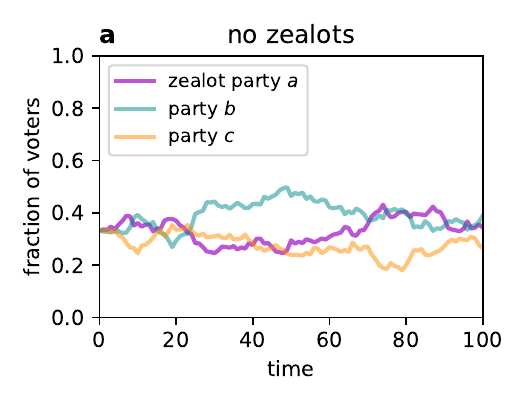}
    \includegraphics[scale=0.65]{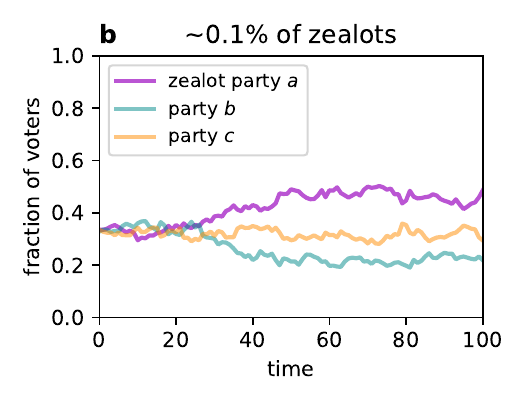}
    \includegraphics[scale=0.65]{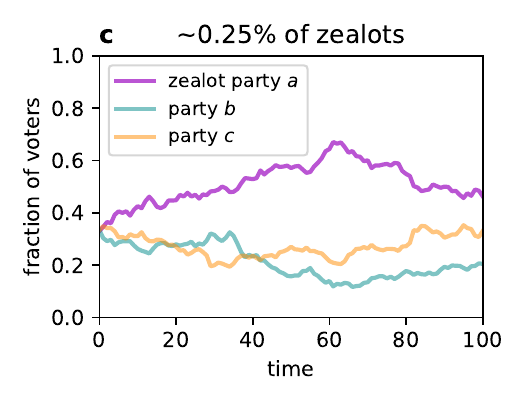}
\end{subfigure}
\begin{subfigure}{\textwidth}
    \includegraphics[scale=0.65]{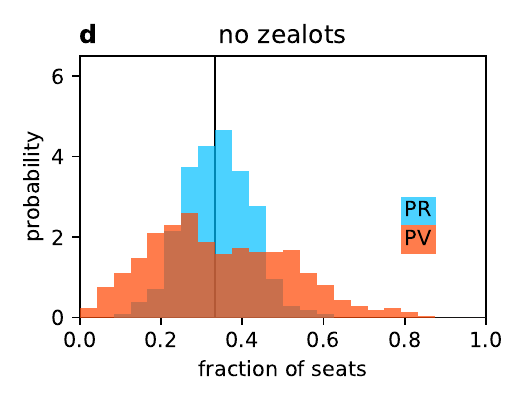}
    \includegraphics[scale=0.65]{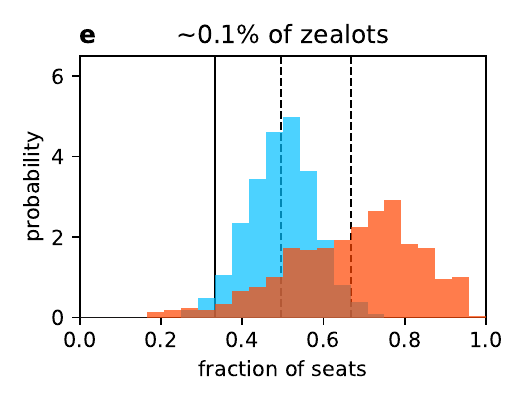}
    \includegraphics[scale=0.65]{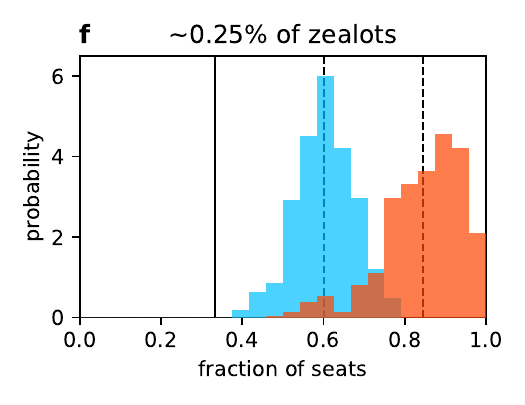}
\end{subfigure}
\begin{subfigure}{0.54\textwidth}
    \includegraphics[scale=0.67]{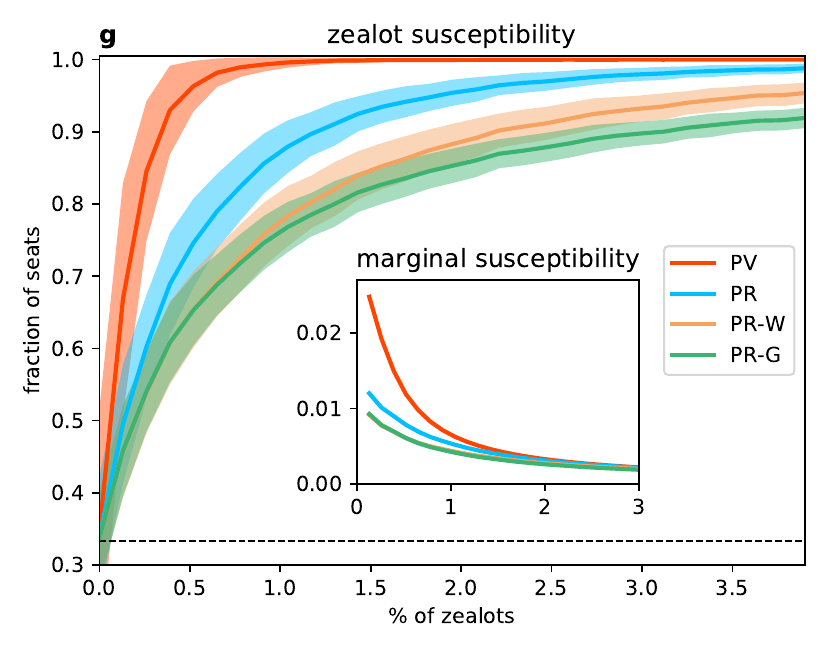}
\end{subfigure}\hfill
\begin{subfigure}{0.445\textwidth}
    \includegraphics[scale=0.61]{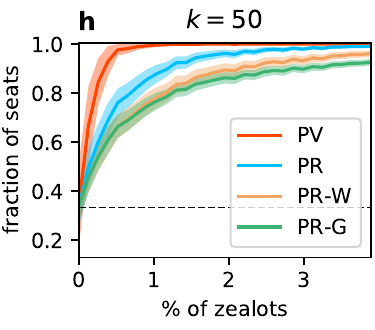}\hspace{0.01cm}
    \includegraphics[scale=0.61]{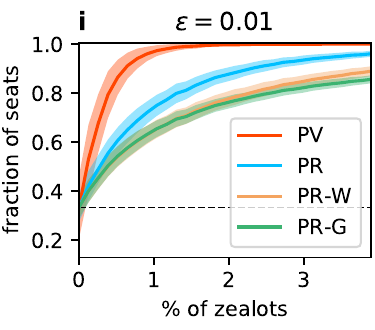}\newline
    \includegraphics[scale=0.61]{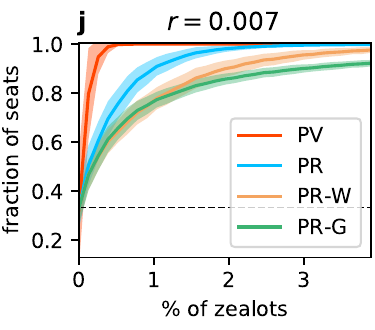}\hspace{0.01cm}
    \includegraphics[scale=0.61]{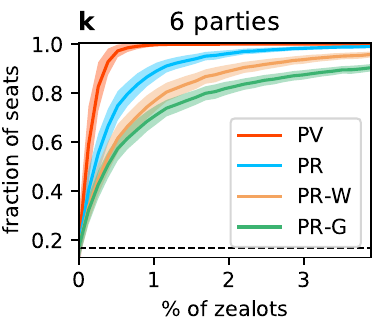}
\end{subfigure}
\caption[Influence of zealots on electoral processes for different ES]{\textbf{Influence of zealots on electoral processes for different ES.}
(a, b, c) Changes over time of the support for three parties with different number of zealots
in the population. Zealots always support party $a$.
(d, e, f) Seat share distribution of the zealot party $a$ under PR (blue) and PV (red) for different
number of zealots. Solid line indicates the average fraction of seats obtained in the symmetrical
case with no zealots which is 1/3, dashed lines indicate average values of shifted distributions.
Note, that the average result increases further for PV.
(g) The average result of the zealot party $a$ vs. percentage of zealots in the population for
four electoral systems: PR, PV, PR-G, PR-W. The colored shade indicates results within
one standard deviation from the average. The dashed line marks the average
fraction of seats obtained in the symmetrical case, i.e. 1/3. Inset plot: marginal susceptibility obtained
by dividing the total change in the result by the number of zealots.
Results obtained over 500 elections with parameters as in Figure~\ref{fig:basic-es_histograms}:
$N=10^4$, $q=100$, $qs=5$, $k=12$, $r=0.002$, $\varepsilon=0.005$, and $n=3$.
(h, i, j, k) Each plot shows zealot susceptibility as in (g), but with a single parameter changed
(indicated above each panel).
Note, that general differences in zealot susceptibility between the four ES are preserved
for varying average degree, noise rate, interconnectedness ratio, and number of parties.
}
\label{fig:basic-zealot_sus}
\end{figure}

To analyze the influence of zealots on elections under various
electoral systems, a straightforward approach involves
introducing a specific number of zealots and assessing the
resulting distribution of seat share. Figure~\ref{fig:basic-zealot_sus}
illustrates this scenario with different numbers of zealots
supporting party $a$. In panels (a) to (c) we can see how a small
percentage of political agitators can shift the trajectory of
opinion dynamics over time. The histograms in panels (d) to (f)
show how strongly it can alter the average fraction of seats
gained in elections and the general seat share distribution.
Already from those few examples, however, it is clear that zealots
have different influence on distinct ES -- PV magnifies the
disturbance when translating the votes into seats. In this sense
we can say that it is a more vulnerable system than PR, at least
in respect to political agitators or internet bots.

In order to comprehensively investigate the vulnerability of
electoral systems to zealot influence, we examine the concept of
zealot susceptibility. It measures the extent to which a system
tends to gravitate towards the zealot's state on average, when
the number of zealots varies. We present the zealot susceptibility
in panel (g) of Figure~\ref{fig:basic-zealot_sus}. Once again,
the disparities between the electoral systems are substantial.
In every system we observe a gradual increase in the mean election
result of the zealot's party towards complete dominance. However,
under PR even with almost 4\% of voters being zealots, the zealot
party does not achieve full victory. Conversely, in PV we observe
a rapid rise in the zealot's party's seat share, leading to
near-total dominance with approximately 1\% of zealots. Consequently,
PV system necessitates considerably less effort and resources to
exert effective influence. Indeed, it was also significantly more
susceptible to random fluctuations in the symmetrical case.

Additionally, we present in Figure~\ref{fig:basic-zealot_sus}~(g)
proportional representation with the Webster method (PR-W) and
with one global district (PR-G). PR-W and PR-G are surprisingly
similar in terms of zealot susceptibility. Both of them are less
susceptible to zealots than PR, but the difference is not as
pronounced as between PR and PV. In the inset plot of panel~(g) we
show the marginal susceptibility, i.e. the increase in the average
fraction of seats gained divided by the number of zealots that
were introduced in the population to achieve it. Notably, when
introducing the first zealots into the symmetrical case they have
much bigger capacity to alter the election results in PV. In this
system the marginal susceptibility is over two times higher than
in PR. It is the lowest in PR-W and PR-G, which are overlapping.
Nonetheless, with increasing influence already exerted we observe
diminishing returns of new zealots being introduced. At 3\% of
zealots all systems have the same marginal susceptibility,
although not all of them fully collapse in support of the zealot
party yet.

\begin{figure}
\centering
\begin{subfigure}{\textwidth}
\centering
    \includegraphics[scale=0.65]{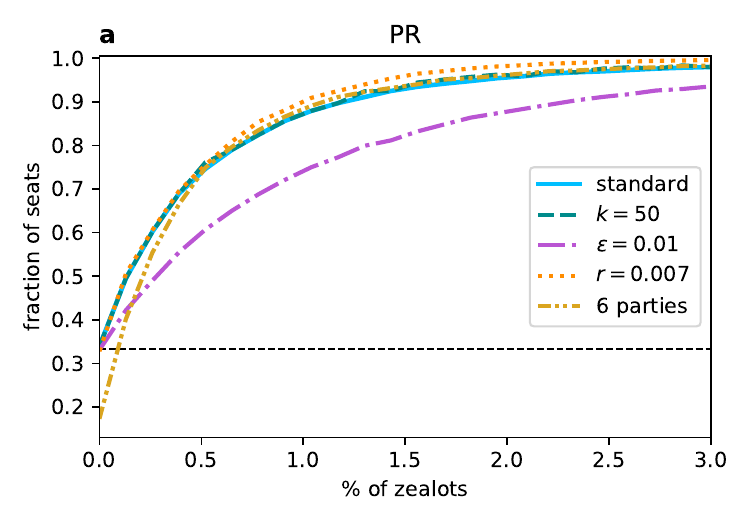}
    \includegraphics[scale=0.65]{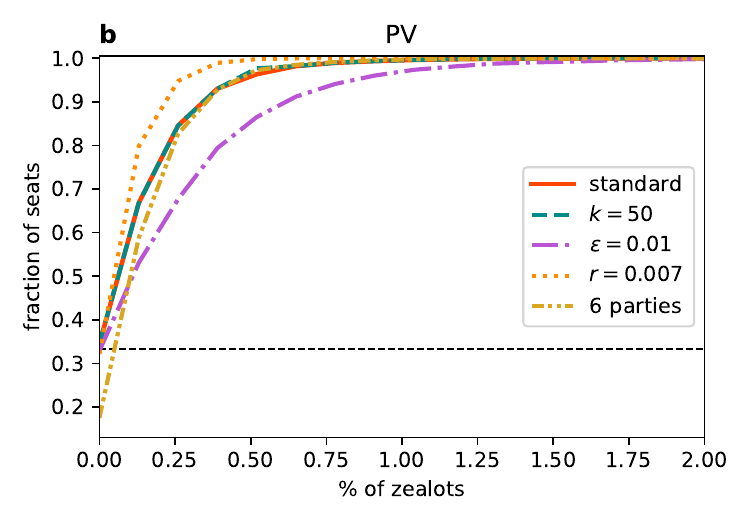}
\end{subfigure}
\begin{subfigure}{\textwidth}
    \includegraphics[scale=0.65]{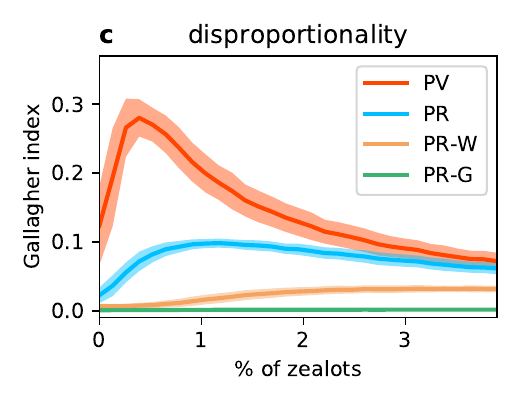}
    \includegraphics[scale=0.65]{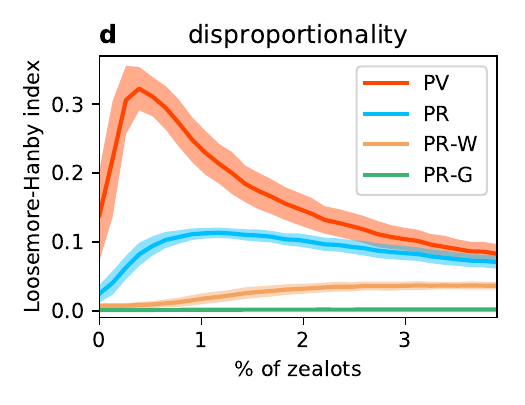}
    \includegraphics[scale=0.65]{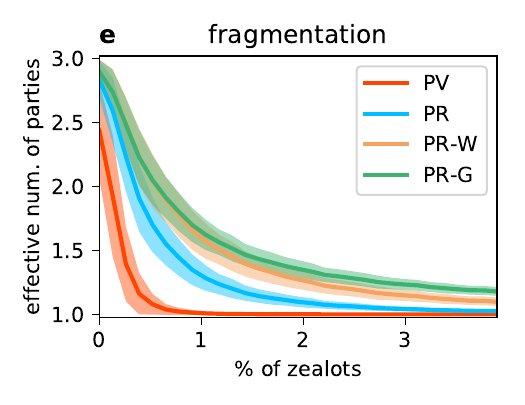}
\end{subfigure}
\caption[Influence of zealots, disproportionality and political fragmentation]{\textbf{Influence of zealots, disproportionality and political fragmentation.}
(a, b) Zealot susceptibility of PR and PV for different parameter values
(see also Figure~\ref{fig:basic-zealot_sus}).
Results obtained over 500 elections with parameters as in Figure~\ref{fig:basic-es_histograms}:
$N=10^4$, $q=100$, $qs=5$, $k=12$, $r=0.002$, $\varepsilon=0.005$, and $n=3$ parties, except for
one parameter that is changed and its value is indicated in the legend.
Change in the ratio $r$ has a significant influence on the  zealot susceptibility of PV
and a very small influence on the zealot susceptibility of PR. The noise rate $\varepsilon$
influences both systems, while the average degree and the number of parties
do not display significant influence on the zealot susceptibility.
(c, d, e) Average values of disproportionality and political fragmentation indexes
vs. percentage of zealots in the population for four electoral systems: PR, PV, PR-G, PR-W.
The colored shade indicates results within one standard deviation from the average.
Averaged over 500 elections with parameters as in Figure~\ref{fig:basic-es_histograms}.
}
\label{fig:basic-zealot_sus_indexes}
\end{figure}

In panels (h) to (k) of Figure~\ref{fig:basic-zealot_sus} we present
the same concept of zealot susceptibility as in panel~(g), but with
one parameter having different value. We can see that for altered
average degree $k$, noise rate $\varepsilon$, interconnectedness
ratio $r$, or number of parties $n$, the qualitative differences
between considered ES do not change -- again PV is the most
susceptible, then PR, PR-W, and PR-G in descending order. The
relative difference in zealot susceptibility between PV and PR is
the same for bigger average degree and more parties, but increases
for higher noise rate or interconnectedness ratio. It can be also
seen from panels~(a) and~(b) in Figure~\ref{fig:basic-zealot_sus_indexes}.
Higher $r$ hardly changes susceptibility of PR, but visibly increases
it for PV. Bigger $\varepsilon$ lowers the susceptibility for both
systems, but more so for PR.

In Figure~\ref{fig:basic-zealot_sus_indexes} panels~(c) to~(e) we also
show how disproportionality and fragmentation indexes change with
varying percentage of zealots. We can put the ES in order of ascending
values of the Gallagher index and the Loosemore-Hanby index and
descending values of the effective number of parties: PR-G, PR-W, PR,
and PV. The relative differences vary, however, and all indexes are
the most similar across the systems for higher numbers of zealots.
Remarkably, the disproportionality indexes reach the maximal value
for intermediate influences of zealots. For around 0.5\% of zealots
in the population, election results become highly disproportional for
PV with the Gallagher index almost reaching a value of 0.3 and the
Loosemore-Hanby index exceeding it. For PR, the maximum is obtained
at approximately 1\% of zealots with the indexes oscillating around 0.1.
It demonstrates that not always vast resources are necessary to damage
proportionality of elections and sway the result in a preferred direction.


\subsection{Media susceptibility}

In democratic societies, the press and media hold significant
influence, often referred to as the \textit{Fourth Power} for
a good reason \cite{anthony2000democracy,curran2011media,kim2022measuring}. The
extent to which newspapers, TV stations, and their online
counterparts can shape public opinion and impact the democratic
system is immense. Recognizing this impact, we incorporate media
as an external influence in our model, affecting all agents
within the system.

In a neutral scenario, each voter has a probability $\varepsilon$
of selecting their state randomly, with an equal chance of choosing each
political party. We introduce the media bias, representing the
effect of propaganda, as a predisposition that increases the
probability of choosing one party over others. Typically, multiple
media conglomerates exist, aligning with different political
orientations and targeting individuals based on particular factors
or seemingly independent attributes. In extreme cases, such as
online news feeds, individuals may find themselves trapped in
filter bubbles that continuously reinforce their existing beliefs
\cite{flaxman2016filter,bozdag2013bias}. While the intricacies
of media influence are a compelling subject in their own right,
our focus here is on a simplified media model to examine its
potential impact on various electoral systems.

\begin{figure}
\centering
\begin{subfigure}{\textwidth}
    \includegraphics[scale=0.65]{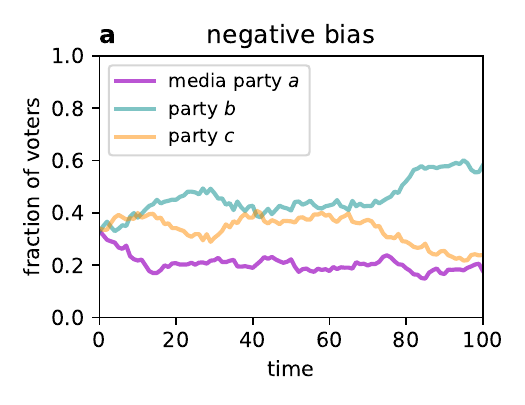}
    \includegraphics[scale=0.65]{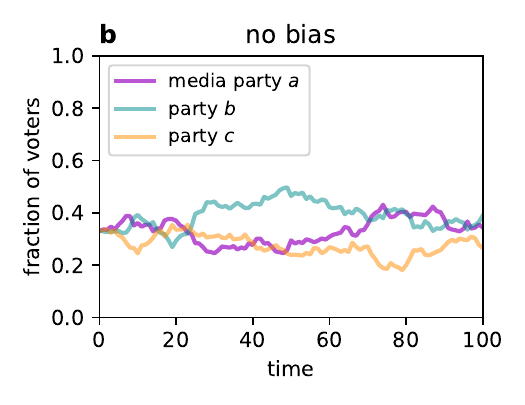}
    \includegraphics[scale=0.65]{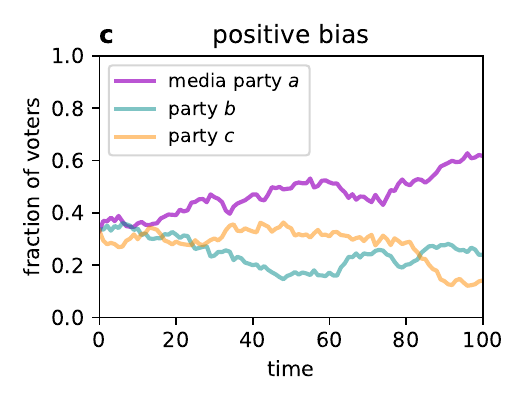}
\end{subfigure}
\begin{subfigure}{\textwidth}
    \includegraphics[scale=0.65]{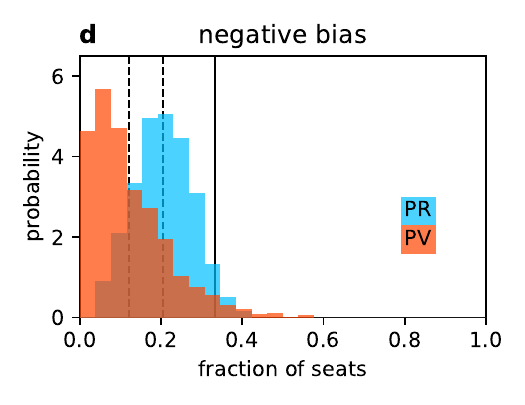}
    \includegraphics[scale=0.65]{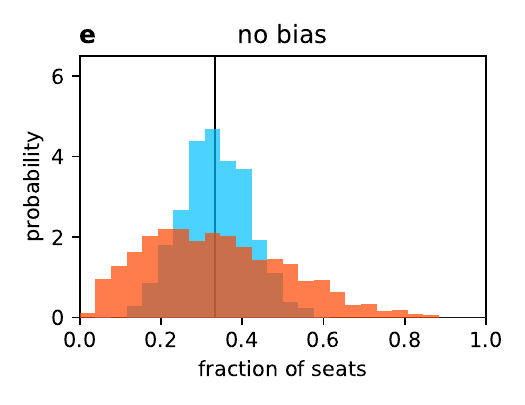}
    \includegraphics[scale=0.65]{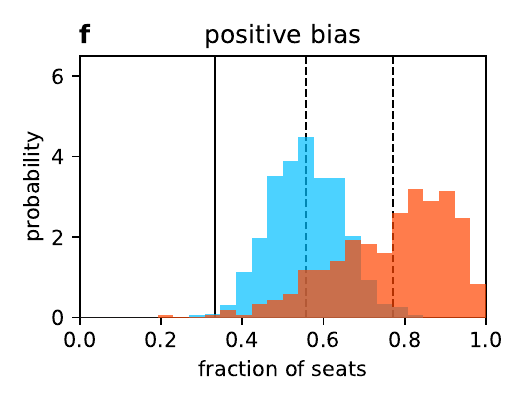}
\end{subfigure}
\begin{subfigure}{0.54\textwidth}
\centering
    \includegraphics[scale=0.67]{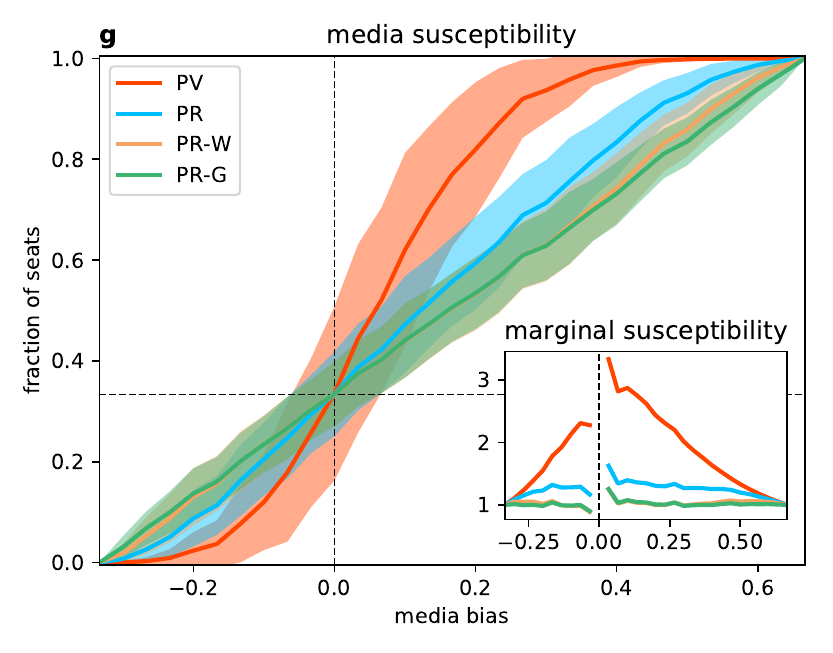}
\end{subfigure}\hfill
\begin{subfigure}{0.445\textwidth}
    \includegraphics[scale=0.61]{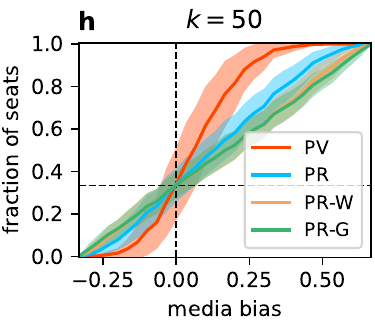}\hspace{0.01cm}
    \includegraphics[scale=0.61]{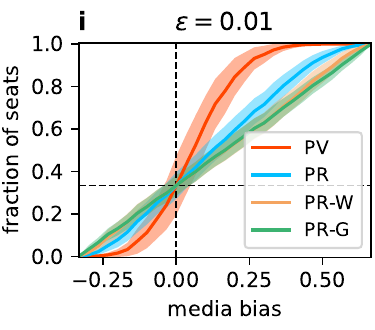}\newline
    \includegraphics[scale=0.61]{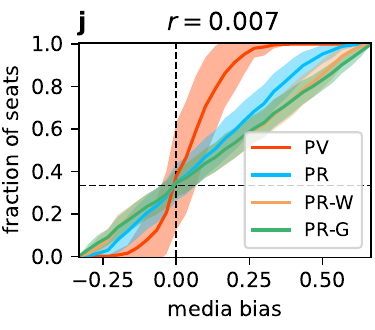}\hspace{0.01cm}
    \includegraphics[scale=0.61]{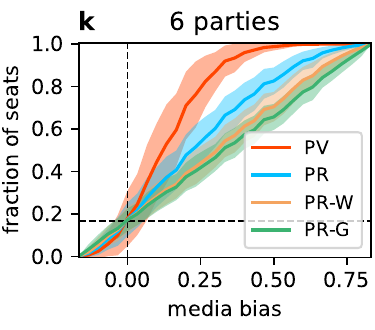}
\end{subfigure}
\caption[Influence of media propaganda on electoral processes for different ES]{\textbf{Influence of propaganda on electoral processes for different ES.}
(a, b, c) Changes over time of the support for three parties with negative bias (-0.13), no bias, and positive bias (0.27)
towards the media party $a$.
(d, e, f) Seat share distribution of the media party $a$ under PR (blue) and PV (red) for 
negative bias (-0.1), no bias, and positive bias (0.17) of media.
Solid line indicates the average fraction of seats obtained in the symmetrical
case with no media bias which is 1/3, dashed lines indicate average values of shifted distributions.
Note, that the average result is shifted more for PV.
(g) The average result of the media party $a$ vs. media bias for
four electoral systems: PR, PV, PR-G, PR-W. The colored shade indicates results within
one standard deviation from the average. The horizontal dashed line marks the average
fraction of seats obtained in the symmetrical case, i.e. 1/3, and the vertical one the case with no bias.
Inset plot: marginal susceptibility obtained
by dividing the total change in the result by the strength of bias.
Results obtained over 1000 elections with parameters as in Figure~\ref{fig:basic-es_histograms}:
$N=10^4$, $q=100$, $qs=5$, $k=12$, $r=0.002$, $\varepsilon=0.005$, and $n=3$.
(h, i, j, k) Each plot shows media susceptibility as in (g), but with a single parameter changed
(indicated above each panel).
Note, that general differences in media susceptibility between the four ES are preserved
for varying average degree, noise rate, interconnectedness ratio, and number of parties.
}
\label{fig:basic-media_sus}
\end{figure}

In our simulations the propaganda is always directed towards the same party
-- party $a$. We then measure the bias as the difference between the
probability of choosing party $a$ and the neutral probability
$1/n$, where $n$ is the number of parties. If the bias is positive,
it means the propaganda is in favor of the party. If it is
negative, media tries to decrease the support for the party. We
illustrate both cases, together with the neutral one, in
Figure~\ref{fig:basic-media_sus}. In panels~(a) to~(c) we present
the impact it has on the trajectory of opinion dynamics over time.
The resulting seat share distributions are shown in panels~(d)
to~(f). Indeed, biased media can shift the average number of seats
gained in both directions and has stronger effect on PV. In this
sense we would say that PV system is more susceptible to media
propaganda than PR.

Figure~\ref{fig:basic-media_sus}~(g) shows media susceptibility as
a function of amount of bias. Additionally, it includes
proportional representation with the Webster method (PR-W) and
with one global district (PR-G), which are very similar in terms
of media susceptibility. Both of them are slightly less
susceptible to media influence than PR, but the biggest difference
can be observed between PV and any proportional system. In the
inset plot of panel~(g) we show the marginal susceptibility,
i.e. the increase (or decrease) in the average fraction of seats
gained divided by the media bias (i.e. the shift from the neutral probability $1/n$
of choosing party $a$). Clearly, the impact of media propaganda is
much higher in PV, especially for relatively small deviations from
the neutral case -- the marginal susceptibility is almost three
times higher than in PR. Nevertheless, with increasing bias we can
see diminishing returns of media influence, however media
susceptibilities of different systems only merge at the extreme
case of full media coverage in favor one party (or against it).

Within Figure~\ref{fig:basic-media_sus}, panels (h) to (k) provide
an exploration of media susceptibility, similar to that shown in
panel~(g), with variations in specific parameters. Despite altering
the values of average degree $k$, noise rate $\varepsilon$,
interconnectedness ratio $r$, or the number of parties $n$, the
qualitative distinctions among the considered electoral systems
remain consistent. Once again, PV exhibits the highest
susceptibility, followed by PR, PR-W, and PR-G in descending order.
Notably, larger values of $\varepsilon$ greatly reduce the
standard deviation of media susceptibility for all systems.

\begin{figure}
\centering
\begin{subfigure}{\textwidth}
\centering
    \includegraphics[scale=0.65]{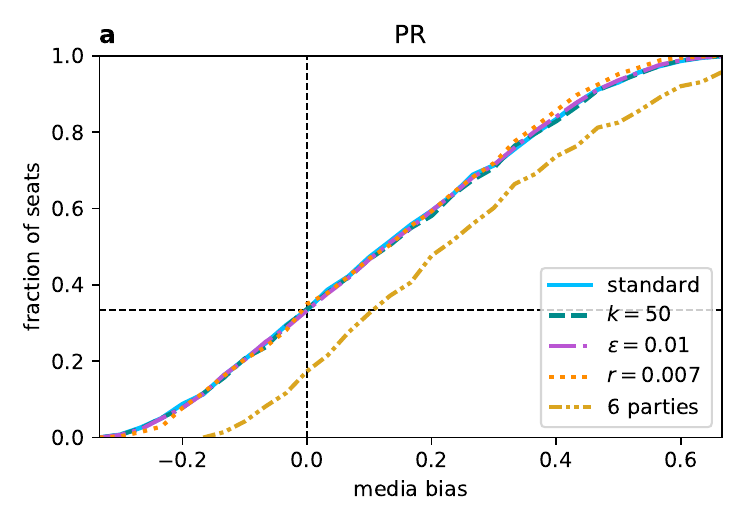}
    \includegraphics[scale=0.65]{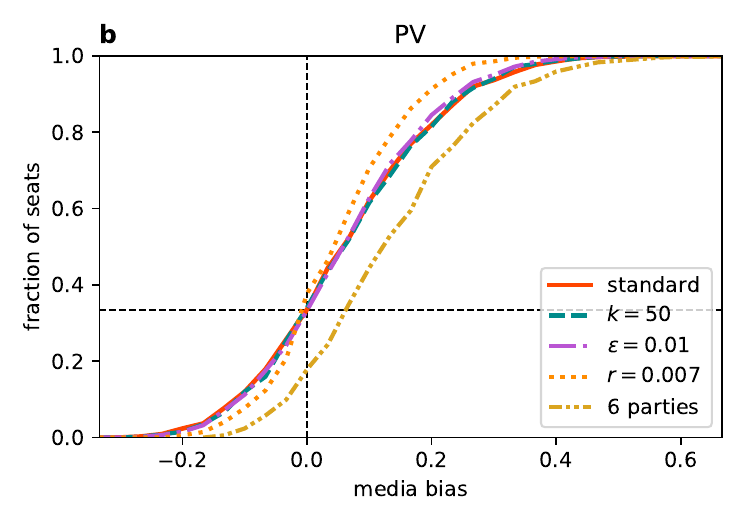}
\end{subfigure}
\begin{subfigure}{\textwidth}
    \includegraphics[scale=0.65]{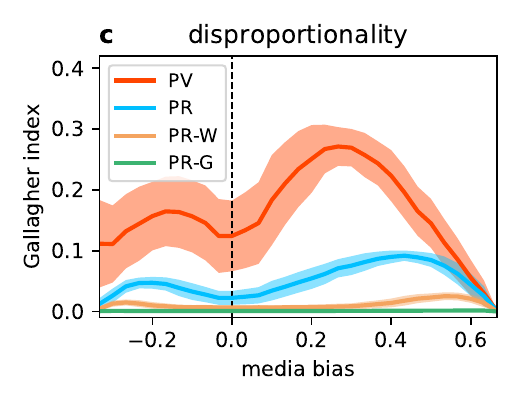}
    \includegraphics[scale=0.65]{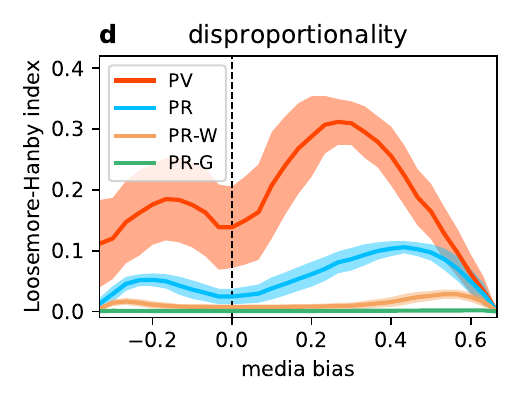}
    \includegraphics[scale=0.65]{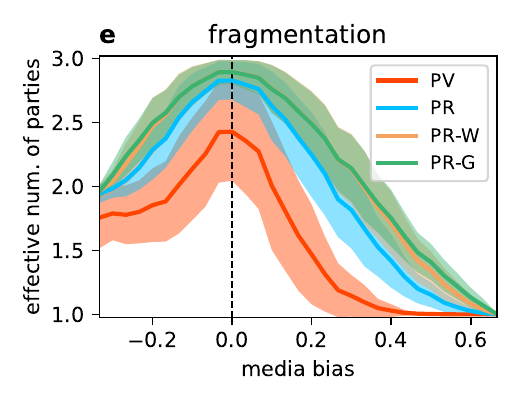}
\end{subfigure}
\caption[Influence of media propaganda, disproportionality and political fragmentation]{\textbf{Influence of propaganda, disproportionality and political fragmentation.}
(a, b) Media susceptibility of PR and PV for different parameter values
(see also Figure~\ref{fig:basic-media_sus}).
Results obtained over 1000 elections with parameters as in Figure~\ref{fig:basic-es_histograms}:
$N=10^4$, $q=100$, $qs=5$, $k=12$, $r=0.002$, $\varepsilon=0.005$, and $n=3$ parties, except for
one parameter that is changed and its value is indicated in the legend.
Change in ratio $r$ has a significant influence on the media susceptibility of PV
and a very small influence on the zealot susceptibility of PR. Noise rate,
average degree, and number of parties
do not display significant influence on the zealot susceptibility.
Note, that the neutral media coverage lays at a different value of x-axis for higher number of parties,
therefore the lines are separated.
(c, d, e) Average values of disproportionality and political fragmentation indexes
vs. media bias for four electoral systems: PR, PV, PR-G, PR-W.
The colored shade indicates results within one standard deviation from the average.
Averaged over 1000 elections with parameters as in Figure~\ref{fig:basic-es_histograms}.
}
\label{fig:basic-media_sus_indexes}
\end{figure}

The relative difference in media susceptibility between PV and PR
maintains the same pattern for larger average degree, higher noise,
or more parties. However, the difference becomes bigger with higher
interconnectedness ratio. This trend is evident in panels~(a) and~(b)
of Figure~\ref{fig:basic-media_sus_indexes}, where an increase in
$r$ has minimal impact on the susceptibility of PR, yet is more
noticeably for PV.

In panels (c) to (e) of Figure~\ref{fig:basic-media_sus_indexes},
we present the variations in disproportionality and fragmentation
indexes in response to different media biases. We arrange the
electoral systems in ascending order of the Gallagher index and
the Loosemore-Hanby index, while in descending order based on the
effective number of parties: PR-G, PR-W, PR, and PV. Nonetheless,
the relative differences between the indexes vary, and for extreme
positive bias all indexes show the highest similarity across the
systems.

Remarkably, the disproportionality indexes reach their maximum
values for intermediate values of negative or positive media bias.
For a positive bias of approximately 0.27 the election results for
PV become highly disproportional, with the Gallagher index almost
reaching a value of 0.3, surpassed by the Loosemore-Hanby index.
For PR, the maximum disproportionality occurs at around 0.45, with
the indexes oscillating around 0.1. The effective number of parties
decreases for extreme negative bias up to 2 (or slightly below for PV),
and up to 1 for extreme positive bias. This is because negative bias
can eliminate one of three parties and positive bias can eliminate two,
by forging a landslide victory of one party. These findings demonstrate
that obtaining high media coverage and using it in favor or against
a political party can drive the election results and disrupt the
proportionality.


\section{Real-world electoral systems}

Studying and understanding the vulnerabilities of electoral systems is of
paramount importance. As we could see in previous sections, they are
susceptible to various influences that can compromise their integrity and
undermine democratic processes. By investigating these vulnerabilities in an
abstract setting, we gain insight into differences between basic types of
electoral systems. We also learn what features of a system can increase its
susceptibility to political agitators or propaganda.

The study of \textit{real-world} electoral systems' vulnerabilities,
however, offers several key benefits. Firstly, it enhances our understanding
of the potential risks and challenges that electoral systems face in practice.
Real-world elections are complex, multifaceted processes influenced by diverse
factors such as technological advancements, social dynamics, and political
strategies. Those factors depend on the design of a specific electoral system.
When studying its properties numerically, we can identify exact points of
weakness that arise from external influences. By examining susceptibility to
zealots and media propaganda, we gain valuable insights into potential risks
and challenges that can compromise the integrity of elections. Understanding
these vulnerabilities enables the development of strategies to
mitigate risks and safeguard the democratic electoral process.

Furthermore, studying vulnerabilities in real-world electoral systems allows
for the development and evaluation of potential reforms and improvements. By
identifying weaknesses and analyzing their impact, we can propose
evidence-based recommendations to enhance the stability of elections and
develop effective measures to protect against manipulation, fraud, and biased
outcomes. This research can inform policymakers, electoral authorities, and
civil society organizations, facilitating informed reforms that strengthen
democratic governance and enhance public trust in electoral systems.


\subsection{Sejm of the Republic of Poland}

\begin{figure}
\centering
\begin{subfigure}{\textwidth}
\centering
    \includegraphics[scale=0.31]{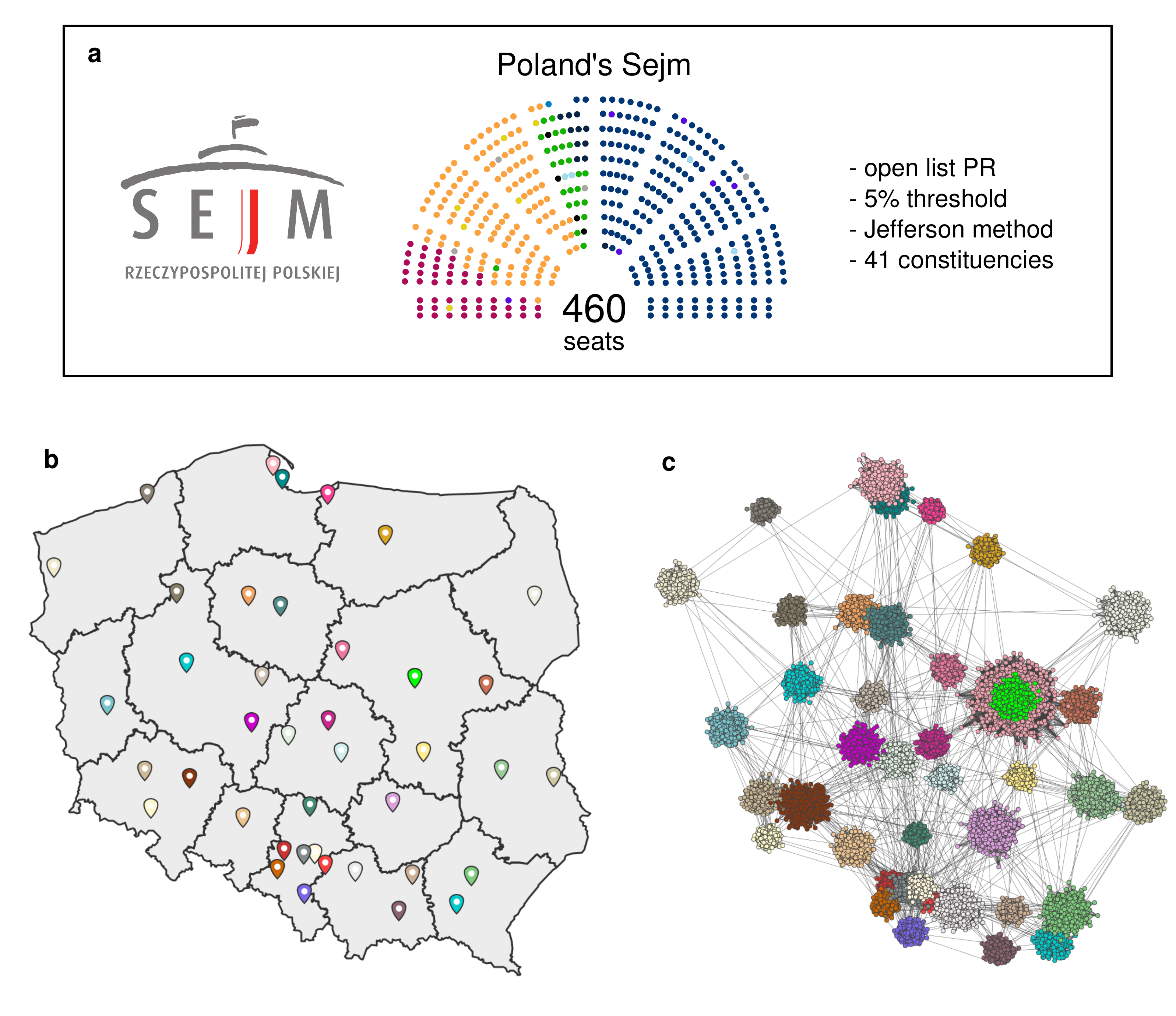}
\end{subfigure}
\begin{subfigure}{\textwidth}
\centering
    \includegraphics[scale=0.65]{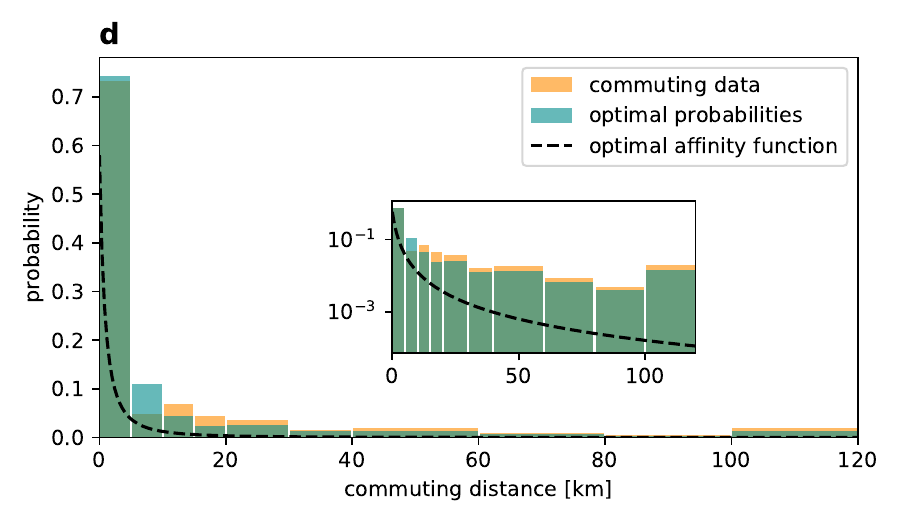}
\end{subfigure}
\caption[Simulation setup for the Poland's Sejm]{\textbf{Simulation setup for the Poland's Sejm}
(a) Logo of the Sejm of the Republic of Poland and basic information about the electoral
system used in electing its members.
(b) Map of Poland with division into 16 voivodeships, i.e. the biggest administrative districts of the country.
The capital of each of 41 electoral districts applied
in elections to the Sejm is indicated with a colored marker.
(c) Network of $41 \times 10^3$ nodes generated for simulations of elections to the Poland's Sejm.
Nodes in different constituencies have different colors (corresponding to the marker on the map)
and are centered around the capital of the constituency.
The bigger the population of a district, the more spread are the nodes.
(d) Probability distribution of home-work commuting distance for Poland. The empirical commuting data is taken from
\cite{statOszacowanieOdlegoci,statPrzepywyLudnoci} (see the supplementary datasets) and the optimal distribution
is found by fitting the affinity function [\ref{eqn:planar_affinity}], with resulting
planar constant $c=1.7237$.
Connections in the network are generated based on the optimal affinity function.
[logo in (a) by Sejm RP, sejm.gov.pl, Public Domain, \href{https://commons.wikimedia.org/w/index.php?curid=46006978}{https://commons.wikimedia.org/w/index.php?curid=46006978}, seats layout in (a) by Szczeszek2035 - Own work, \href{https://creativecommons.org/licenses/by-sa/4.0/}{CC BY-SA 4.0}, \href{https://commons.wikimedia.org/w/index.php?curid=100983133}{https://commons.wikimedia.org/w/index.php?curid=100983133}]
}
\label{fig:pl_sejm-basic_info}
\end{figure}

The first real-world electoral system that we study is the one used in elections to the Poland's Sejm,
which is the lower house of the bicameral parliament of Poland. The system is a party-list
proportional representation with open lists -- voters select a specific candidate
from those put on the list by the party they support, influencing not only the number of seats
the party will obtain, but also which candidates will enter the parliament. There are 460 seats in total
in the Sejm. To assign them, the country is divided into 41 constituencies with
the number of seats varying from 7 to 20 in each. Parties obtaining less than 5\% of all votes nation-wide
are excluded and do not enter the parliament (for coalitions the threshold is 8\%,
and for ethnic minorities its 5\% district-wide, but in the simulations
we use the global 5\%). The seats are assigned in every district to the parties passing over this threshold
using the Jefferson seat assignment method (also known as the D'Hondt method). See Figure~\ref{fig:pl_sejm-basic_info}
for a summary of the Sejm's electoral system. In the 2019 elections to the Poland's Sejm
there was approximately 30 million people able to vote with an average of $733\pm219$ thousands 
of voters per constituency. In every district there were at least 5 political parties registered
\cite{pkwWyborySejmu}.

For the simulation of elections to the Poland's Sejm we create a network with
$41 \times 10^3$ nodes divided into 41 topological communities using the spatial version of the SBM
(see subsection \textit{Network generation}). Those communities correspond to real electoral
districts and their relative sizes are based on the actual populations of voters
in the districts. Each community is
assigned a geographical location, more precisely the coordinates of the district's capital
(see panels~(b) and~(c) of Figure~\ref{fig:pl_sejm-basic_info}).
The connectivity between nodes decreases as a function of the geographical distance between them,
where the function is obtained by fitting [\ref{eqn:planar_affinity}] to empirical data
on work commuting distance \cite{statOszacowanieOdlegoci,statPrzepywyLudnoci},
as demonstrated in Figure~\ref{fig:pl_sejm-basic_info}~(d).
We include all the relevant details of the electoral system in the simulation, including
the 5\% entry threshold, the Jefferson seat assignment method, and the exact number of seats in
each district. We run the simulation for $n=5$ parties, an average degree of $k=12$,
and various noise rates $\varepsilon$. See supplementary datasets or the configuration files 
in our online code repository \cite{github_repo} for the exact data used in simulations.

In addition to the actual electoral system used in elections to the Poland's Sejm 
(labeled as ,,41 dist. PR*'' in the figures),
we study four alternative systems that either could be applied in Poland or are theoretically meaningful.
The first alternative system (,,41 dist. PV'') is a PV system with the same division
into districts and the same number of seats
assigned to each district, but the party with plurality in a given district obtains all seats
(effectively it is a PBV system). Plurality voting was actually considered during 2011 electoral reform
in Poland. That year the system of the Poland's Senate changed from a PR to FPTP voting.
Similar transformation was considered also for the Sejm, but was deemed unconstitutional
due to a provision stating that
,,Elections to the Sejm shall be universal, equal, direct, and \textit{proportional},
and shall be conducted by secret ballot'' \cite{sejmAktyPrawne}.
We then consider two alternative systems with electoral districts corresponding to 16
voivodeships, i.e. the highest-level administrative districts of Poland
(see Figure~\ref{fig:pl_sejm-basic_info}~(b) for a map). We merge all electoral districts
within the same voivodeship into one, accumulating the seats into a single pool per voivodeship.
On top of this new division we consider the original PR system (,,16 dist. PR'') and
a PBV system with the winner-takes-all approach in each district (,,16 dist. PV'').
Finally, we examine a PR system where all 460 seats are assigned within a singe nation-wide 
district (,,1 dist. PR''). We do not include the entry threshold in the PV alternative systems,
and the PR alternative systems have the same 5\% threshold and the seat assignment method as the original one.

\begin{figure}
\centering
\begin{subfigure}{0.672\textwidth}
\centering
    \includegraphics[scale=0.67]{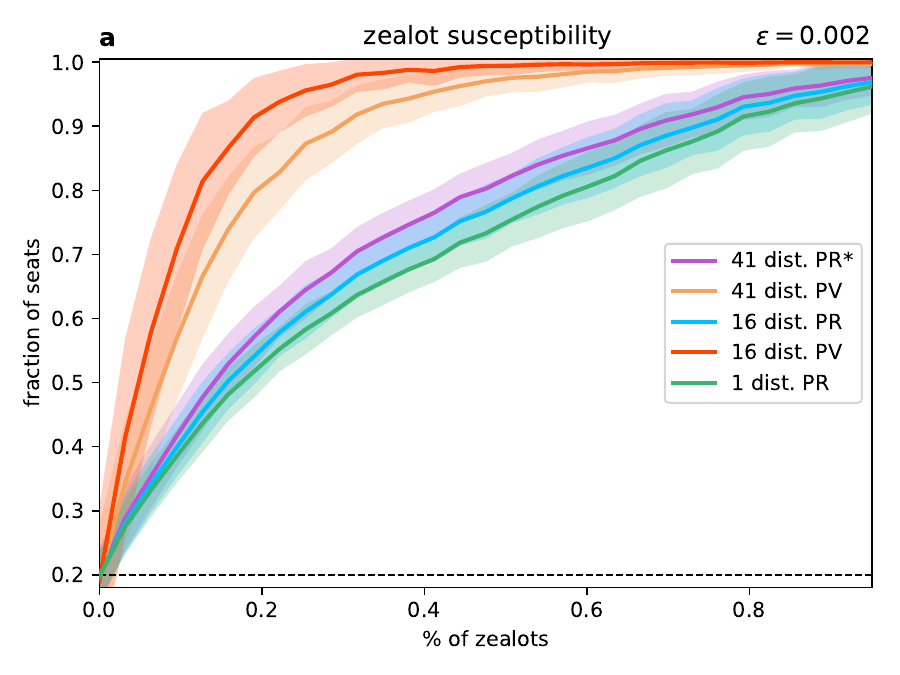}
\end{subfigure}\hfill
\begin{subfigure}{0.328\textwidth}
    \includegraphics[scale=0.65]{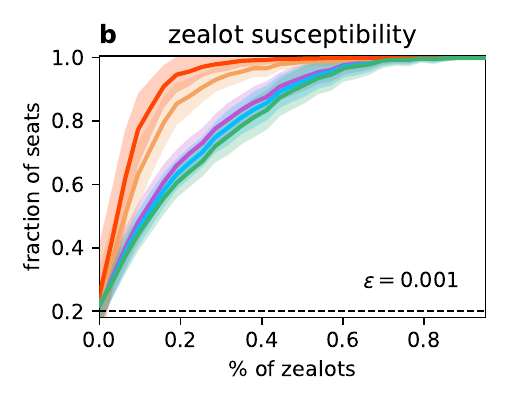}
    \includegraphics[scale=0.65]{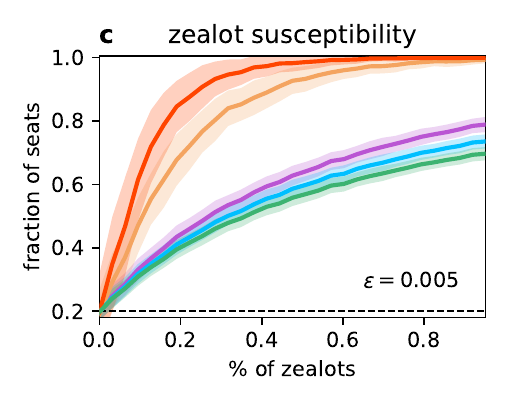}
\end{subfigure}
\begin{subfigure}{\textwidth}
    \includegraphics[scale=0.65]{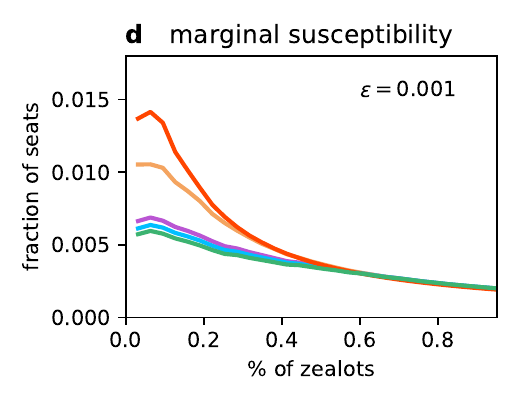}
    \includegraphics[scale=0.65]{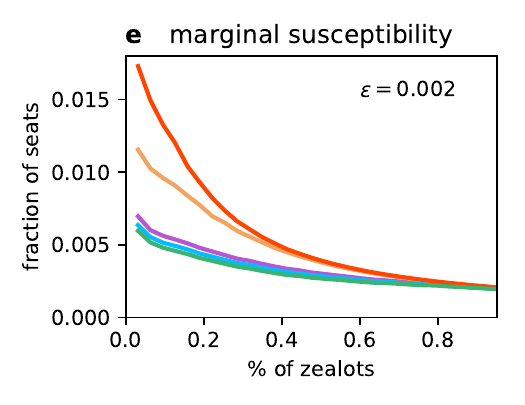}
    \includegraphics[scale=0.65]{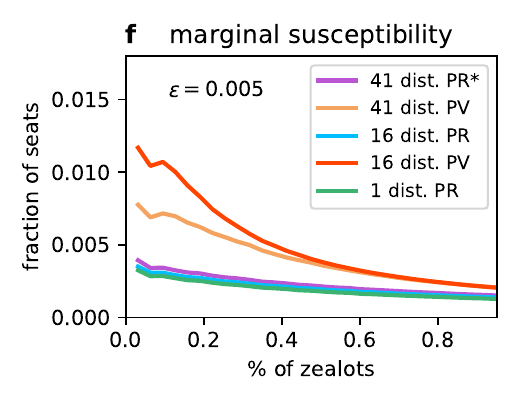}
\end{subfigure}
\begin{subfigure}{\textwidth}
    \includegraphics[scale=0.65]{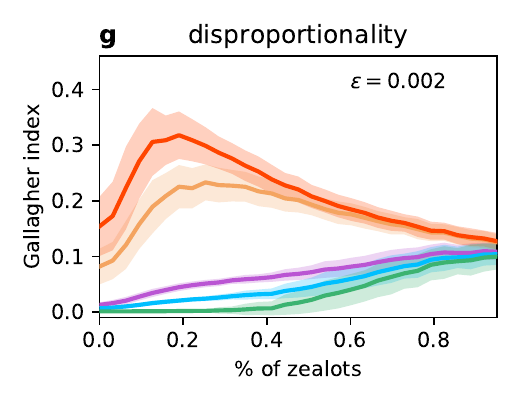}
    \includegraphics[scale=0.65]{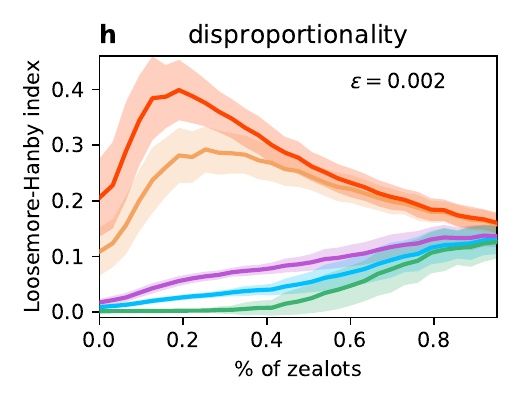}
    \includegraphics[scale=0.65]{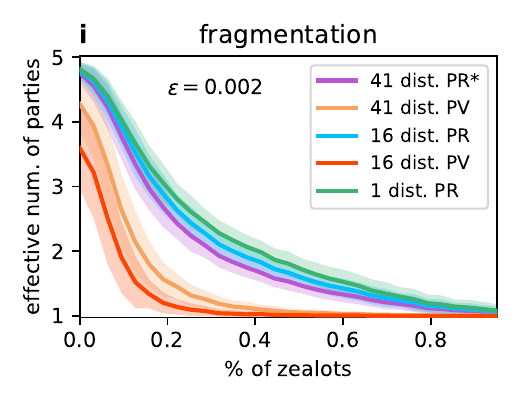}
\end{subfigure}
\caption[Zealot susceptibility of elections to the Poland's Sejm]{\textbf{Zealot susceptibility of elections to the Poland's Sejm.}
All plots compare results for the actual electoral system used in elections to the Sejm
(marked with an asterisk) with 4 alternative systems with different numbers of districts.
See the main text for details on ES used.
(a, b, c)  The average result of the zealot party $a$ vs. percentage of zealots in the population.
The dashed line marks the average
fraction of seats obtained in the symmetrical case, i.e. 1/5.
(d, e, f) Marginal zealot susceptibility  obtained
by dividing the total change in the fraction of seats won by the number of zealots.
(g, h, i) Average values of disproportionality and political fragmentation indexes
vs. percentage of zealots in the population.
The colored shade indicates results within
one standard deviation from the average.
All results were obtained over
200 elections with 5 parties. The network has $41 \times 10^3$ nodes, with the average degree
$k=12$, divided into 41 communities corresponding to the actual electoral districts
(see Figure~\ref{fig:pl_sejm-basic_info}).
There is 460 seats to gain in total.
The noise rate $\varepsilon$ is given in each panel.
}
\label{fig:pl_sejm-zealot_sus}
\end{figure}

In Figure~\ref{fig:pl_sejm-zealot_sus} we demonstrate an analysis of the susceptibility to zealots
for all ES considered for the Sejm.
Panels~(a) to~(c) show the average fraction of seats won by the zealot party
versus the percentage of zealots in the population. The first important result is the difference in
zealot susceptibility between PR and PV systems. All three PR systems are relatively similar
in their response to zealots and much less susceptible than both PV systems. This observation
holds for any value of the noise rate $\varepsilon$, although with bigger noise zealots are
less influential in general. Looking at the marginal susceptibility in panels~(d) to~(f),
we can see that investing in political agitators pays much better dividends in PV
especially at the beginning -- influence of each zealot can be three times bigger than in PR.

In panels (g) to (i) of Figure~\ref{fig:pl_sejm-zealot_sus} we show the Gallagher index, 
the Loosemore-Hanby index, and the effective number of parties. Strikingly, under PV systems
elections can become highly disproportional for very small percentage of zealots in the population
with the zealot party still far away from a landslide victory. The maximum value
of disproportionality indexes in PV is reached below 0.2\% of zealots for $\varepsilon=0.002$,
while in PR systems disproportionality becomes maximal when the zealot party is close to
a complete domination. The political fragmentation decreases with increasing number of zealots,
as expected, and is generally lower for PV systems.

\begin{figure}
\centering
\begin{subfigure}{0.66\textwidth}
\centering
    \includegraphics[scale=0.67]{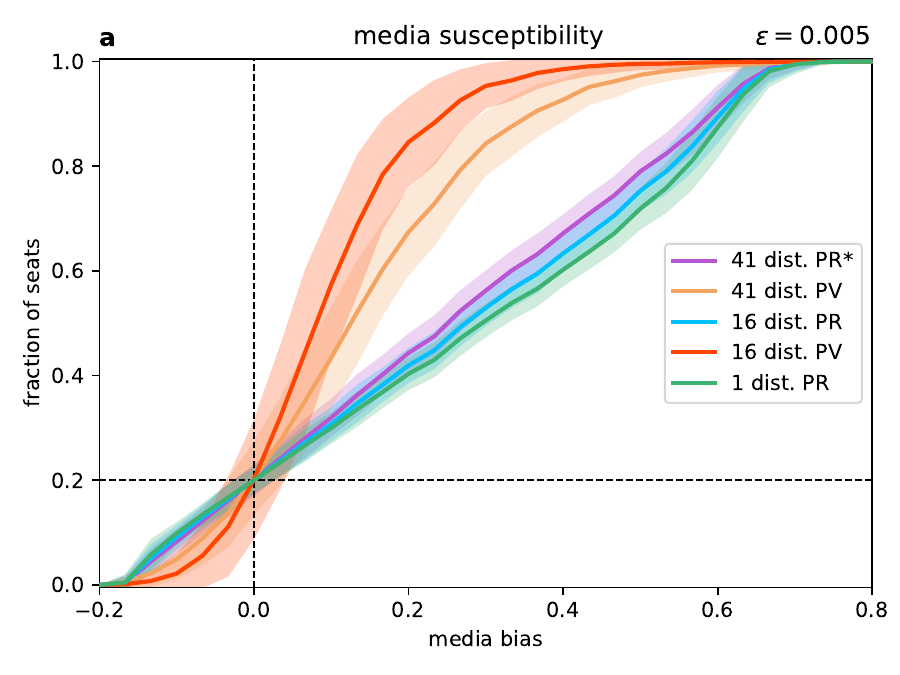}
\end{subfigure}\hfill
\begin{subfigure}{0.34\textwidth}
    \includegraphics[scale=0.65]{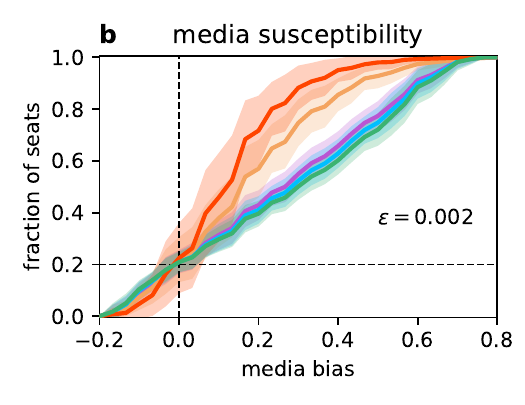}
    \includegraphics[scale=0.65]{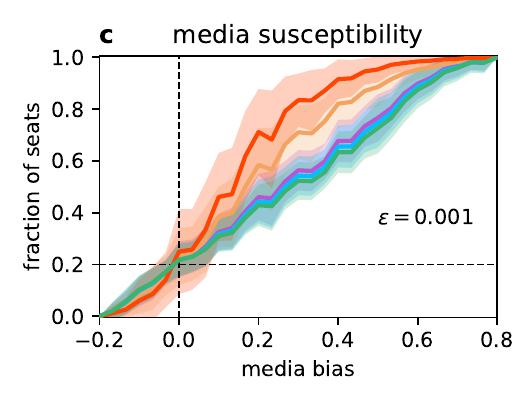}
\end{subfigure}
\begin{subfigure}{\textwidth}
    \includegraphics[scale=0.65]{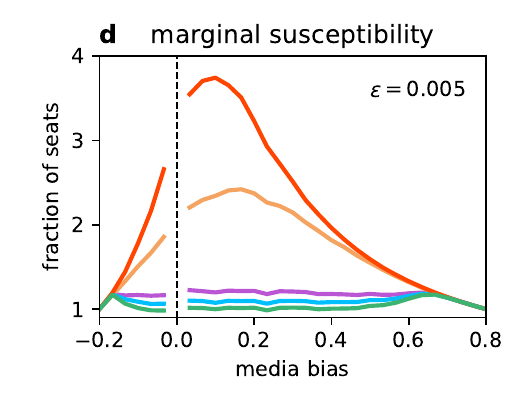}
    \includegraphics[scale=0.65]{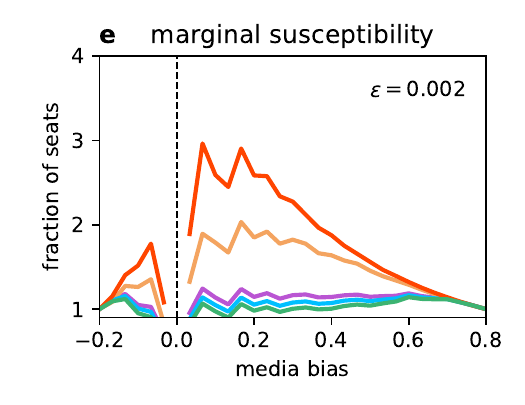}
    \includegraphics[scale=0.65]{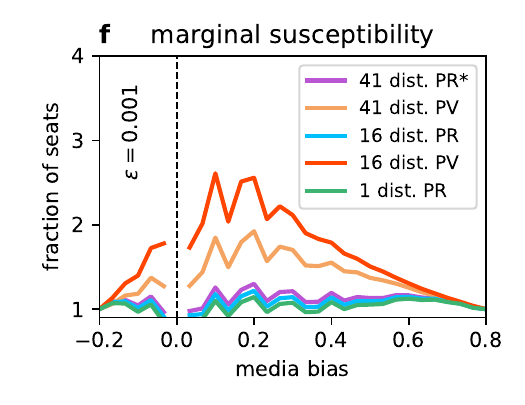}
\end{subfigure}
\begin{subfigure}{\textwidth}
    \includegraphics[scale=0.65]{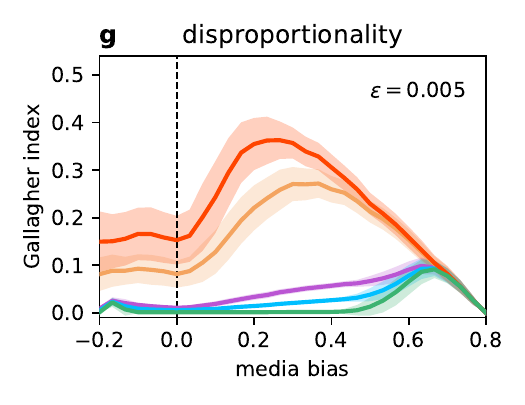}
    \includegraphics[scale=0.65]{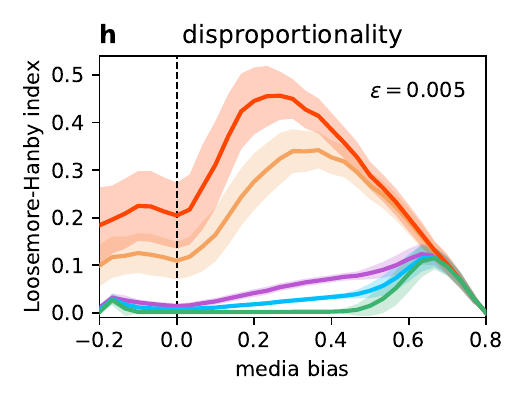}
    \includegraphics[scale=0.65]{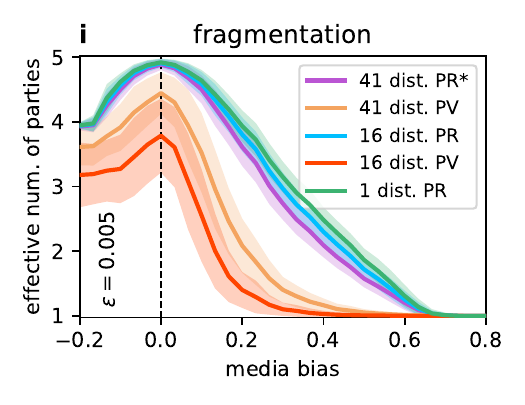}
\end{subfigure}
\caption[Media susceptibility of elections to the Poland's Sejm]{\textbf{Media susceptibility of elections to the Poland's Sejm.}
All plots compare results for the actual electoral system used in elections to the Sejm
(marked with an asterisk) with 4 alternative systems with different numbers of districts.
See the main text for details on ES used.
(a, b, c)  The average result of the media party $a$ vs. media bias.
The horizontal dashed line marks the average
fraction of seats obtained in the symmetrical case, i.e. 1/5, and the vertical one the case with no bias.
(d, e, f) Marginal media susceptibility  obtained
by dividing the total change in the fraction of seats won by the strength of bias.
(g, h, i) Average values of disproportionality and political fragmentation indexes
vs. media bias.
The colored shade indicates results within
one standard deviation from the average.
All results were obtained over
1000 elections with 5 parties. The network has $41 \times 10^3$ nodes, with the average degree
$k=12$, divided into 41 communities corresponding to the actual electoral districts
(see Figure~\ref{fig:pl_sejm-basic_info}).
There is 460 seats to gain in total.
The noise rate $\varepsilon$ is given in each panel.
}
\label{fig:pl_sejm-media_sus}
\end{figure}

Figure~\ref{fig:pl_sejm-media_sus} illustrates the second external influence that we study,
namely the susceptibility to media propaganda, as a function of amount of bias.
The conclusions are similar to the ones in respect to zealots -- PV systems are generally
more susceptible to biased media than PR systems. All PR systems, although having
very different district divisions, display a very similar media susceptibility
within one standard deviation from each other. In panels~(a) to~(c) we can see that
for lower value of the nose rate $\varepsilon$ ES become less susceptible
and the general pattern becomes more irregular. This is also confirmed by the marginal 
susceptibility in panels~(d) to~(f). Notably, in PV media bias provides the highest
returns not at the very beginning, like zealots, but after already exerting some influence
-- the maximum of marginal susceptibility is shifted to the right from the zero bias.

Similarly to zealots, media propaganda causes the highest disproportionality for intermediate
values of influence, as visible in panels~(g) to~(i) of Figure~\ref{fig:pl_sejm-media_sus}.
It happens much earlier, however, in PV systems than in PR ones. Again, the disparity between
the two types of ES is much bigger in terms of proportionality than political fragmentation,
but PR systems maintain higher values of the effective number of parties most of the time.

If the stability of elections to the Poland's Sejm was the main concern of policymakers,
based on our analysis we can say, that an electoral reform changing the system to plurality
voting would be detrimental. Conversely, decreasing the number of districts under the
current electoral system would marginally improve the robustness against external influences,
however the effect is so small that most likely it would not justify a reform.
The current electoral system, therefore, is a prudent choice in the context of vulnerabilities.


\subsection{Senate of the Republic of Poland}

\begin{figure}
\centering
\begin{subfigure}{\textwidth}
\centering
    \includegraphics[scale=0.31]{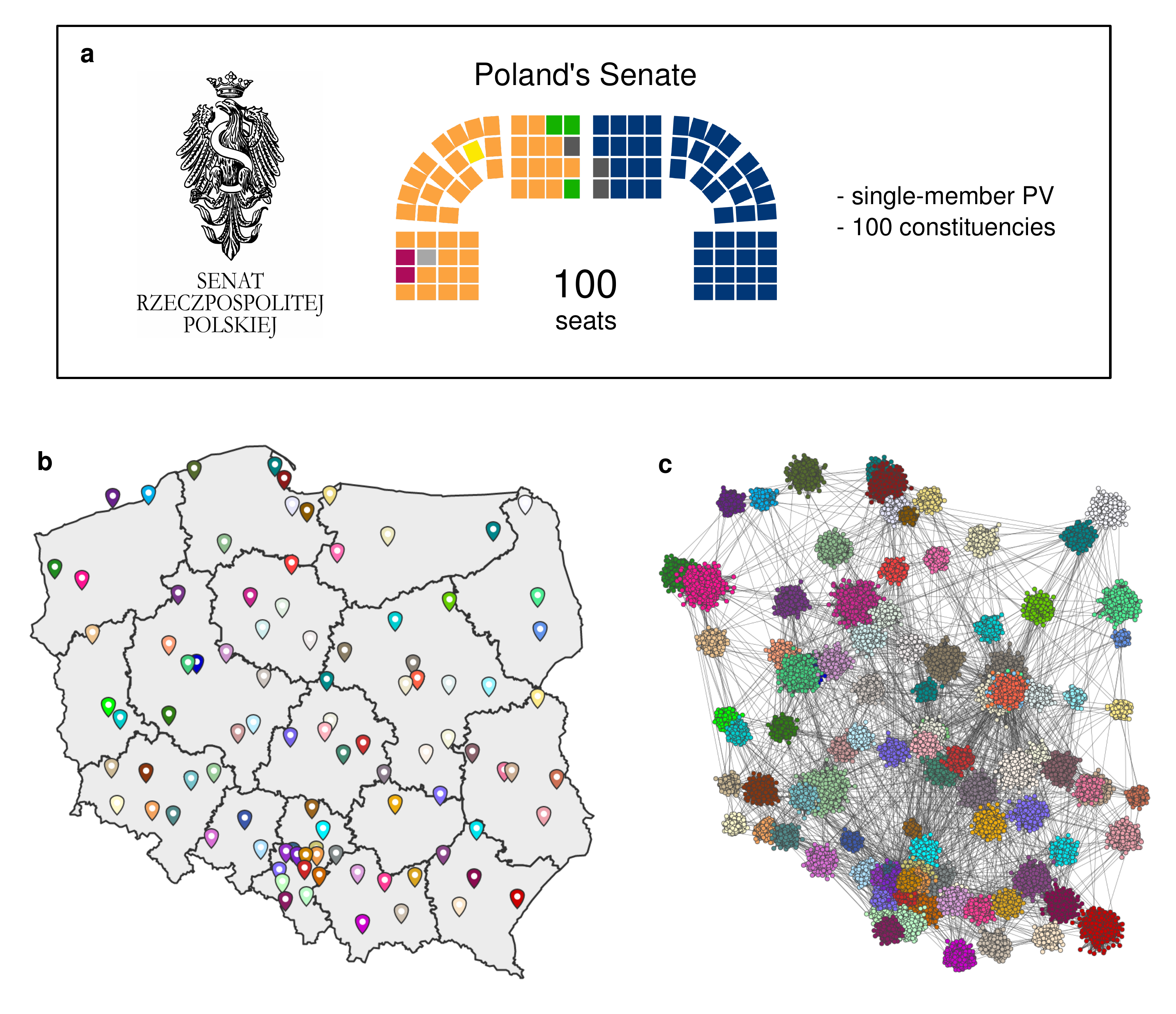}
\end{subfigure}
\begin{subfigure}{\textwidth}
\centering
    \includegraphics[scale=0.65]{figures/pl_sejm/diagram/pl_planar_c.pdf}
\end{subfigure}
\caption[Simulation setup for the Poland's Senate]{\textbf{Simulation setup for the Poland's Senate.}
(a) Logo of the Senate of the Republic of Poland and basic information about the electoral
system used in electing its members.
(b) Map of Poland with division into 16 voivodeships, i.e. the biggest administrative districts of the country.
The capital of each of 100 electoral districts applied in elections to
the Senate is indicated with a colored marker.
(c) Network of $5 \times 10^4$ nodes generated for simulations of elections to the Poland's Senate.
Nodes in different constituencies have different colors (corresponding to the marker on the map)
and are centered around the capital of the constituency.
The bigger the population of a district, the more spread are the nodes.
(d) Probability distribution of home-work commuting distance for Poland. The empirical commuting data is taken from
\cite{statOszacowanieOdlegoci,statPrzepywyLudnoci} (see the supplementary datasets) and the optimal distribution
is found by fitting the affinity function [\ref{eqn:planar_affinity}], with resulting
planar constant $c=1.7237$.
Connections in the network are generated based on the optimal affinity function.
[logo in (a) by Senate RP, senat.gov.pl, Public Domain, \href{https://commons.wikimedia.org/w/index.php?curid=84036616}{https://commons.wikimedia.org/w/index.php?curid=84036616}, seats layout in (a) by Szczeszek2035 - Own work, \href{https://creativecommons.org/licenses/by-sa/4.0/}{CC BY-SA 4.0}, \href{https://commons.wikimedia.org/w/index.php?curid=100921879}{https://commons.wikimedia.org/w/index.php?curid=100921879}]
}
\label{fig:pl_senate-basic_info}
\end{figure}

The second electoral system we investigate is the one employed in elections
to the Poland's Senate, the upper house of the 
bicameral parliament of Poland. It is a plurality voting system with 100 single-member districts,
i.e. first-past-the-post voting.
Each party can put forward only one candidate per constituency (independent candidates can 
also participate, each candidacy has to be supported by at least 2 thousand voters' signatures).
The candidate with the biggest number of votes wins the seat, and if there is a draw
the number of electoral committees won is decisive (if the draw still exists, the winner
is chosen randomly).
See Figure~\ref{fig:pl_senate-basic_info} for a summary of the Senate's electoral system.
In the 2019 elections, approximately 
30 million Poles were eligible to vote, with an average of $297\pm68$ thousand voters 
per district. There were 3 candidates registered in each district on average \cite{pkwWyborySejmu}.

For the simulation of elections to the Poland's Senate, a network is constructed consisting
of $5 \times 10^4$ nodes divided into 100 topological communities, analogous to real
electoral districts. The relative sizes of these communities are based on the actual 
populations of voters in the districts. Each community is assigned a geographical location,
corresponding to the coordinates of the district's capital
(see panels~(b) and~(c) of Figure~\ref{fig:pl_senate-basic_info}). The connectivity between nodes 
decreases with the geographical distance between them, as determined by fitting 
[\ref{eqn:planar_affinity}] to empirical data on work commuting distance 
\cite{statOszacowanieOdlegoci,statPrzepywyLudnoci}, as shown in
Figure~\ref{fig:pl_senate-basic_info}~(d). The simulations are conducted for a scenario 
involving $n=3$ parties, an average degree of $k=12$, and various noise rates $\varepsilon$.
The supplementary datasets and the 
configuration files in our online code repository \cite{github_repo} contain the exact 
data used in the simulations.

In addition to the actual electoral system employed in elections to the Poland's Senate 
(labeled as ,,100 dist. PV*'' in the figures),
we also explore five alternative systems
that are either applicable to Poland or hold theoretical significance. The first alternative
system (,,40 dist. PR'') is a PR system with a division into 40 multi-member constituencies,
with the number of seats varying between 2 and 4 and the Jefferson method of seat assignment.
This is the electoral system that was used in elections to the Poland's Senate before
the 2011 electoral reform changing it to FPTP. We merge all single-member districts
of the current system within the same before-the-reform constituency,
accumulating the seats into a single pool, identical to the one existing before 2011
\cite{sejmUstawaDnia}. The second alternative system (,,40 dist. PV'') uses the same division into
40 constituencies, but applies PV where the party with the most votes in a district
secures all the seats (effectively it is a PBV system). 
We then investigate two alternative systems based on the 16 voivodeships (administrative 
districts) of Poland, where all original electoral districts within a voivodeship are merged into 
a single district, consolidating the seats within each voivodeship. Under this new division,
we examine a PR system (,,16 dist. PR''), as well as a PV system 
(,,16 dist. PV'') where the winner in each district claims all the seats. Lastly, we analyze
a PR system in which all 100 seats are assigned within a single nation-wide district
(,,1 dist. PR''). The alternative systems do not include an entry threshold (as the real one),
and the PR alternative systems use the Jefferson seat assignment method, which was used before the reform.

\begin{figure}
\centering
\begin{subfigure}{0.665\textwidth}
\centering
    \includegraphics[scale=0.67]{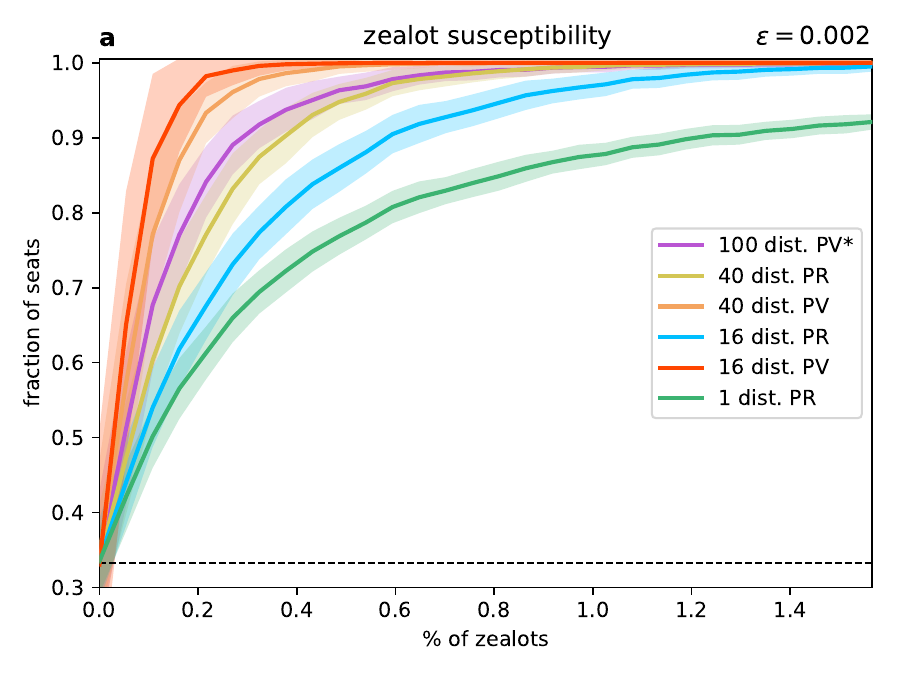}
\end{subfigure}\hfill
\begin{subfigure}{0.335\textwidth}
    \includegraphics[scale=0.65]{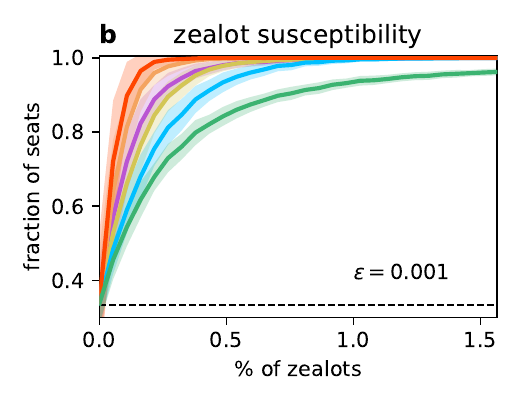}
    \includegraphics[scale=0.65]{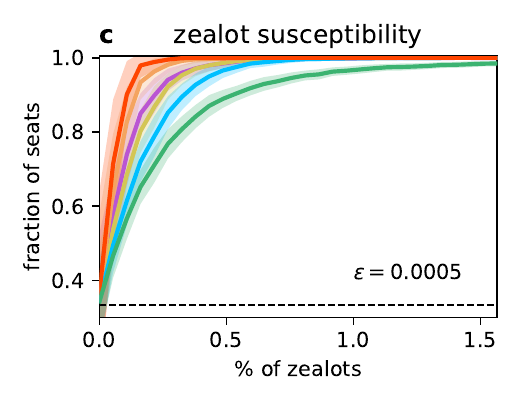}
\end{subfigure}
\begin{subfigure}{\textwidth}
    \includegraphics[scale=0.65]{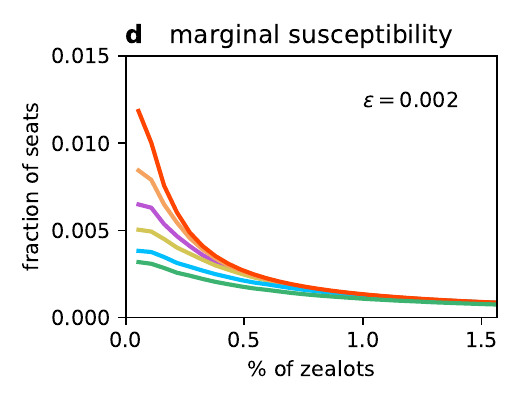}
    \includegraphics[scale=0.65]{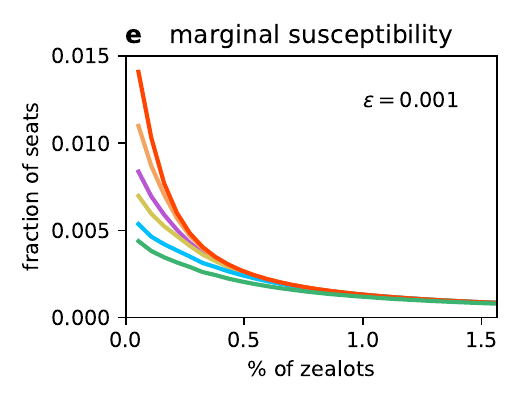}
    \includegraphics[scale=0.65]{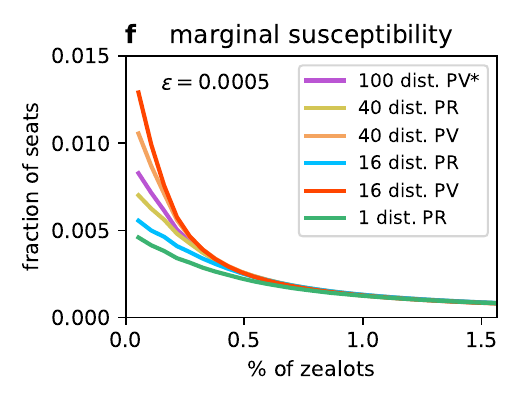}
\end{subfigure}
\begin{subfigure}{\textwidth}
    \includegraphics[scale=0.65]{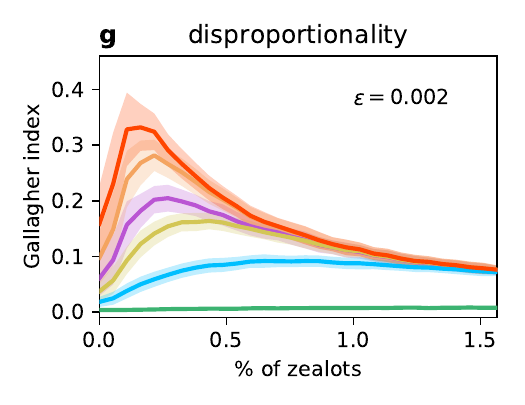}
    \includegraphics[scale=0.65]{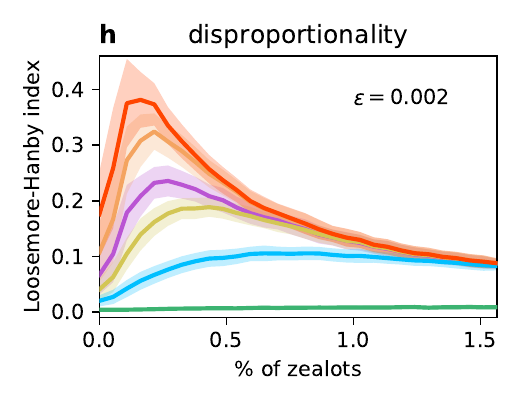}
    \includegraphics[scale=0.65]{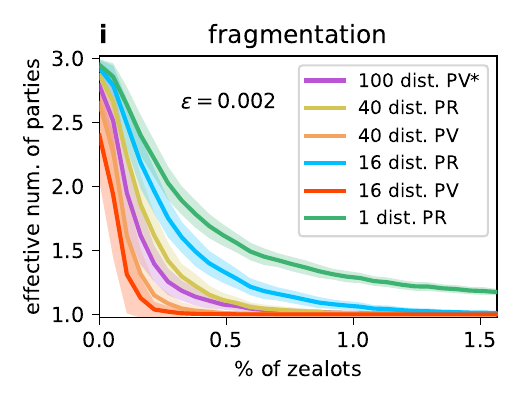}
\end{subfigure}
\caption[Zealot susceptibility of elections to the Poland's Senate]{\textbf{Zealot susceptibility of elections to the Poland's Senate.}
All plots compare results for the actual electoral system used in elections to the Senate
(marked with an asterisk) with 5 alternative systems with different numbers of districts.
See the main text for details on ES used.
(a, b, c)  The average result of the zealot party $a$ vs. percentage of zealots in the population.
The dashed line marks the average
fraction of seats obtained in the symmetrical case, i.e. 1/3.
(d, e, f) Marginal zealot susceptibility  obtained
by dividing the total change in the fraction of seats won by the number of zealots.
(g, h, i) Average values of disproportionality and political fragmentation indexes
vs. percentage of zealots in the population.
The colored shade indicates results within
one standard deviation from the average.
All results were obtained over
200 elections with 3 parties. The network has $5 \times 10^4$ nodes, with the average degree
$k=12$, divided into 100 communities corresponding to the actual electoral districts
(see Figure~\ref{fig:pl_senate-basic_info}).
There is 100 seats to gain in total.
The noise rate $\varepsilon$ is given in each panel.
}
\label{fig:pl_senate-zealot_sus}
\end{figure}

In Figure~\ref{fig:pl_senate-zealot_sus}, we present an analysis of zealot susceptibility across all
ES considered for the Senate. Panels (a) to (c) depict the average fraction of seats 
won by the zealot party in relation to the percentage of 
zealots in the population. Similarly to the Sejm, the divergence in zealot susceptibility between PR 
and PV systems is clear. In case of the Senate, however, the three PR systems exhibit 
distinct responses to zealots. The system with one constituency is the least susceptible to
zealots, then 16 constituencies, and for 40 constituencies it in the most susceptible.
The differences in the average seat share are significantly bigger than one standard deviation.
This discrepancy between the Sejm and the Senate simulations exists due to the difference in the total
number of seats, 460 and 100 respectively. The average district magnitude for the Sejm, for example for the 
division into 41 constituencies, is much higher than in the similar division for the Senate
or even that for 16 districts in the Senate.
The differences between zealot susceptibility of the PV systems are smaller, but
the actual electoral system of the Senate (,,100 dist. PV*'') is markedly the least susceptible.
Examining the marginal susceptibility in panels
(d) to (f), we observe that investing in zealots yields a higher profit 
in PV systems, particularly at the initial stages where the influence of each zealot can be
almost three times greater than in PR systems.
This observation holds true for any noise rate $\varepsilon$, albeit with higher noise 
levels leading to reduced zealot influence overall. 

In panels (g) to (i) of Figure~\ref{fig:pl_senate-zealot_sus}, we present the Gallagher index, the 
Loosemore-Hanby index, and the effective number of parties. Again, under PV systems, elections 
can become highly disproportional even with a minuscule percentage of zealots in the population.
The maximum level of 
disproportionality in PV systems is reached when the zealot percentage falls below 0.2\% for 
$\varepsilon=0.002$. As expected, an increase in the number of zealots leads to 
a decrease in political fragmentation, with PV systems generally exhibiting lower levels of 
fragmentation.
The most notable result, from the applied point of view, is the very similar scoring 
in all measures presented in Figure~\ref{fig:pl_senate-zealot_sus} for the actual
system of the Poland's Senate (,,100 dist. PV*'') and the system used before
the 2011 reform (,,40 dist. PR''). Zealot susceptibility, disproportionality, and
fragmentation of these two systems are mostly within a single standard deviation from each other.
This is partially because the old system has low district magnitude on average --
in Figure~\ref{fig:hist-box_seats_per_dist} we could see that the lower the magnitude
the more similar PR and FPTP are.

\begin{figure}
\centering
\begin{subfigure}{0.665\textwidth}
\centering
    \includegraphics[scale=0.67]{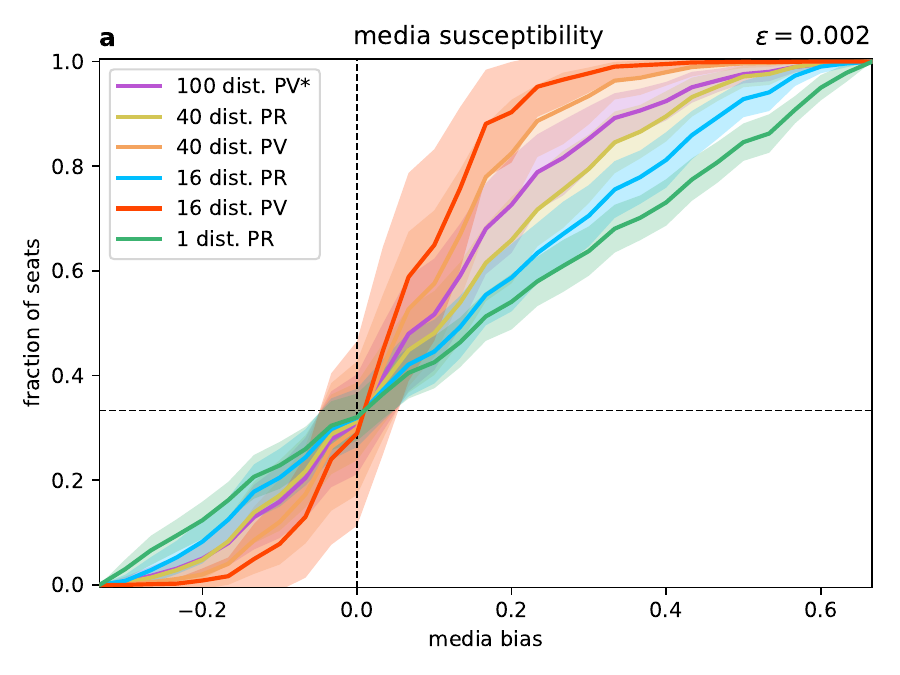}
\end{subfigure}\hfill
\begin{subfigure}{0.335\textwidth}
    \includegraphics[scale=0.65]{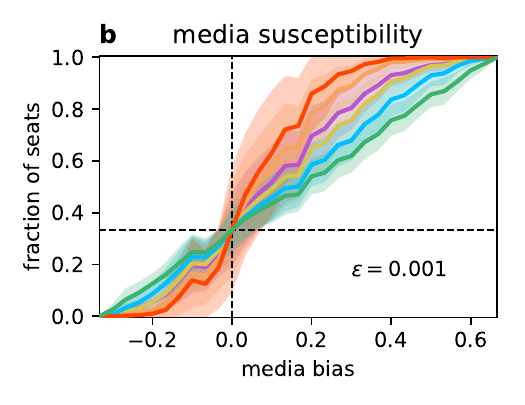}
    \includegraphics[scale=0.65]{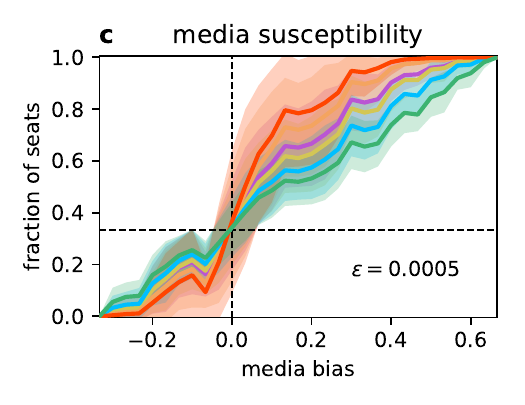}
\end{subfigure}
\begin{subfigure}{\textwidth}
    \includegraphics[scale=0.65]{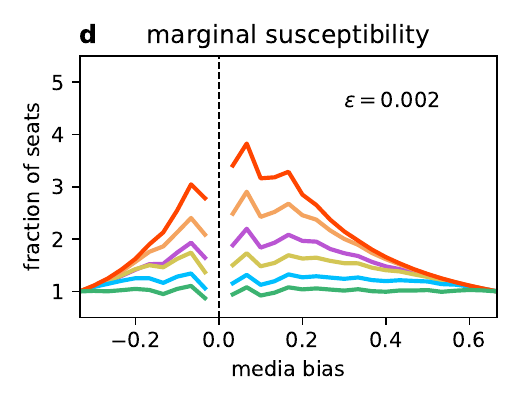}
    \includegraphics[scale=0.65]{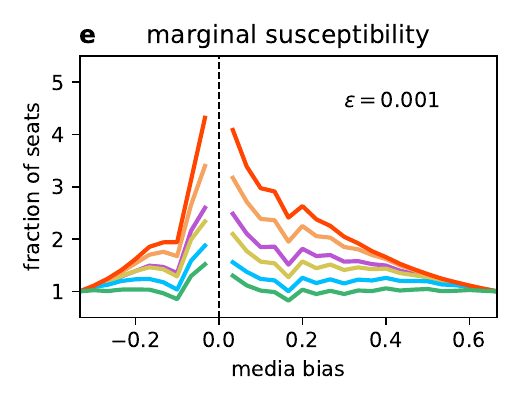}
    \includegraphics[scale=0.65]{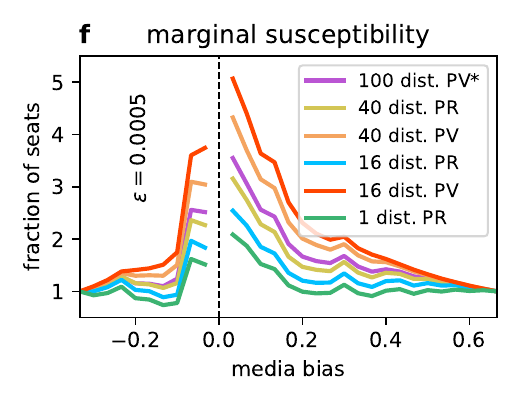}
\end{subfigure}
\begin{subfigure}{\textwidth}
    \includegraphics[scale=0.65]{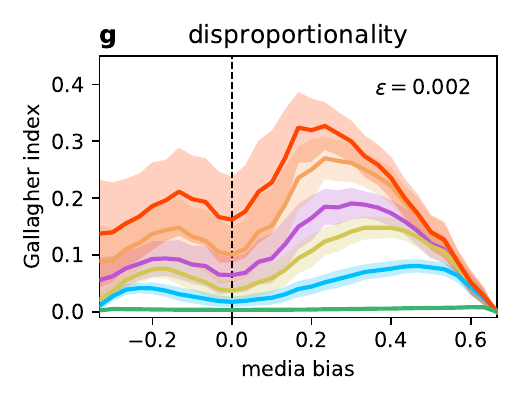}
    \includegraphics[scale=0.65]{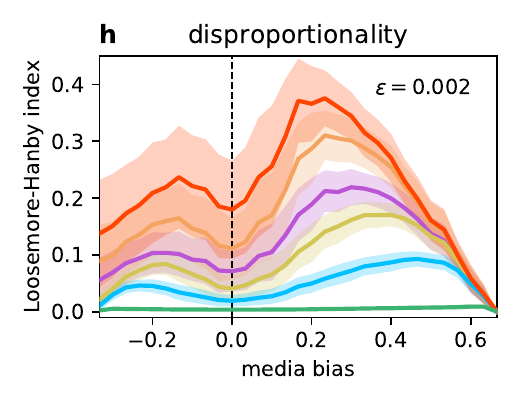}
    \includegraphics[scale=0.65]{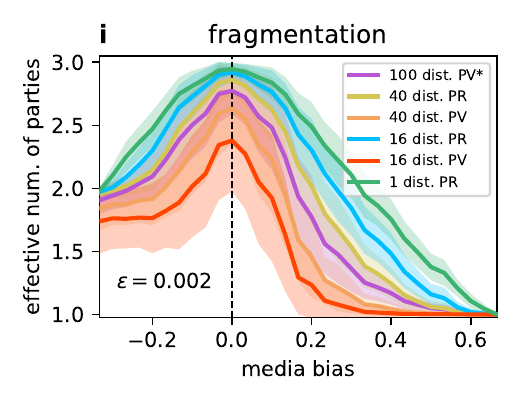}
\end{subfigure}
\caption[Media susceptibility of elections to the Poland's Senate]{\textbf{Media susceptibility of elections to the Poland's Senate.}
All plots compare results for the actual electoral system used in elections to the Senate
(marked with an asterisk) with 5 alternative systems with different numbers of districts.
See the main text for details on ES used.
(a, b, c)  The average result of the media party $a$ vs. media bias.
The horizontal dashed line marks the average
fraction of seats obtained in the symmetrical case, i.e. 1/3, and the vertical one the case with no bias.
(d, e, f) Marginal media susceptibility  obtained
by dividing the total change in the fraction of seats won by the strength of bias.
(g, h, i) Average values of disproportionality and political fragmentation indexes
vs. media bias.
The colored shade indicates results within
one standard deviation from the average.
All results were obtained over
1000 elections with 3 parties. The network has $5 \times 10^4$ nodes, with the average degree
$k=12$, divided into 100 communities corresponding to the actual electoral districts
(see Figure~\ref{fig:pl_senate-basic_info}).
There is 100 seats to gain in total.
The noise rate $\varepsilon$ is given in each panel.
}
\label{fig:pl_senate-media_sus}
\end{figure}

Figure~\ref{fig:pl_senate-media_sus} shows the second external influence, i.e
the susceptibility to media propaganda as a function of bias magnitude. Similar to zealots, PV systems 
tend to be more susceptible to biased media compared to PR systems. Surprisingly,
PR systems exhibit more comparable levels of media susceptibility within one
standard deviation of each other. In panels (a) to (c), we can see that lower noise rates 
result in a slightly reduced susceptibility for PV systems, and the general pattern being more irregular.
This observation is also confirmed by the marginal susceptibility displayed in panels (d) to (f). 
In contrast to the Sejm, media bias yields the highest returns at the  very beginning, as observed 
with zealots.

Analogous to zealots, media propaganda causes the highest level of disproportionality for 
intermediate levels of influence, as depicted in panels (g) to (i) of
Figure~\ref{fig:pl_senate-media_sus}. However, disproportionality raises earlier in PV systems compared 
to PR systems. The discrepancy between the two types of ES is more pronounced in
terms of proportionality rather than political fragmentation, although PR systems generally 
maintain higher values of the effective number of parties throughout most of the analysis.
Once again, the similarity between the actual and the old electoral system of the Senate
is striking.

Given the above results, we can say that the 2011 electoral reform changing the electoral system
of the Poland's Senate did not significantly increase the vulnerability of electoral processes.
In the context of stability of elections the shift to plurality voting did not
impair democratic governance. To meaningfully improve the robustness against external influences
the PR system would need to involve much fewer districts or much more seats.


\subsection{Knesset of the State of Israel}

\begin{figure}
\centering
\begin{subfigure}{\textwidth}
\centering
    \includegraphics[scale=0.31]{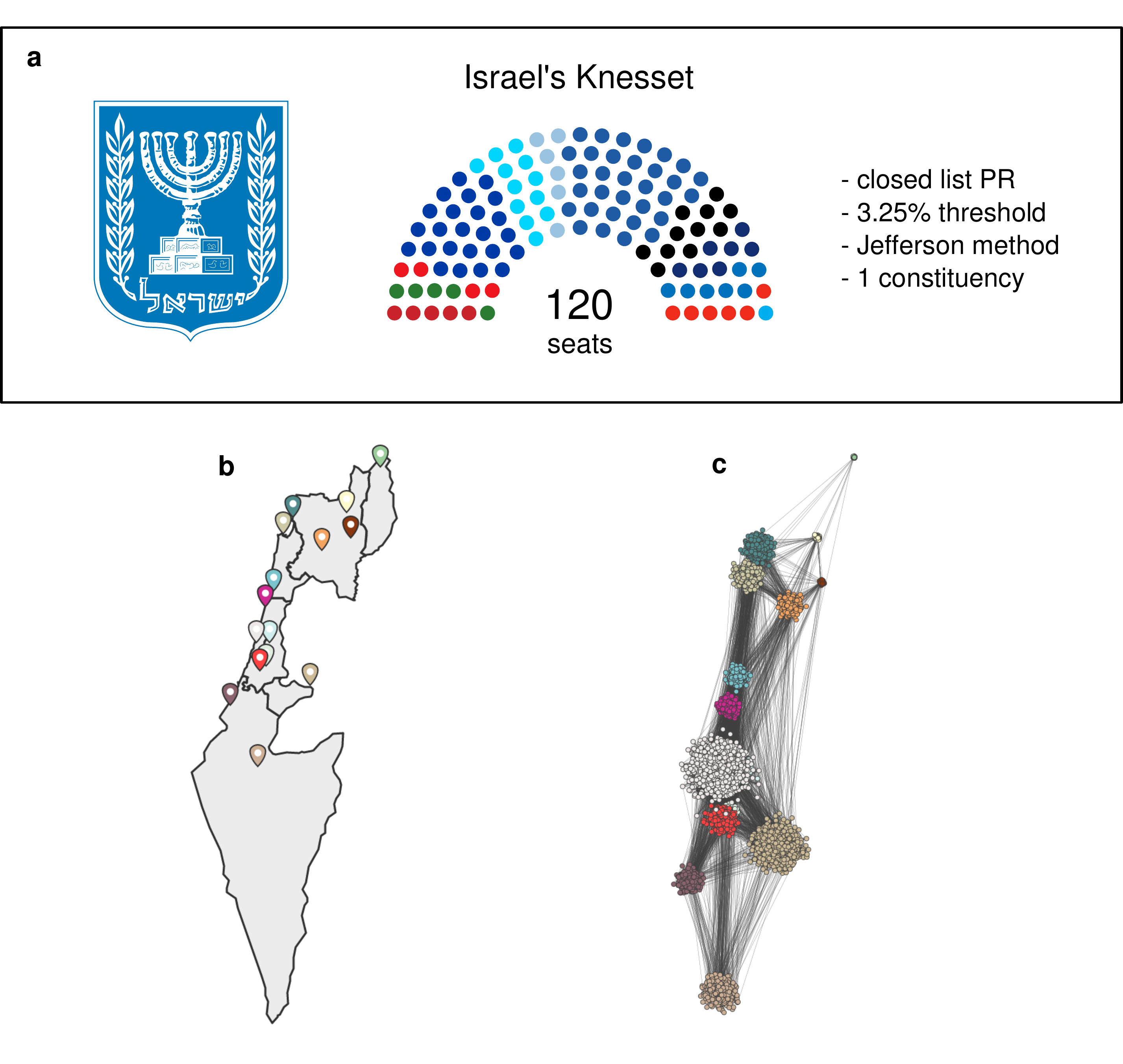}
\end{subfigure}
\begin{subfigure}{\textwidth}
\centering
    \includegraphics[scale=0.65]{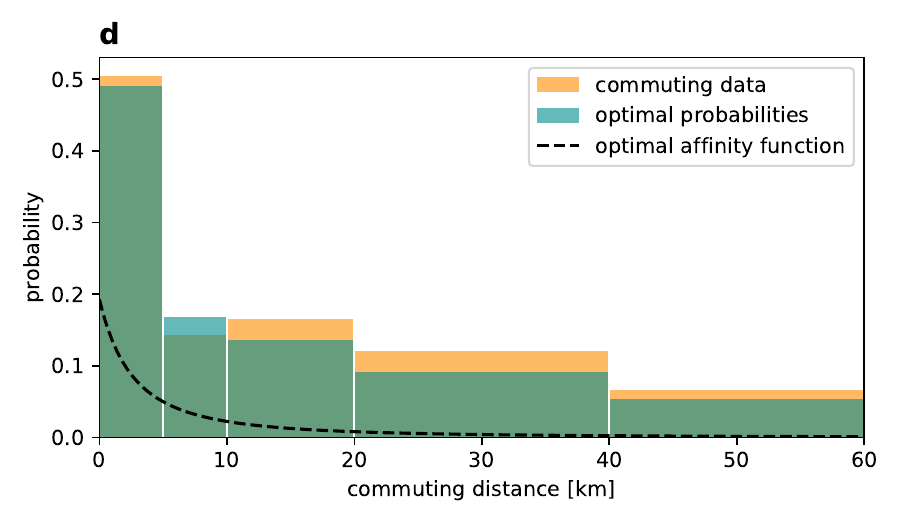}
\end{subfigure}
\caption[Simulation setup for the Israel's Knesset]{\textbf{Simulation setup for the Israel's Knesset.}
(a) Logo of the Knesset of the State of Israel and basic information about the electoral
system used in electing its members.
(b) Map of Israel with division into 6 main administrative districts of the country.
They are further divided into 15 administrative sub-districts and their capitals are indicated with a colored 
marker.
(c) Network of $9 \times 10^3$ nodes generated for simulations of elections to the Israel's Knesset.
Nodes in different administrative sub-districts have different colors (corresponding to the marker on the map)
and are centered around the capital of the sub-district.
The bigger the population of a sub-district, the more spread are the nodes.
(d) Probability distribution of home-work commuting distance for Israel. The empirical commuting data is taken from
\cite{bleikh2018back} (see the supplementary datasets) and the optimal distribution
is found by fitting the affinity function [\ref{eqn:planar_affinity}], with resulting
planar constant $c=5.1874$.
Connections in the network are generated based on the optimal affinity function.
[logo in (a) by Max and Gabriel Shamir, Public Domain, \href{https://commons.wikimedia.org/w/index.php?curid=3498966}{https://commons.wikimedia.org/w/index.php?curid=3498966}, seats layout in (a) by Parliament diagram tool, Public Domain, \href{https://commons.wikimedia.org/w/index.php?curid=125744500}{https://commons.wikimedia.org/w/index.php?curid=125744500}]
}
\label{fig:il_knesset-basic_info}
\end{figure}

The third real-world electoral system that we study is the one used in elections to the Knesset, 
which is the unicameral legislature of Israel. The system is a party-list proportional 
representation with closed lists. In this system, voters select a preferred party
list, influencing the number of seats the party 
will obtain but not which candidates from the list will enter the parliament. The Knesset comprises 
a total of 120 seats, which are assigned within a single nation-wide constituency. Parties and electoral alliances
that receive less than 3.25\% of the overall vote are excluded and do not enter the legislature. 
The seats are allocated to the parties passing this threshold using the Jefferson seat assignment 
method, also known as the D'Hondt method. See Figure~\ref{fig:il_knesset-basic_info} for
a summary of the Knesset’s electoral system. All Israeli citizens over 18 years old may vote
in legislative elections. In 2018 there was approximately 9 million citizens of Israel
with an average of $589 \pm 379$ thousands of people per administrative sub-district \cite{StatisticalIsrael}.
In 2020 elections there were 8 party lists registered (without minority parties),
with other recent elections even surpassing this number.

For the simulation of elections to the Israel's Knesset, we create a network with $9 \times 10^3$ 
nodes, divided into 15 topological communities using the spatial version of the SBM. These
communities correspond to the administrative sub-districts of Israel,
and their relative sizes are based on the actual 
populations in the sub-districts. Each community is assigned a geographical location given
by the coordinates of the sub-district's capital 
(see panels (b) and (c) of Figure~\ref{fig:il_knesset-basic_info}).
The connectivity between nodes decreases with geographical distance, as determined by empirical data 
on work commuting distance \cite{bleikh2018back}, what is demonstrated in
Figure~\ref{fig:il_knesset-basic_info}~(d).
We include all relevant details of the electoral system in the simulation,
including the 3.25\% entry threshold and the Jefferson seat assignment method. 
We run the simulations for $n=8$ parties,
the average degree of $k = 12$,
and various noise rates $\varepsilon$. For precise data used in the simulations, refer to
the supplementary datasets or the configuration files available in our online code repository
\cite{github_repo}.

In addition to the actual electoral system employed in elections to the Israel's Knesset
(labeled as "1 dist. PR*" in the figures), we also explore four alternative systems that are either applicable to
Israel or hold theoretical significance. The first two alternative system, ,,6 dist. PR''
and ,,15 dist. PR'', are PR systems with a division into 6 or 15 multi-member constituencies
corresponding to Israel's administrative districts and sub-districts. Both systems
have the same 3.25\% entry threshold as in the actual system and use the Jefferson seat assignment 
method. The other two alternative system, ,,6 dist. PV'' and ,,15 dist. PV'', use the same division into
constituencies but apply the winner-takes-all rule, where the party with the most
votes in a district secures all the seats (effectively being a PBV).
Remarkably, dividing the country into several or more electoral districts was considered in Israel
and the discussion on its advantages and disadvantages has a long history \cite{idiDistrictElections}.

\begin{figure}
\centering
\begin{subfigure}{0.66\textwidth}
\centering
    \includegraphics[scale=0.67]{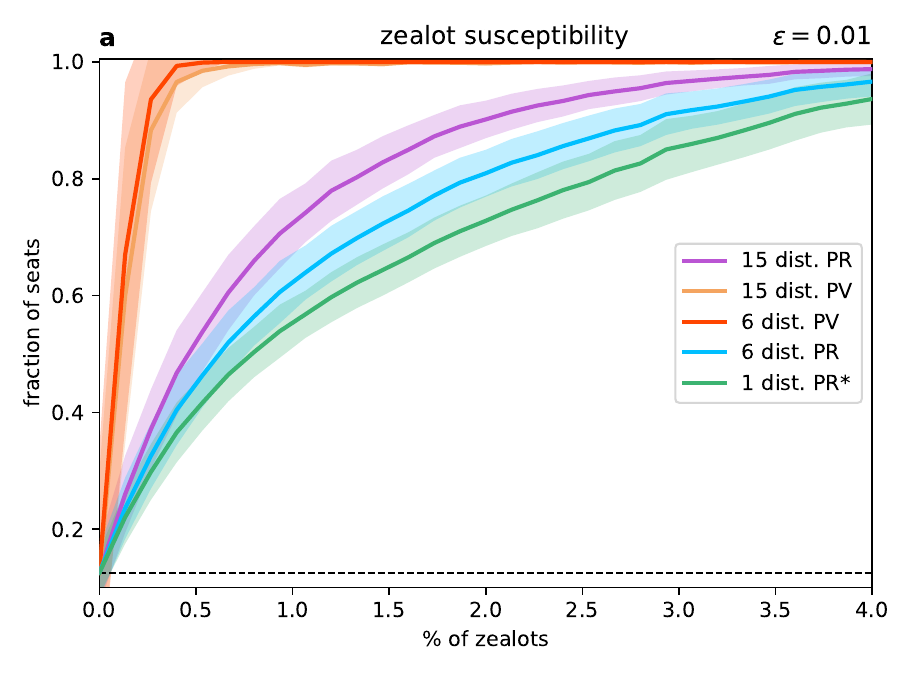}
\end{subfigure}\hfill
\begin{subfigure}{0.34\textwidth}
    \includegraphics[scale=0.65]{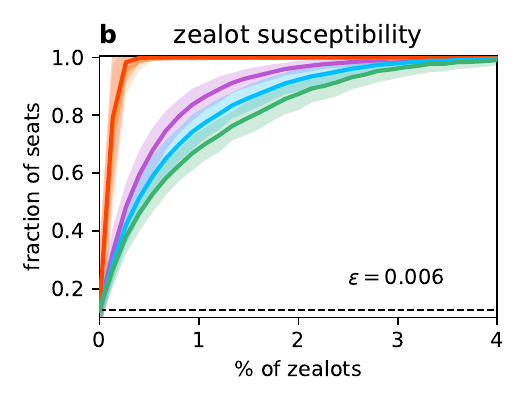}
    \includegraphics[scale=0.65]{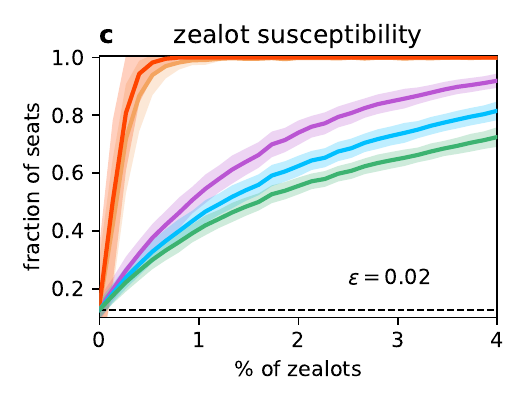}
\end{subfigure}
\begin{subfigure}{\textwidth}
    \includegraphics[scale=0.65]{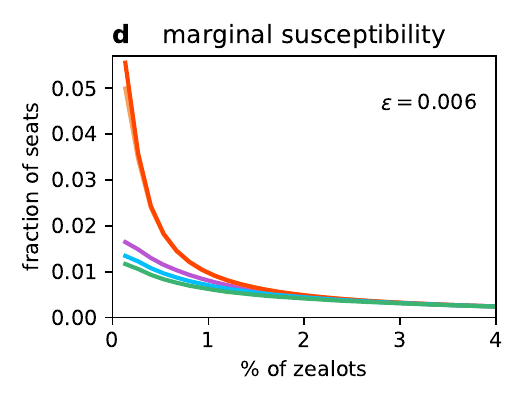}
    \includegraphics[scale=0.65]{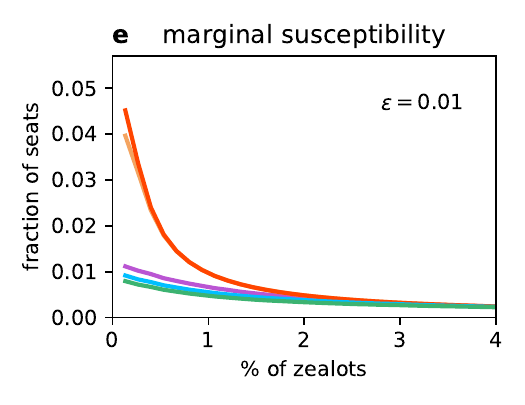}
    \includegraphics[scale=0.65]{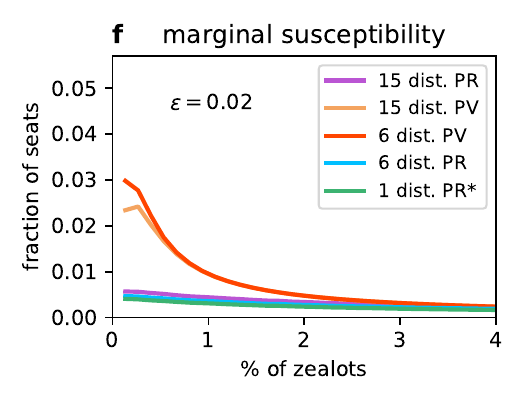}
\end{subfigure}
\begin{subfigure}{\textwidth}
    \includegraphics[scale=0.65]{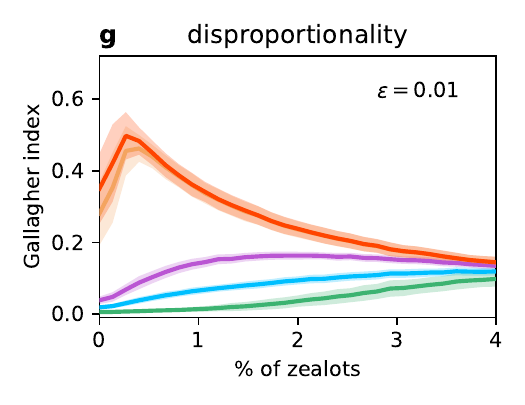}
    \includegraphics[scale=0.65]{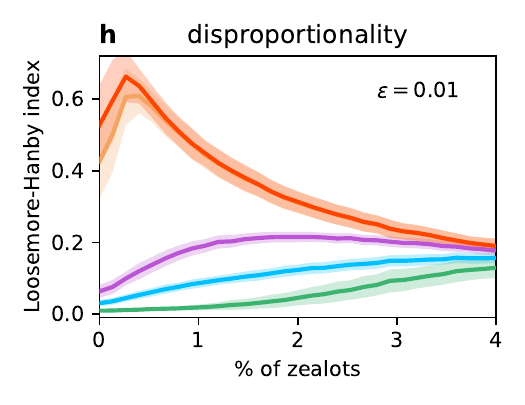}
    \includegraphics[scale=0.65]{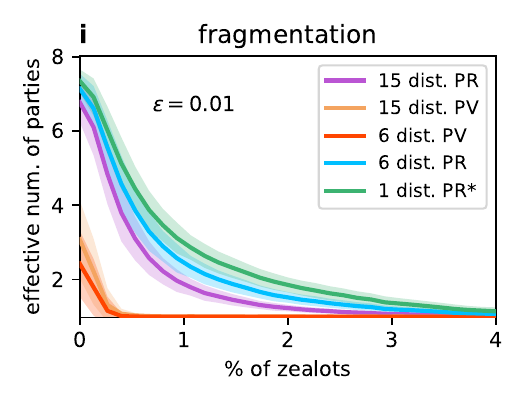}
\end{subfigure}
\caption[Zealot susceptibility of elections to the Israel's Knesset]{\textbf{Zealot susceptibility of elections to the Israel's Knesset.}
All plots compare results for the actual electoral system used in elections to the Knesset
(marked with an asterisk) with 4 alternative systems with different numbers of districts.
See the main text for details on ES used.
(a, b, c)  The average result of the zealot party $a$ vs. percentage of zealots in the population.
The dashed line marks the average
fraction of seats obtained in the symmetrical case, i.e. 1/8.
(d, e, f) Marginal zealot susceptibility  obtained
by dividing the total change in the fraction of seats won by the number of zealots.
(g, h, i) Average values of disproportionality and political fragmentation indexes
vs. percentage of zealots in the population.
The colored shade indicates results within
one standard deviation from the average.
All results were obtained over
400 elections with 8 parties. The network has $9 \times 10^3$ nodes, with the average degree
$k=12$, divided into 15 communities corresponding to administrative sub-districts
(see Figure~\ref{fig:il_knesset-basic_info}).
There is 120 seats to gain in total.
The noise rate $\varepsilon$ is given in each panel.
}
\label{fig:il_knesset-zealot_sus}
\end{figure}

In Figure~\ref{fig:il_knesset-zealot_sus}, we present an analysis of susceptibility to zealots 
for the original and all alternative ES considered for the Knesset. Panels (a) to (c) depict the average 
fraction of seats won by the zealot party as a function of the percentage of zealots in the 
population. The discrepancy in zealot susceptibility between 
PR and PV systems is considerable, as in the case of the Poland's Sejm. All three PR systems exhibit 
relatively similar responses to zealots and are considerably less susceptible than both PV 
systems. This observation holds true for any value of the noise rate $\varepsilon$, although 
zealots have diminished overall influence in the presence of higher noise levels.
Also, the difference between PR systems' susceptibility is greater for higher noise.
Examining the marginal susceptibility in panels (d) to (f), we observe that introducing zealots
is highly more profitable in PV systems, particularly in the initial stages, where 
the influence of each zealot can be up to five times greater than in PR.

In panels (g) to (i) of Figure~\ref{fig:il_knesset-zealot_sus}, we demonstrate the Gallagher index, 
the Loosemore-Hanby index, and the effective number of parties. Again, under PV systems, 
elections can become highly disproportional even with a very small percentage of zealots in the 
population. The maximum 
level of disproportionality in PV systems is reached with zealot percentages around 0.3\% for 
$\varepsilon=0.01$, whereas in PR systems, disproportionality peaks much later, when the zealot party nears 
complete domination. Political fragmentation decreases with an increasing number of 
zealots, with PV systems generally displaying much lower levels of fragmentation.

\begin{figure}
\centering
\begin{subfigure}{0.66\textwidth}
\centering
    \includegraphics[scale=0.67]{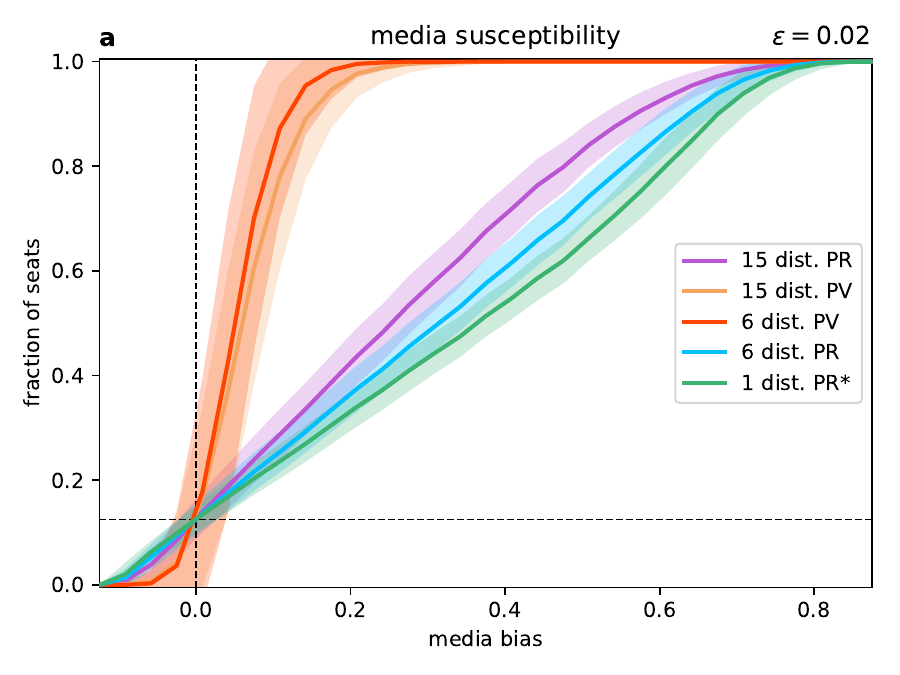}
\end{subfigure}\hfill
\begin{subfigure}{0.34\textwidth}
    \includegraphics[scale=0.65]{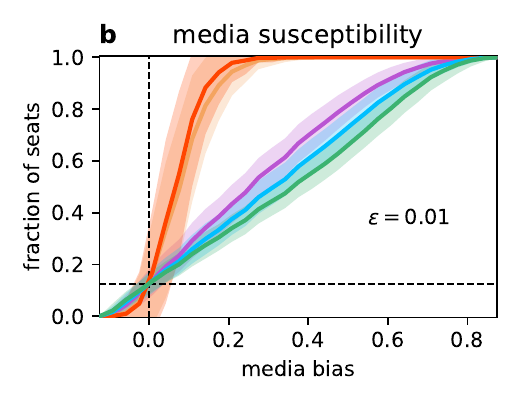}
    \includegraphics[scale=0.65]{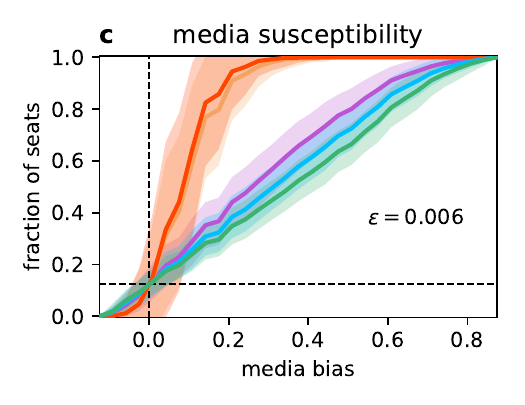}
\end{subfigure}
\begin{subfigure}{\textwidth}
    \includegraphics[scale=0.65]{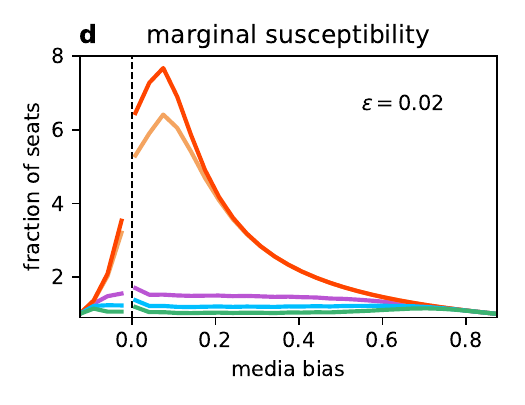}
    \includegraphics[scale=0.65]{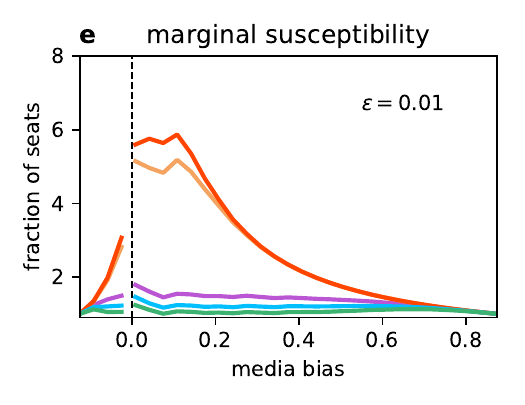}
    \includegraphics[scale=0.65]{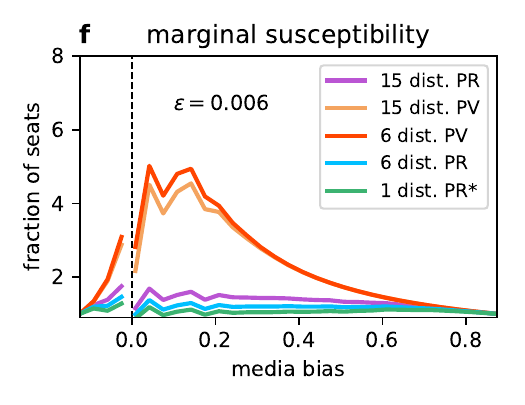}
\end{subfigure}
\begin{subfigure}{\textwidth}
    \includegraphics[scale=0.65]{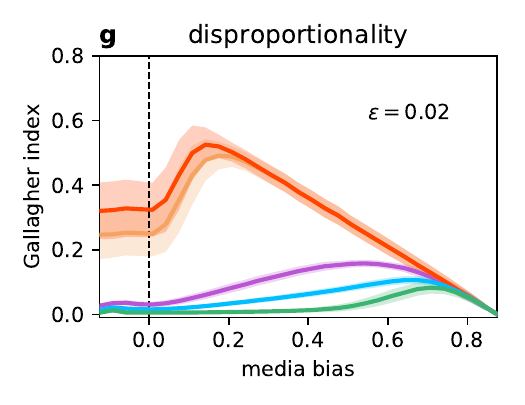}
    \includegraphics[scale=0.65]{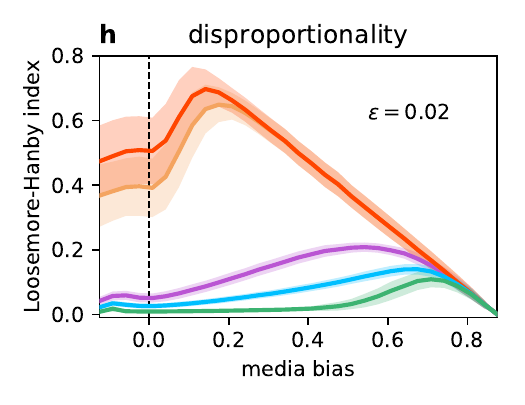}
    \includegraphics[scale=0.65]{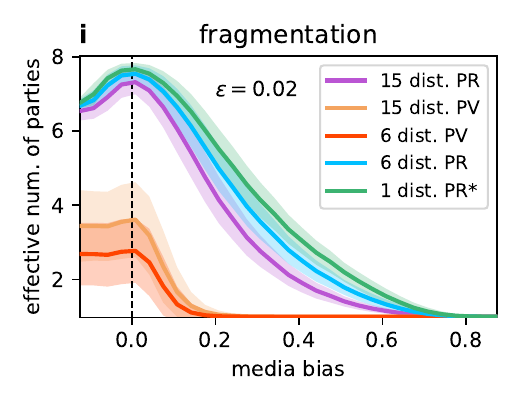}
\end{subfigure}
\caption[Media susceptibility of elections to the Israel's Knesset]{\textbf{Media susceptibility of elections to the Israel's Knesset.}
All plots compare results for the actual electoral system used in elections to the Knesset
(marked with an asterisk) with 4 alternative systems with different numbers of districts.
See the main text for details on ES used.
(a, b, c)  The average result of the media party $a$ vs. media bias.
The horizontal dashed line marks the average
fraction of seats obtained in the symmetrical case, i.e. 1/8, and the vertical one the case with no bias.
(d, e, f) Marginal media susceptibility  obtained
by dividing the total change in the fraction of seats won by the strength of bias.
(g, h, i) Average values of disproportionality and political fragmentation indexes
vs. media bias.
The colored shade indicates results within
one standard deviation from the average.
All results were obtained over
2000 elections with 8 parties. The network has $9 \times 10^3$ nodes, with the average degree
$k=12$, divided into 15 communities corresponding to administrative sub-districts
(see Figure~\ref{fig:il_knesset-basic_info}).
There is 120 seats to gain in total.
The noise rate $\varepsilon$ is given in each panel.
}
\label{fig:il_knesset-media_sus}
\end{figure}

Figure~\ref{fig:il_knesset-media_sus} illustrates the second external influence under investigation: 
susceptibility to media propaganda as a function of bias intensity. Similar to zealots, PV 
systems are highly more susceptible to biased media compared to PR systems.
Although the PR systems employed different district divisions, they display comparable levels 
of media susceptibility, except ,,15 dist. PR'' and ,,1 dist. PR*'', which are further than
one standard deviation apart. Panels (a) to (c) reveal that for lower 
noise rates $\varepsilon$ the general susceptibility 
pattern becomes more irregular. This observation is further confirmed by the marginal 
susceptibility depicted in panels (d) to (f), which also shows that ES exhibit reduced susceptibility
for higher noise. Notably, in PV systems, media bias yields the 
highest returns not at the onset but after exerting some influence, as evidenced by the rightward 
shift of the maximum marginal susceptibility from zero bias.

Media propaganda causes the highest disproportionality at intermediate levels 
of influence, as evident in panels (g) to (i) of Figure~\ref{fig:il_knesset-media_sus}. However, 
this disproportionality occurs much earlier in PV systems compared to PR systems. The 
disparity between the two types of ES is also strongly pronounced in terms of 
political fragmentation -- PV systems consistently maintain much lower values of the effective 
number of parties, even in the neutral case of no bias. This is caused by a relatively low number
of constituencies, what creates fewer possibilities to win seats under PBV. 

Our results add an important argument to the discussion about possible divisions
of the country into multiple electoral districts,
given that the stability of elections to the Israel's Knesset is a serious matter.
An electoral reform creating 6 constituencies corresponding to the administrative districts would
only slightly increase the susceptibility to external influences. On the other hand,
creating 15 constituencies corresponding to the administrative sub-districts would have
a considerable effect on the vulnerability of the electoral system. Therefore, if the policymakers
want to maintain other aspect of the current ES, division into a smaller number of districts
is advisable in the context of mitigating susceptibilities to zealots and media propaganda.


\subsection{House of the People of the Republic of India}

\begin{figure}
\centering
\begin{subfigure}{\textwidth}
\centering
    \includegraphics[scale=0.31]{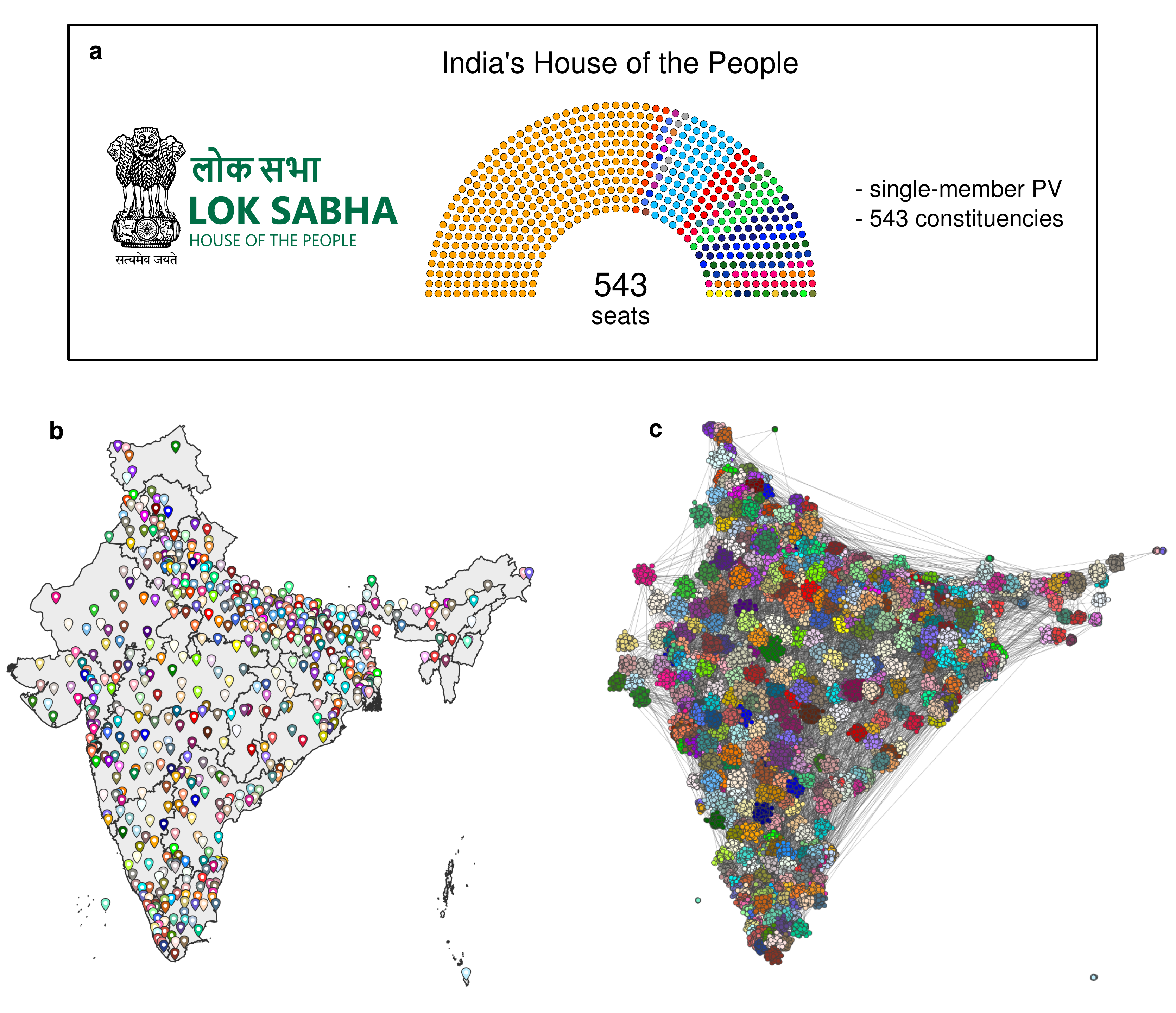}
\end{subfigure}
\begin{subfigure}{\textwidth}
\centering
    \includegraphics[scale=0.65]{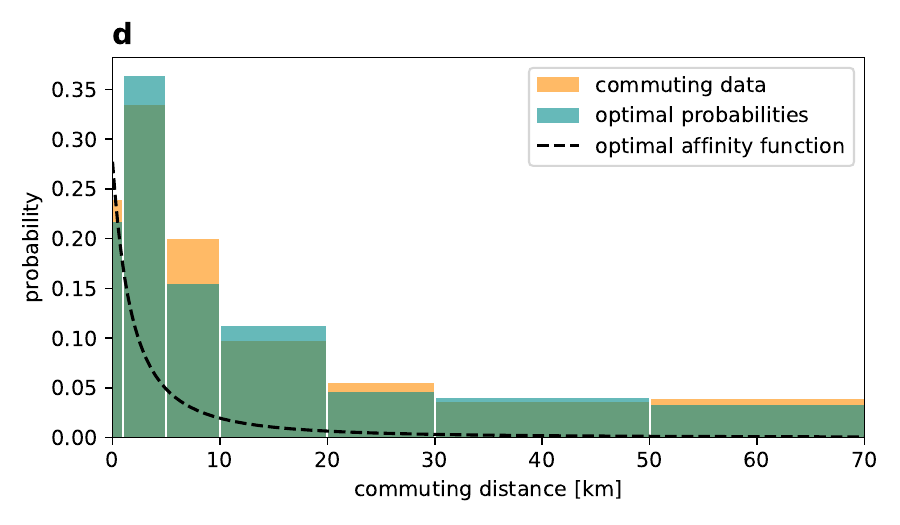}
\end{subfigure}
\caption[Simulation setup for the India's House of the People]{\textbf{Simulation setup for the India's House of the People.}
(a) Logo of the House of the People of the Republic of India and basic information about the electoral
system used in electing its members.
(b) Map of India with division into 36 administrative districts of the country
(28 states and 8 union territories).
The capital of each of 542 electoral districts applied in elections to
the House of the People is indicated with a colored marker (the district of Vellore is not included).
(c) Network of $10^5$ nodes generated for simulations of elections to the India's House of the People.
Nodes in different constituencies have different colors (corresponding to the marker on the map)
and are centered around the capital of the constituency.
The bigger the population of a district, the more spread are the nodes.
(d) Probability distribution of home-work commuting distance for India. The empirical commuting data is taken from
\cite{censusindiaHomeGovernment} (see the supplementary datasets) and the optimal distribution 
is found by fitting the affinity function [\ref{eqn:planar_affinity}], with resulting
planar constant $c=3.6068$.
Connections in the network are generated based on the optimal affinity function.
[logo in (a) by Swapnil1101, Public Domain, \href{https://commons.wikimedia.org/w/index.php?curid=116559677}{https://commons.wikimedia.org/w/index.php?curid=116559677}, seats layout in (a) by Lok Sabha - Own work, \href{https://creativecommons.org/licenses/by-sa/4.0/}{CC BY-SA 4.0}, \href{https://commons.wikimedia.org/w/index.php?curid=121501360}{https://commons.wikimedia.org/w/index.php?curid=121501360}]
}
\label{fig:ind_house_otp-basic_info}
\end{figure}

The electoral system used in elections to the India's House of the People is the fourth one we investigate.
The House of the People is the lower house of the bicameral parliament of India and is also known
as Lok Sabha. The system employed to select its members is a plurality voting system with 543 
single-member districts, where the candidate with the biggest number of votes wins the seat
(first-past-the-post voting). The constitution of India allows for
a maximum of 550 elected members in the House, with 530 members representing the states and
20 representing the union territories. Additionally, the president can appoint two members
from the Anglo-Indian community if he believes that community is under-represented.
See Figure~\ref{fig:ind_house_otp-basic_info} for a summary of the Lok Sabha's electoral system.
In the 2019 elections, approximately 910 million of Indian citizens were eligible to vote, with an average of 
$1.679\pm0.317$ million voters per district.
There is a big number of active political parties in India, however in 2019 only 9 parties obtained
more than 2\% of votes \cite{indiaGeneralElection}. The same year election in the constituency
of Vellore in Tamil Nadu was countermanded with president's approval due to alleged use of
money to abuse the election process. We use the 2019 election data in our simulations and therefore
we also exclude Vellore.

For the simulation of elections to the India's House of the People, a network is constructed consisting of 
$10^5$ nodes divided into 542 topological communities, analogous to real electoral 
districts (excluding Vellore). The relative sizes of these communities are based on the actual populations of voters 
in the districts. Each community is assigned a geographical location, corresponding to the 
coordinates of the district's capital (see panels~(b) and~(c) of 
Figure~\ref{fig:ind_house_otp-basic_info}). The connectivity between nodes decreases with the 
geographical distance between them, as determined by fitting [\ref{eqn:planar_affinity}] to 
the empirical data on work commuting distance \cite{censusindiaHomeGovernment}, as shown in 
Figure~\ref{fig:ind_house_otp-basic_info}~(d). The simulations are conducted for a scenario
involving $n=9$ parties, an average degree of $k=12$, and various noise rates $\varepsilon$.
The supplementary datasets and the configuration files in our online code repository
\cite{github_repo} contain the exact data used in the simulations.

In addition to the actual electoral system employed in elections to the lower house of
the India's parliament (labeled as ,,542 dist. PV*'' in the figures), we also explore four
alternative systems that are either applicable to India or hold theoretical significance.
The first alternative system (,,132 dist. PV'') is a PV system with a division into 132 multi-member
constituencies corresponding to India's divisions, i.e. administrative units bigger than districts
and smaller than states. There are 102 divisions in India -- 10 states and 5 union territories
are not divided into divisions. However, for a more homogeneous distribution of district magnitude,
we divide bigger states according to the geographical proximity of districts resulting
in 132 constituencies in total in the alternative system. Under ,,132 dist. PV''
system all seats in a constituency are assigned to the party with the best score
(effectively it is a PBV system).
Then, we consider a PR system with an identical division of the country and
Jefferson method of seat assignment (,,132 dist. PR'').
Additionally, we consider two more PR systems with electoral districts corresponding to 
the 28 states and 8 union territories of India (,,36 dist. PR'') and with one
nation-wide district (,,1 dist. PR'').
We merge all single-member districts of the original system within the same 
alternative constituency, accumulating the seats into a single pool. 
The alternative systems do not include an entry threshold, and
the PR alternative systems use the Jefferson seat assignment method.

\begin{figure}
\centering
\begin{subfigure}{0.66\textwidth}
\centering
    \includegraphics[scale=0.67]{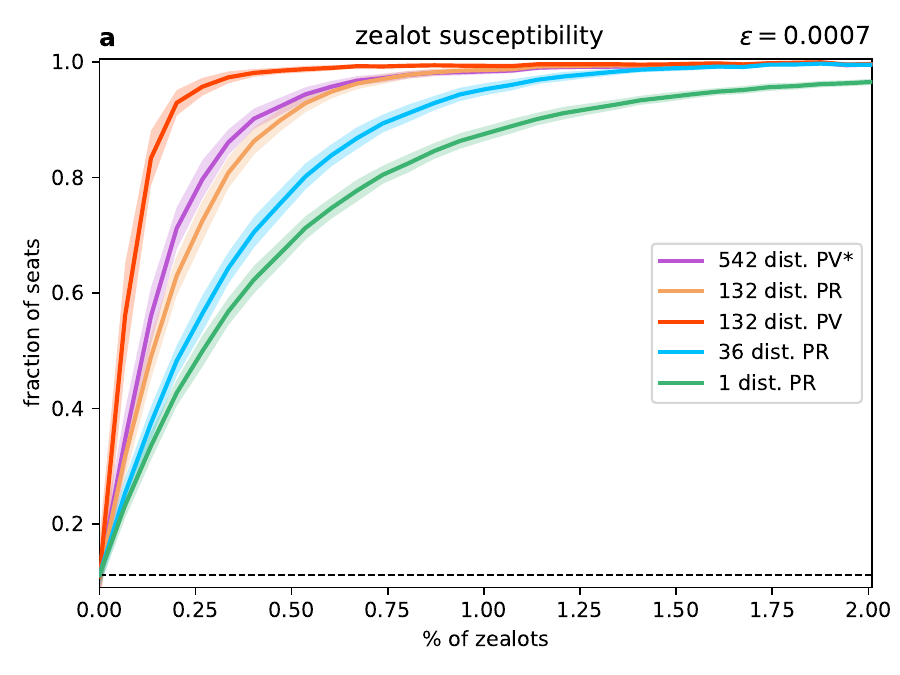}
\end{subfigure}\hfill
\begin{subfigure}{0.34\textwidth}
    \includegraphics[scale=0.65]{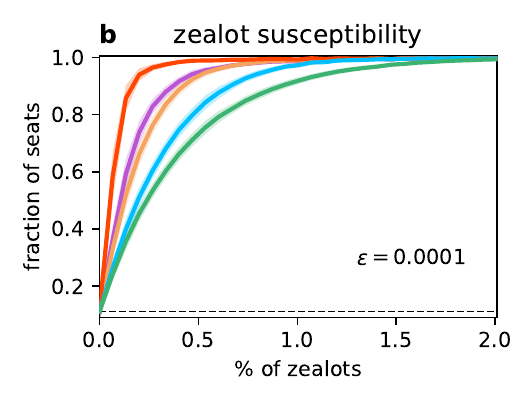}
    \includegraphics[scale=0.65]{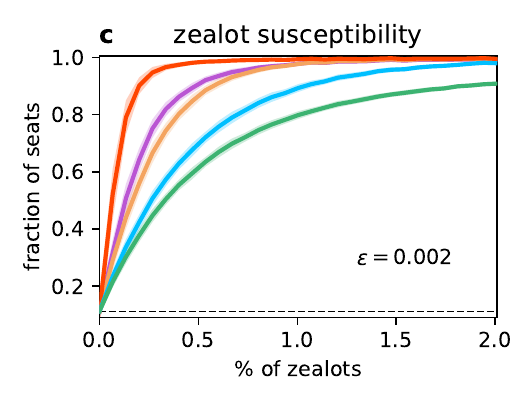}
\end{subfigure}
\begin{subfigure}{\textwidth}
    \includegraphics[scale=0.65]{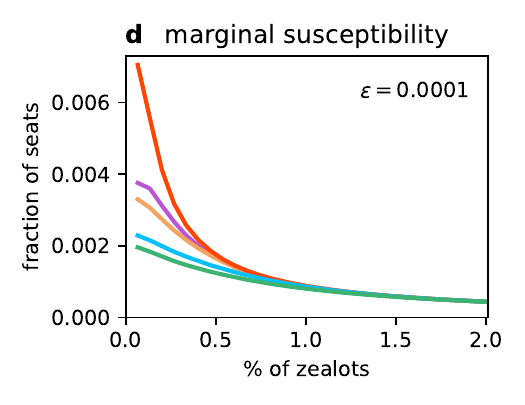}
    \includegraphics[scale=0.65]{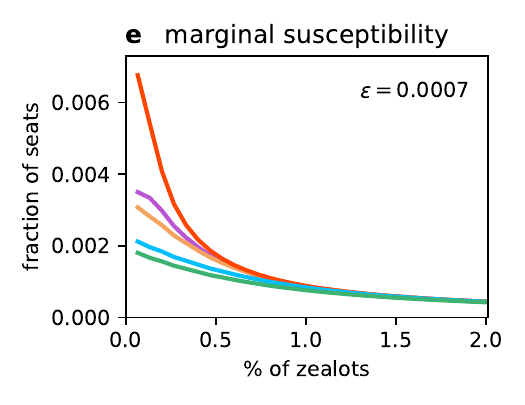}
    \includegraphics[scale=0.65]{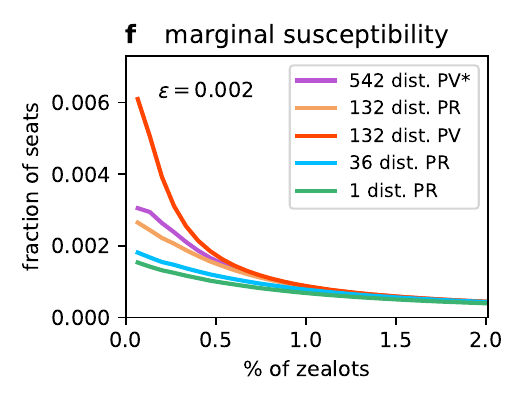}
\end{subfigure}
\begin{subfigure}{\textwidth}
    \includegraphics[scale=0.65]{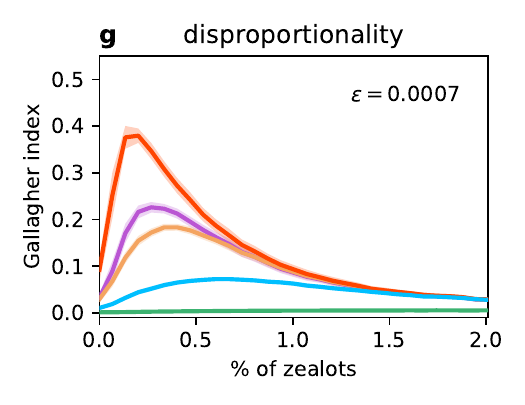}
    \includegraphics[scale=0.65]{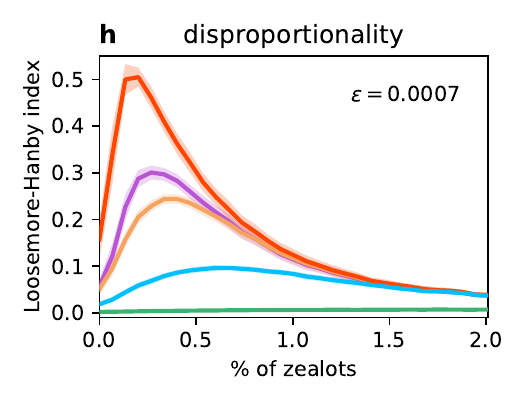}
    \includegraphics[scale=0.65]{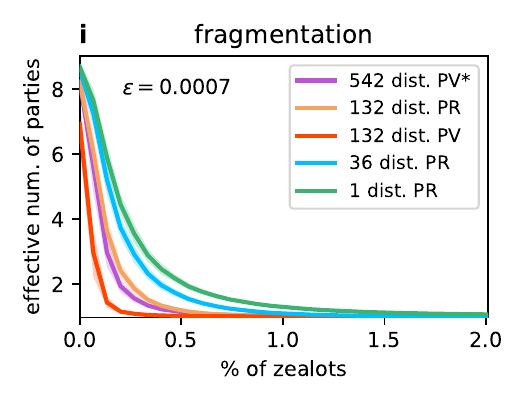}
\end{subfigure}
\caption[Zealot susceptibility of elections to the India's House of the People]{\textbf{Zealot susceptibility of elections to the India's House of the People.}
All plots compare results for the actual electoral system used in elections to the House of the People
(marked with an asterisk) with 4 alternative systems with different numbers of districts.
See the main text for details on ES used.
(a, b, c)  The average result of the zealot party $a$ vs. percentage of zealots in the population.
The dashed line marks the average
fraction of seats obtained in the symmetrical case, i.e. 1/9.
(d, e, f) Marginal zealot susceptibility  obtained
by dividing the total change in the fraction of seats won by the number of zealots.
(g, h, i) Average values of disproportionality and political fragmentation indexes
vs. percentage of zealots in the population.
The colored shade indicates results within
one standard deviation from the average.
All results were obtained over
200 elections with 9 parties. The network has $10^5$ nodes, with the average degree
$k=12$, divided into 542 communities corresponding to the actual electoral districts
(see Figure~\ref{fig:pl_sejm-basic_info}).
There is 542 seats to gain in total.
The noise rate $\varepsilon$ is given in each panel.
}
\label{fig:ind_house_otp-zealot_sus}
\end{figure}

In Figure~\ref{fig:ind_house_otp-zealot_sus} we present an analysis of zealot susceptibility across 
all electoral systems considered for the House of the People. Panels (a) to (c) depict the average fraction 
of seats won by the zealot party in relation to the percentage of zealots in the population. 
In general, there is a clear divergence in zealot susceptibility between proportional 
representation and plurality voting systems. However, the three PR 
systems exhibit distinct responses to zealots. The system with one constituency is the least 
susceptible to zealots, followed by the one with 36 constituencies, and the most susceptible 
system is the one with 132 constituencies. The differences in the average seat share are 
significantly larger than one standard deviation. 
The PR system with 132 constituencies (,,132 dist. PR'') is surprisingly similar, in terms
of zealot susceptibility, to the real system (,,542 dist. PV*'').
Examining the marginal susceptibility in panels (d) to (f) of Figure~\ref{fig:ind_house_otp-zealot_sus},
we observe that the influence exerted by one zealot is higher in PV systems,
particularly at the initial stages where the marginal susceptibility
can be over three times greater than in PR systems. Those observations 
hold true for any noise rate $\varepsilon$, although higher noise levels lead to reduced zealot 
influence overall.

\begin{figure}
\centering
\begin{subfigure}{0.66\textwidth}
\centering
    \includegraphics[scale=0.67]{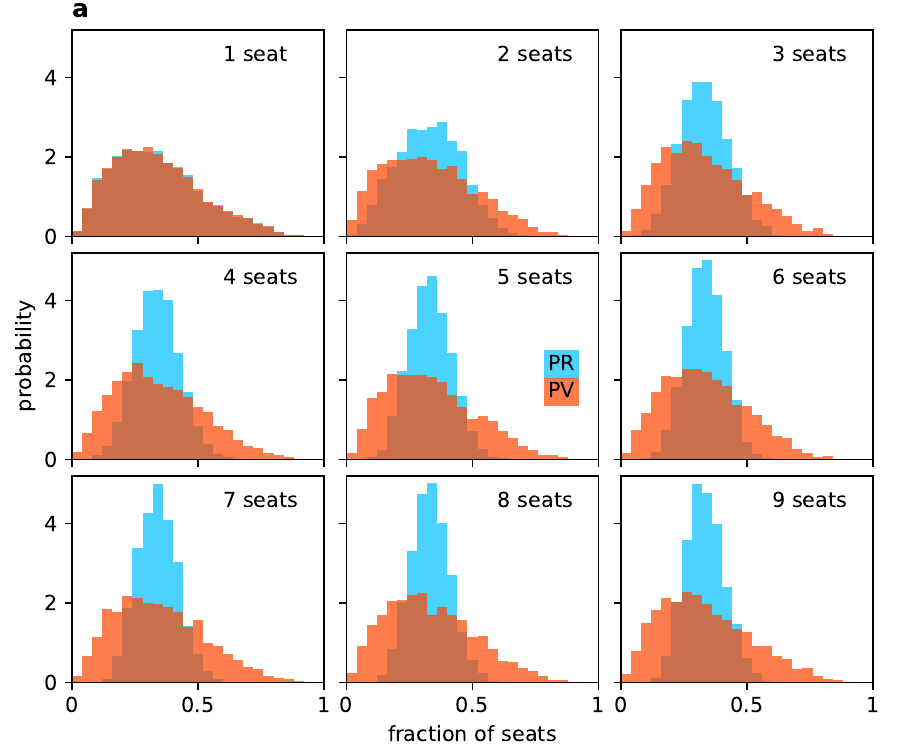}\hfill
\end{subfigure}\hfill
\begin{subfigure}{0.33\textwidth}
    \includegraphics[scale=0.65]{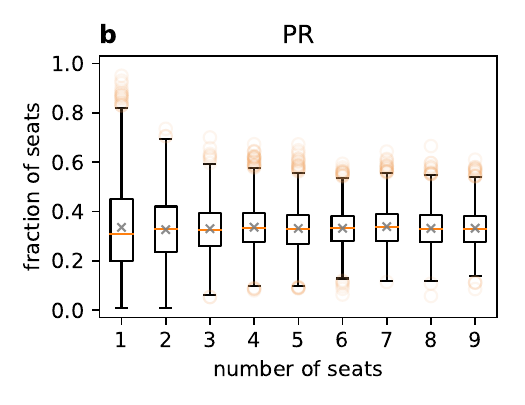}
    \includegraphics[scale=0.65]{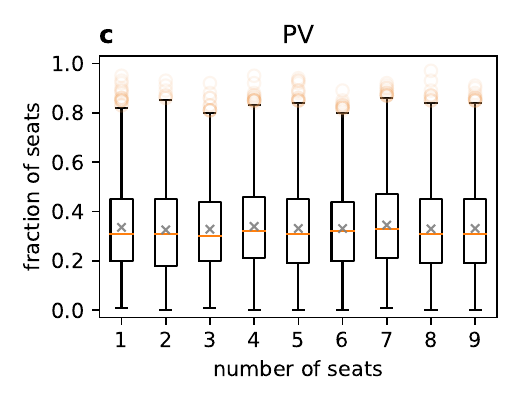}
\end{subfigure}\hfill
\begin{subfigure}{1\textwidth}
\centering
    \includegraphics[scale=0.65]{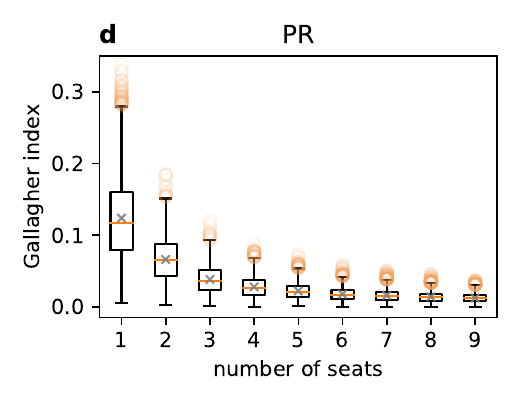}
    \includegraphics[scale=0.65]{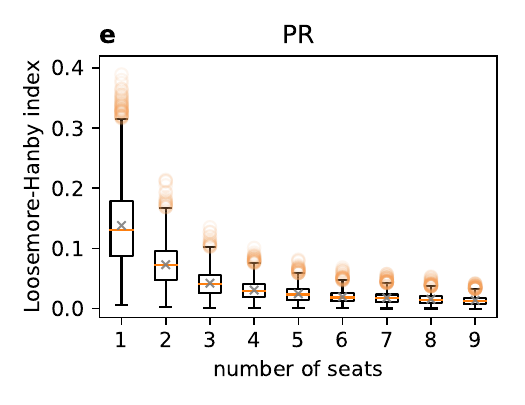}
    \includegraphics[scale=0.65]{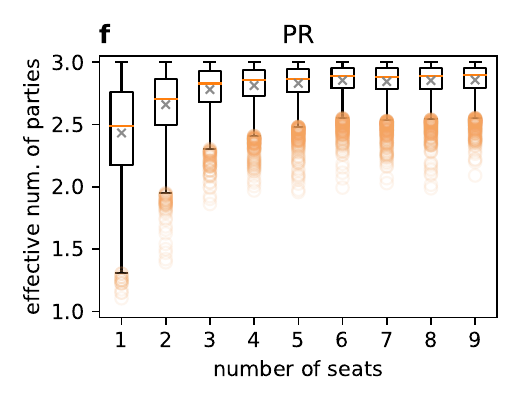}
    \includegraphics[scale=0.65]{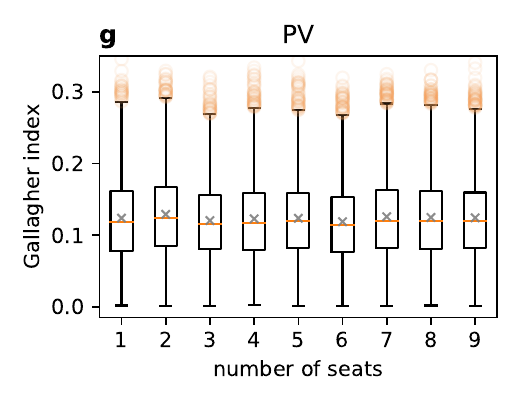}
    \includegraphics[scale=0.65]{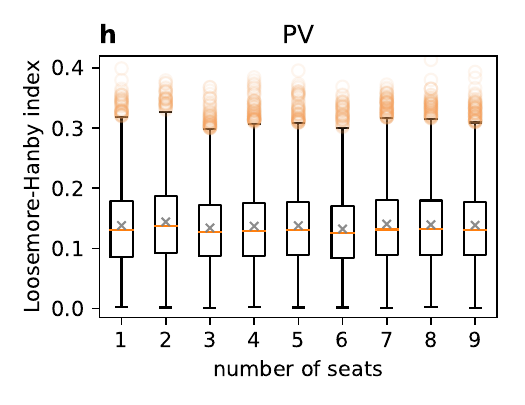}
    \includegraphics[scale=0.65]{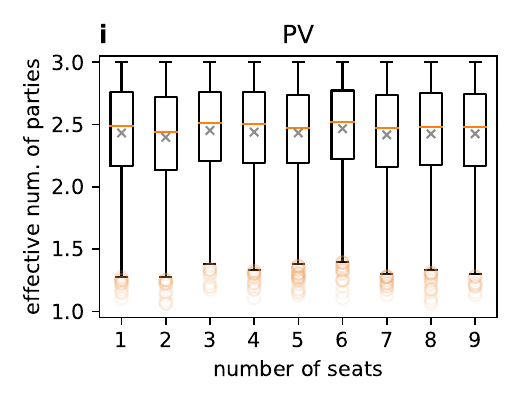}
\end{subfigure}
\caption[Comparison between PR and PV for different numbers of seats per district]{\textbf{Comparison between PR and PV for different numbers of seats per district.}
(a) Histograms of the seat share distribution of one of three parties when using
proportional representation (PR, blue) and plurality voting (PV, red) for varying number of seats.
The number of seats per one district $qs$ changes from 1 to 9, therefore the total number of seats 
changes from 100 to 900.
Note how the the variance of the distribution decreases for PR with increasing number of seats,
but stays the same for PV.
For one seat per district both ES become effectively the same FPTP system.
(b, c) Box plots of the same results highlighting the differences in the distribution.
(d, e, f, g, h, i) Box plots of disproportionality and political fragmentation indexes for both ES.
The average value in box plots is indicated by an \textit{x}, the median by an orange line,
the width of the box is given by the first and third quartile,
the whiskers indicate values within 1.5 times the IQR,
and outliers are marked by a circle.
Other simulation parameters are the same as in Figure~\ref{fig:basic-es_histograms}:
$N=10^4$, $q=100$, $k=12$, $r=0.002$, $\varepsilon=0.005$, and $n=3$ parties, results obtained over 4000 elections.
}
\label{fig:hist-box_seats_per_dist}
\end{figure}

In panels (g) to (i) of Figure~\ref{fig:ind_house_otp-zealot_sus} we present the Gallagher index, the 
Loosemore-Hanby index, and the effective number of parties. Again, under PV systems, elections can 
become highly disproportional even with a minuscule percentage of zealots in the population.
Surprisingly, it also the case for the PR system with 132 districts (,,132 dist. PR'').
The maximum level of disproportionality in PV systems is reached for 0.2-0.3\% of zealots 
for $\varepsilon=0.0007$, and for approximately 0.4\% for ,,132 dist. PR'' system.
As expected, an increase in the number of zealots leads to a decrease 
in political fragmentation, with PV systems generally exhibiting lower levels of fragmentation.
The discrepancy between PV and PR in the effective number of parties, however, is much smaller
than for Israel.
In particular, the actual ES of Lok Sabha (,,542 dist. PV*'') has a very similar scoring 
across all measures to the PR system with 132 constituencies (,,132 dist. PR'').
Zealot susceptibility, disproportionality, and fragmentation of these two systems are in many cases 
within a single standard deviation from each other. This is partially an effect of
relatively small district magnitude for 132 constituencies -- on average each district
has only 4 seats assigned in this case.
As shown in Figure~\ref{fig:hist-box_seats_per_dist},
for lower district magnitude PR systems become similar to FPTP.

\begin{figure}
\centering
\begin{subfigure}{0.66\textwidth}
\centering
    \includegraphics[scale=0.67]{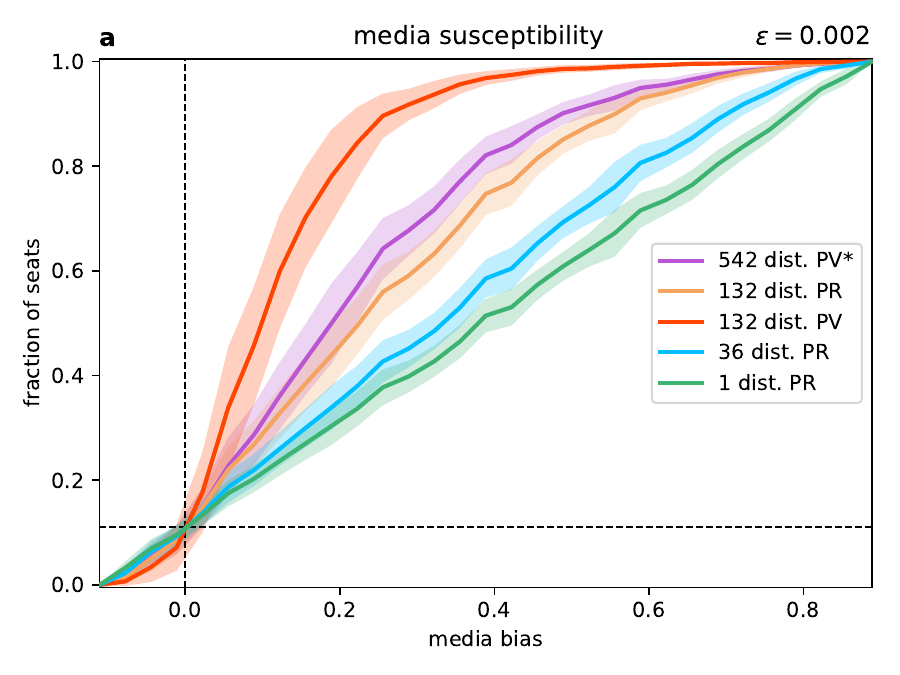}
\end{subfigure}\hfill
\begin{subfigure}{0.34\textwidth}
    \includegraphics[scale=0.65]{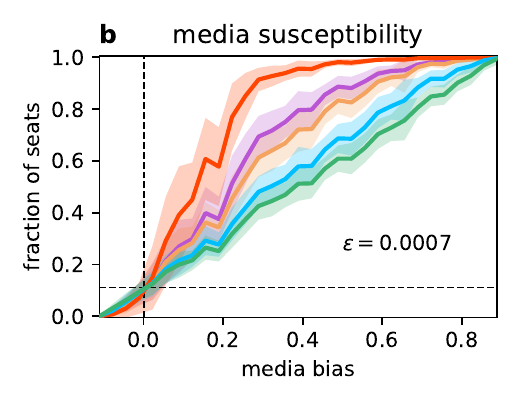}
    \includegraphics[scale=0.65]{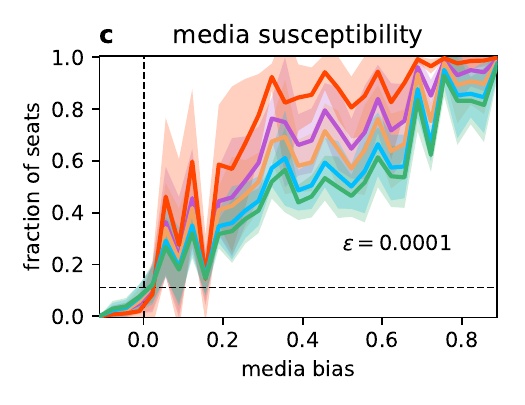}
\end{subfigure}
\begin{subfigure}{\textwidth}
    \includegraphics[scale=0.65]{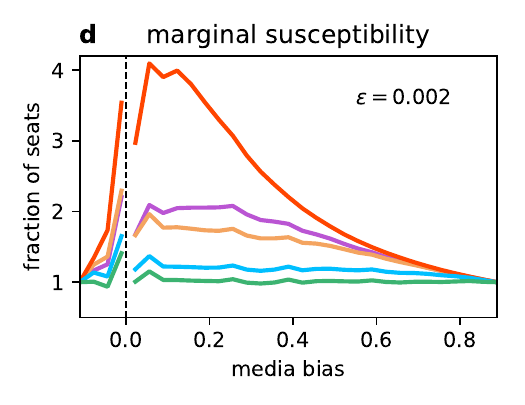}
    \includegraphics[scale=0.65]{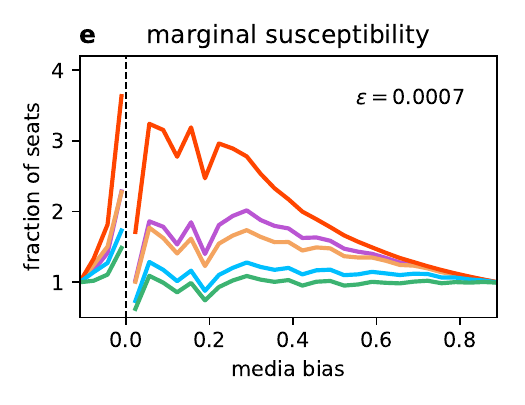}
    \includegraphics[scale=0.65]{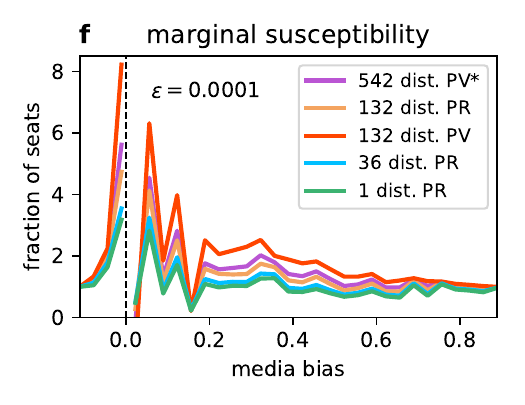}
\end{subfigure}
\begin{subfigure}{\textwidth}
    \includegraphics[scale=0.65]{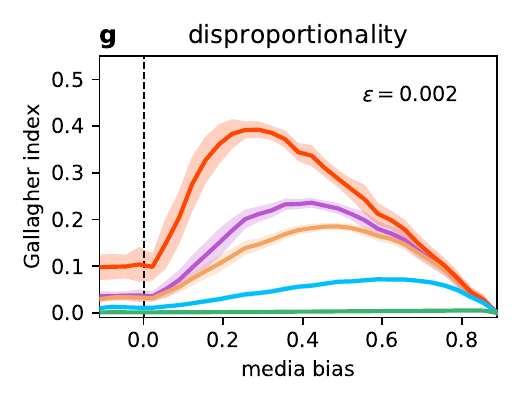}
    \includegraphics[scale=0.65]{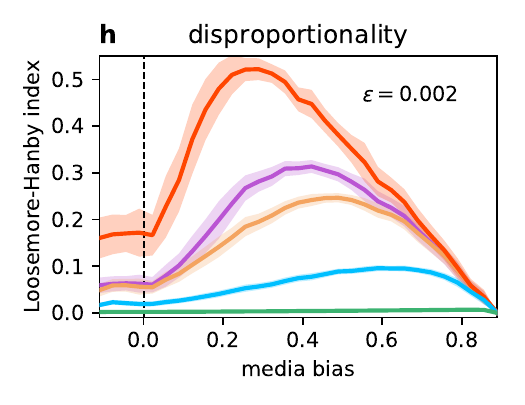}
    \includegraphics[scale=0.65]{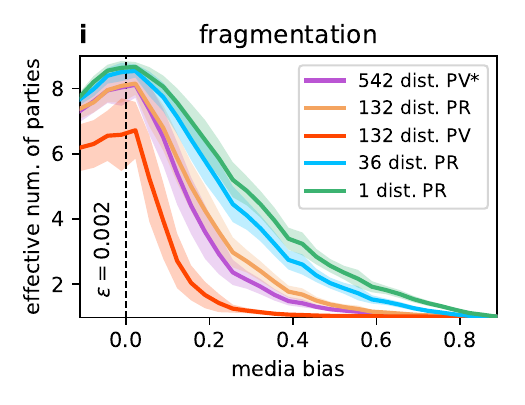}
\end{subfigure}
\caption[Media susceptibility of elections to the India's House of the People]{\textbf{Media susceptibility of elections to the India's House of the People.}
All plots compare results for the actual electoral system used in elections to the House of the People
(marked with an asterisk) with 4 alternative systems with different numbers of districts.
See the main text for details on ES used.
(a, b, c) The average result of the media party $a$ vs. media bias.
The horizontal dashed line marks the average
fraction of seats obtained in the symmetrical case, i.e. 1/9, and the vertical one the case with no bias.
(d, e, f) Marginal media susceptibility  obtained
by dividing the total change in the fraction of seats won by the strength of bias.
(g, h, i) Average values of disproportionality and political fragmentation indexes
vs. media bias.
The colored shade indicates results within
one standard deviation from the average.
All results were obtained over
500 elections with 9 parties. The network has $10^5$ nodes, with the average degree
$k=12$, divided into 542 communities corresponding to the actual electoral districts
(see Figure~\ref{fig:ind_house_otp-basic_info}).
There is 542 seats to gain in total.
The noise rate $\varepsilon$ is given in each panel.
}
\label{fig:ind_house_otp-media_sus}
\end{figure}

Figure~\ref{fig:ind_house_otp-media_sus} shows the susceptibility to media propaganda as a function of 
bias magnitude. As one might predict, PV systems tend to be more susceptible to biased media compared 
to PR systems. But in contrast to the zealot susceptibility,
PR systems exhibit more comparable levels of media susceptibility to each other.
In panels (a) to (c), we can see that lower noise rates 
result in reduced media susceptibility, however the general pattern becomes highly irregular. 
Analogously, the marginal susceptibility displayed in panels (d) to (f) is lower (especially for
PV systems) and more random for a smaller $\varepsilon$. 
In PV systems, media bias seems to give the 
highest returns after already exerting some influence, as indicated by the 
shift of the maximal marginal susceptibility to the right from zero bias.

Media propaganda causes the highest level of disproportionality for 
intermediate levels of influence, just like the zealots (see panels (g) to (i) of 
Figure~\ref{fig:ind_house_otp-media_sus}). 
Maximal disproportionality arises slightly earlier in PV systems compared to PR systems. The discrepancy 
between the two types of ES is more pronounced in terms of proportionality rather than political 
fragmentation. PR systems maintain higher level of the effective number of 
parties, but the difference is not as pronounced as in the case of Israel.
The similarity between the actual ES and the PR system with 132 districts is again striking,
with the PR system only marginally less susceptible to media propaganda,
more proportional, and more fragmented.

As in many countries, there is a discussion in India about possible improvements
of the electoral system, in particular the one used in the House of the People elections, possibly
transforming it into a proportional representation \cite{misra2018shift}.
If one of the goals were minimizing potential threats to democratic processes,
it would be more advisable to create 36 multi-member constituencies based
on the states, rather than 132 constituencies based on the divisions.
It would also decrease the disproportionality further, although at the cost of a higher
political fragmentation. Finally, the lowest
susceptibility to zealots and media propaganda is achievable for one nation-wide district,
however it could yield other issues in the biggest democracy in the world.


\FloatBarrier



